\documentclass[review]{elsarticle}

\usepackage{hyperref}
\usepackage{subfigure}
\usepackage{color,soul} % highlight
\usepackage{siunitx}
\usepackage{amsmath, amssymb}
\usepackage{cases}
\usepackage{nccmath} % small fraction below fraction
\usepackage{stackengine} % stack figures without label a, b, c..

\newcommand{\mosize}{0.32}
\newcommand{\mosizeb}{0.18}
\newcommand{\plasize}{0.58}
\newcommand{\plasizec}{0.5}

\journal{Advances in Space Research}

\biboptions{authoryear}

\begin{document}

\begin{frontmatter}

\title{Packaging of Thick Membranes using a Multi-Spiral Folding Approach: Flat and Curved Surfaces}

\author[A]{Victor Parque\corref{mycorrespondingauthor}}
\cortext[mycorrespondingauthor]{Corresponding author}
\ead{parque@aoni.waseda.jp}

\author[A]{Wataru Suzaki}
\author[A]{Satoshi Miura}
\author[C]{Ayako Torisaka}
\author[A]{Tomoyuki Miyashita}
\author[B]{Natori Michihiro}

\address[A]{Department of Modern Mechanical Engineering, Waseda University, Tokyo, 169-8555, Japan}
\address[C]{Department of Aeronautics and Astronautics, Tokyo Metropolitan University, Tokyo, 191-0065, Japan}
\address[B]{Institute of Space and Astronautical Science (ISAS)/JAXA (retired), Kanagawa, 252-5210, Japan}

\begin{abstract}
Elucidating versatile configurations of spiral folding, and investigating the deployment performance is of relevant interest to extend the applicability of deployable membranes towards large-scale and functional configurations.

In this paper we propose new schemes to package flat and curved membranes of finite thickness by using multiple spirals, whose governing equations render folding lines by juxtaposing spirals and by accommodating membrane thickness. Our experiments using a set of topologically distinct flat and curved membranes deployed by tensile forces applied in the radial and circumferential directions have shown that (1) the multi-spiral approach with prismatic folding lines offered the improved deployment performance, and (2) the deployment of curved surfaces progresses rapidly within a finite load domain. Furthermore, we confirmed the high efficiency of membranes folded by multi-spiral patterns.

From viewpoints of configuration and deployment performance, the multi-spiral approach is potential to extend the versatility and maneuverability of spiral folding mechanisms.
\end{abstract}


\end{frontmatter}

%\linenumbers

\section{Introduction}

Membrane packaging schemes with low weight storage and high deployment efficiency are significant to the operation and the development of large-scale solar sail structures \citep{state16}. As such, the study of foldable and deployable membrane structures has attracted the attention of the community owing much to low bending rigidity, high contractibility and recent materials with excellent tolerance for space environments. Generally speaking, since the packaging methods and the operational wrinkles have a direct effect on the membranes's deployment dynamics, the selection of the packaging schemes becomes relevant to attain the utmost deployment efficiency.

% Uniaxial
In line of the above, a number of methods realizing the deployable membrane models have been proposed in the last decades. Well-known methods for membrane packaging and folding usually involve some form of creasing and distribution of bends, such as the \emph{z-folding}, the \emph{wrapping} (rolling) and the \emph{fan-folding} approaches. In the above-mentioned schemes, deployment is usually performed by \emph{uniaxial} mechanisms, such as a telescopic boom; and the folding mechanism is not only adaptable to the thickness of the membrane, but also able to avoid the plastic deformations \citep{state16}. Examples include the rolling scheme used in the Hubble Space Telescope \citep{pelle} and in the Roll-Out Solar Array (ROSA) \citep{rosa}.

% zfolding: cite{biddy}

% Bi axial
To further compress the package efficiency of the above mentioned mechanisms, a number of alternative approaches have been proposed. For instance, the \emph{Miura-folding} scheme modifies the double \emph{z-folding} by skewing parallel fold lines \citep{miura}, in which empirical tests confirmed the deployment effectiveness \citep{nato85} and the linear elasticity of membranes \citep{pelle08}. Also, other approaches inspired by nature were proposed: the folding into \emph{polygonal membranes} with discretized curvature, such as the folding mechanisms observed on tree leaves \citep{smith02}, and the folding based on the \emph{frog-leg} pattern using triangular membranes, such as the ones found in the four-quadrant solar sails \citep{frog11,cube11}.

Also, being studied since the early 60's by \cite{huso}, an alternative folding mechanism is the \emph{spiral folding} approach, which consists of wrapping a membrane (flat sail) around a polygonal central hub. In this scheme, the folding lines are tangential to the core polygonal hub\citep{scheel74}, and byproducts such as the coordinates of the folding lines, crease lengths, folding angles, and hill-valley folds are usually computed by considering both the flat and the folded configuration of the membrane. In particular, the first spiral folding for a solar sail was proposed in the late 80's \citep{solar89}, and the approach was extended to consider variants with zero thickness \citep{pelle92,pelle01,nojima01a,nojima01b}, non-zero thickness \citep{pelle92,pelle01,spiral07,thick13,thick13b,wata,nato07,nojimavipsi,samu15}, curved membranes with spherical, conical and parabolic wrapping \citep{nojima01a,nojimavipsi}, and rotationally skew fold with offset leaves\citep{furuya12}. Hybrid schemes were also proposed, such as the simultaneous deployment of mast and sail by taking advantage from storing the strain energy stored into elastic spar members \citep{spiral07}.

It is also possible to combine folding schemes to enable the hybrid packaging and deployment in more than one dimension. For instance, the hybrid between the \emph{z-folding} and the \emph{wrapping} schemes \citep{cube08,biddy}, and the hybrid between the \emph{star folding} and the \emph{wrapping} schemes \citep{furu11}. Often, the crease patterns that consider the thickness of the membranes are computed by numerical methods, such as the \emph{spiral-wrapped square membrane} \citep{thick13b} and the \emph{curved beech leaf creases} \citep{thickbleech14}; and the folding lines are determined by theoretical analysis of the mechanical properties of creases subject to tensile forces \citep{furu13}.

%% added
In the approach proposed by \cite{thick13b}, leaves are shifted in offset position at each quadrant \citep{furuya12}, and point away from the centre of the polygon hub following the \emph{leaf-out} scheme \citep{smith02,kobayaashi98}), thus its folding pattern follows the \emph{z-fold} and \emph{wrap fold} to pack a flat sheet into a strip around a polygonal hub. In particular, the crease patterns are equally spaced and lie on a horizontal plane when wrapped around the central polygonal hub. Crease patterns are modeled to be initialized with an Archimedes' spiral for one round, then allowed the gradual increase in the radius of curvature to consider the thickness of the sheet and to balance the tension of outer layers with the compression of inner layers. Subsequently, \cite{thickbleech14} extended to creases that may not lie on the horizontal plane.

An alternative approach to folding membranes lies in the introduction of cuts and slits. Examples of such schemes include the spiral cuts of the membrane in 6-8 petals \citep{pelle04}, the division into manufacturable surfaces (gores) \citep{cut11}, and the square solar sail with separate parallel strips \citep{gre02}. In a recent study, cuts and slits were used to devise new packaging for membranes with finite thickness, and rigid and curved membranes \citep{arya17}.

% added: deployment methods
Along with research endeavours on folding, the study on membrane deployment strategies has attracted the attention of the community. As such, well-known deployment mechanisms of thin membrane and solar sails comprise the extendable masts at NASA\citep{tale05,spiral07}, the deployable booms and sails in Europe\citep{leipold12}, the deployable boom in NASA\citep{nanosaild,nanosaild2} and the centrifugal force in Japan\citep{kawa04,furu11}. Among these approaches the centrifugal force renders a spin-type deployment mechanism which can be reproduced by the rotation of actuators\citep{salama03,furu11,furuya12,hib20}. Often, the centrifugal force is accompanied by concepts of retraction with tensile forces, such as the ones proposed by \cite{lanford61}, \cite{pelle92} and \cite{manan14}. In particular, \cite{fururet12} and \cite{yasutaka14} used a retractive mechanism to allow the seamless folding of double corrugated areas observed in petals of skew membrane folds, and \cite{manan14} used a quasi-static approach in which two equal forces were applied at two opposing points along the periphery of the folded membrane.

% spiral and spin
From a practical perspective, the membranes with spiral folding are potential due to their suitability to realize spin-deployable mechanisms. Here, potential benefits over membranes deployed by rigid support structures, e.g. mast sails, include (1) the favorable mass to membrane surface area ratio allowing to render large accelerations, (2) the small angular momentum useful for attitude stabilization due to the disturbances generated from spin axis, and (3) the simplicity, flexibility and favorable deployment performance over the entire spin periods \citep{sola16,hib20,spiral09}. Indeed, one of the successful endeavours using the spin type deployment is the one used in the Interplanetary Kite-craft Accelerated by Radiation Of the Sun (IKAROS), the solar power sail developed by JAXA in Japan \citep{hirotaka,ikaros14}, whose shape is square and consists of one core and four independent trapezoidal membranes. Basically the approach used in IKAROS consists of \emph{bellow folding} all the sides of the membrane towards the core side, letting the remaining four parts be wrapped around the core. Whereas the \emph{bellow folding} deploys while extending, the \emph{spiral folding} deploys while rotating.

From an efficiency viewpoint, considering the thickness of membranes in the folding dynamics of membranes was shown significant to ensure the high efficiency of deployment and packaging \citep{satou,arya17}. As such, although the study of the thickness of membranes of rigid origami patterns \citep{thickorigami15,flatcurved,tachi} and of spiral folding schemes \citep{pelle92,pelle01,thick13,thick13b,wata,samu15,nato07} has attracted the attention of the community, the study of thick spiral folding patterns under higher degrees of versatility has remained elusive in the literature. Elucidating the governing equations of spiral folding schemes with improved versatility and with finite thickness considerations, and investigating the deployment performance, is potential to extend the applicability of the spin-type deployable membranes towards adaptive and functional configurations. In this paper, inspired by the approach of \cite{wata}, we extend the spiral folding scheme to consider the multi-spiral patterns in flat and curved membranes with finite thickness. Concretely speaking, our contributions are as follows:

\begin{itemize}
  \item the schemes to package flat and curved membranes of finite thickness by using folding with multiple spirals. The governing equations rendering the folding lines are determined by the juxtaposition of concentric spirals and by accommodating the folding lines to consider the membranes' thickness.
  \item the set of experiments using the tensile deployment based on force applied in the radial and circumferential directions on paper-based membranes have confirmed that (1) the multi-spiral with prismatic folding lines offered the improved deployment performance when compared to other spiral approaches, and (2) the deployment in curved surfaces progressed rapidly within a finite load domain.
\end{itemize}

In the next sections, we describe our proposed approach, our experiments and obtained insights.

\begin{figure}
%\hfill
\begin{center}
{\includegraphics[width=0.95\textwidth]{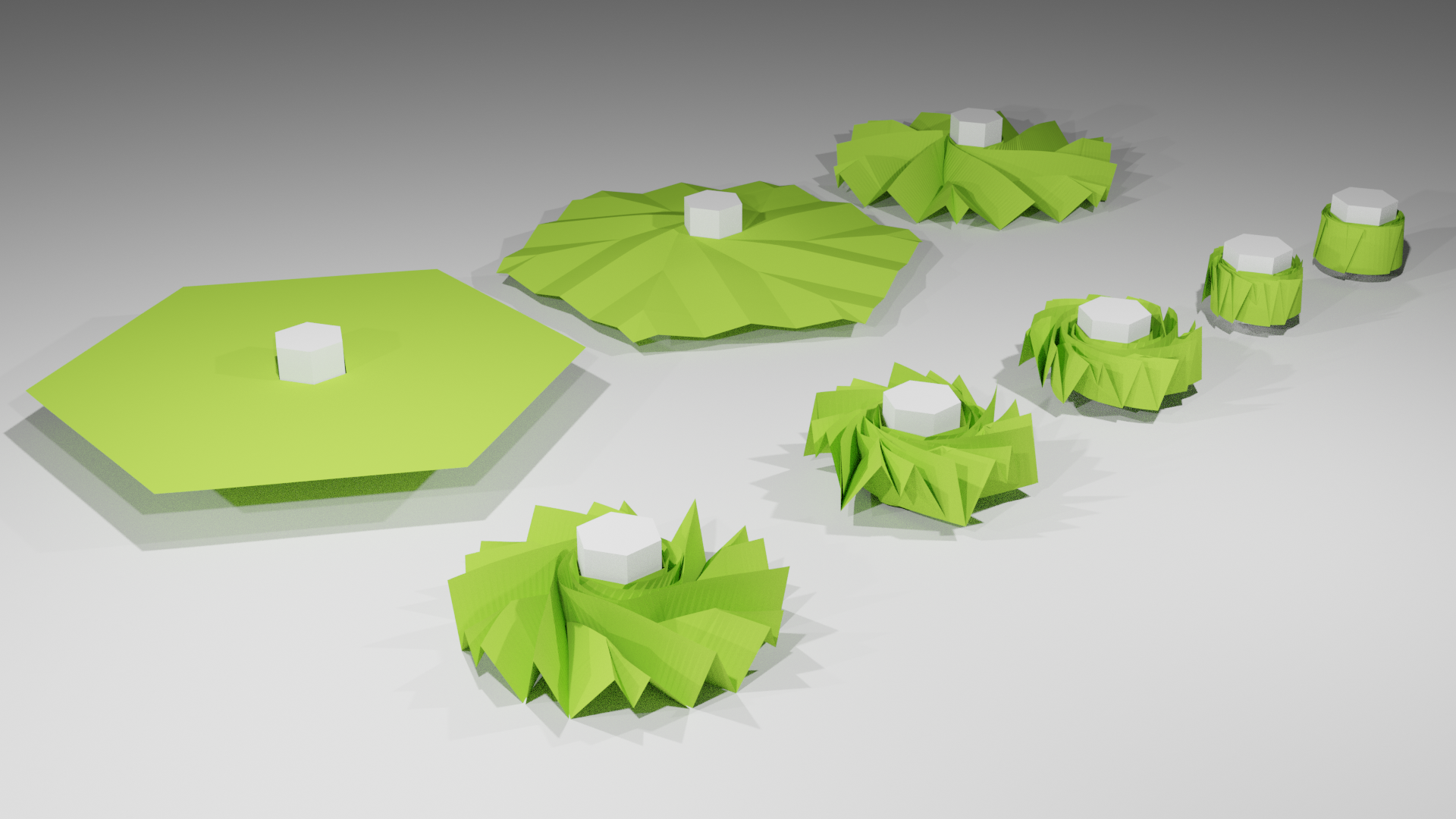}}
\end{center}
\caption{Basic idea of the folding pattern of a flat membrane with hexagonal boundaries.}
\label{foldscn}
\end{figure}

\section{Multi-Spiral Folding Pattern of Flat Membranes}

In this section, we describe the governing equations of the multi-spiral folding pattern on a flat membrane.

\begin{figure}[t]
%\hfill
\begin{center}
{\includegraphics[width=1\textwidth]{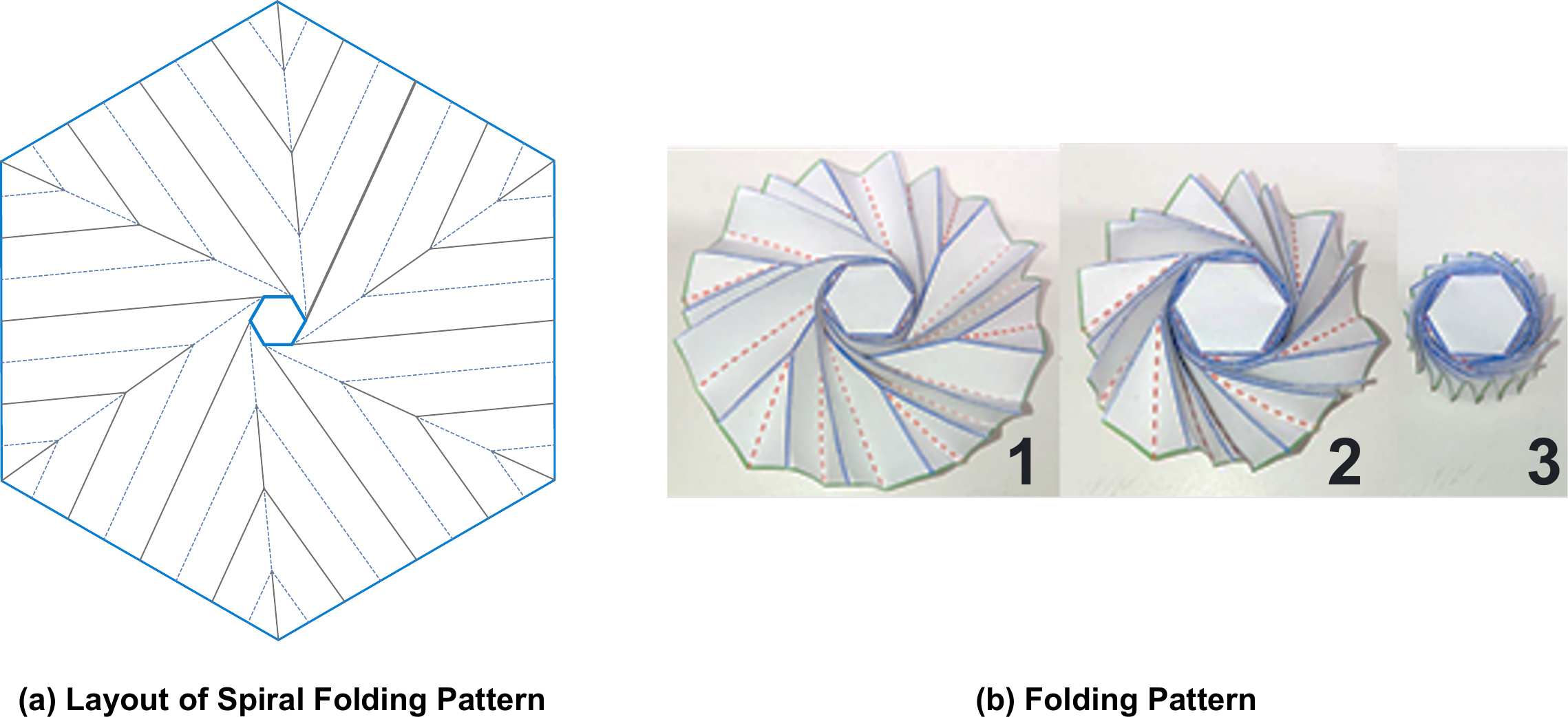}}
\end{center}
\caption{Layout of the spiral folding pattern on a \emph{flat membrane} and its folding process.}
\label{flatspiral}
\end{figure}

\begin{figure}[ht!]
%\hfill
\begin{center}
{\includegraphics[width=0.55\textwidth]{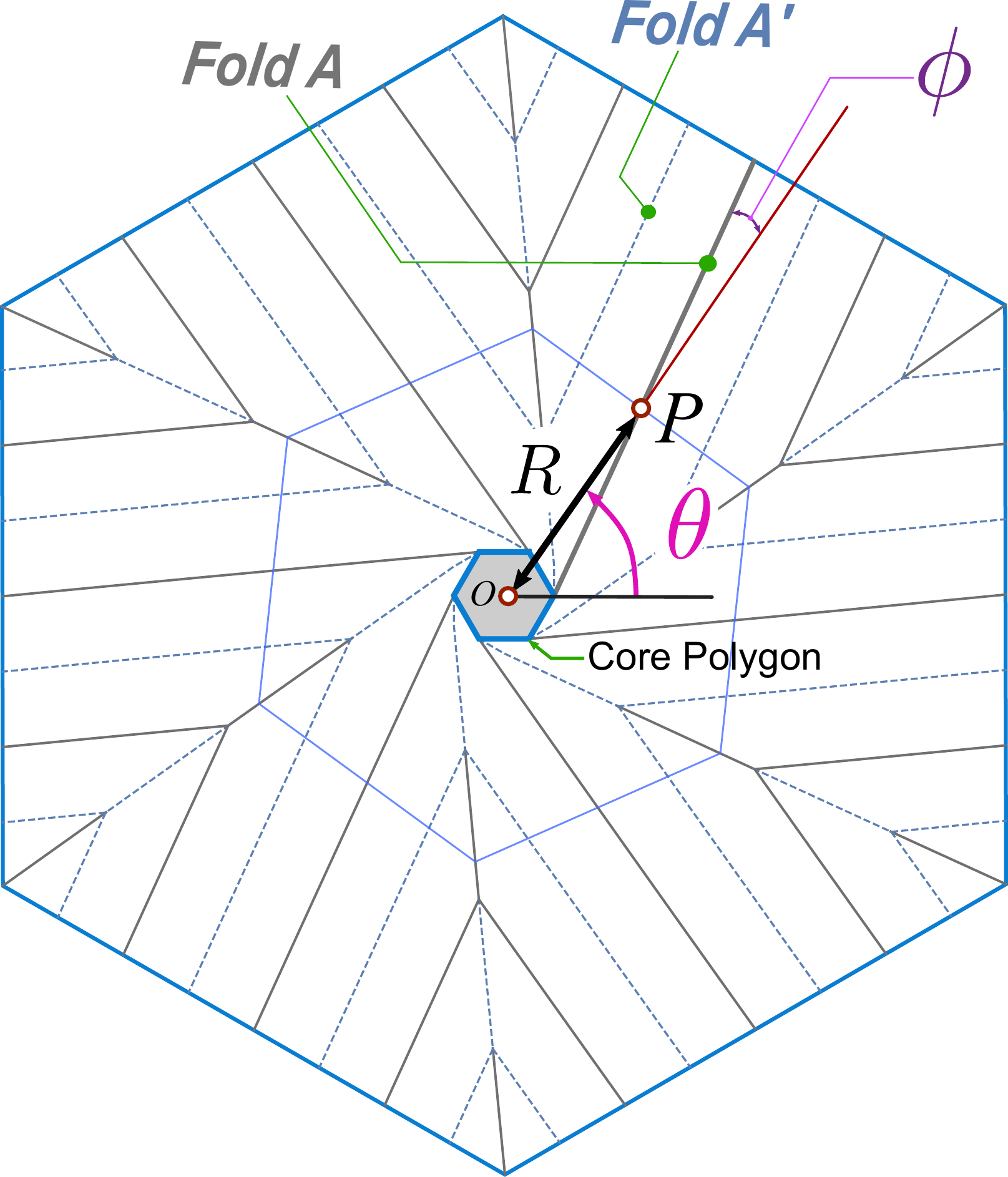}}
\end{center}
\caption{Main elements in the spiral folding pattern. The center of the polygon hub is located at $O$,  \emph{Fold A} is tangential to edges of the polygon hub, $P$ is a point in \emph{Fold A}, $R$ is the length of segment $\overline{OP}$, $\theta$ is the angle between the $x$-axis and the segment $\overline{OP}$, and $\phi$ is the acute angle between the segment $\overline{OP}$ and \emph{Fold} $A$ at point $P$.}
\label{flatelem}
\end{figure}

\subsection{Preliminaries}

The spiral folding pattern wraps a flat membrane into a cylinder as basically portrayed by Fig. \ref{foldscn}. Here we describe the basic idea of the spiral folding pattern and its governing equations. Generally speaking, given a flat membrane and a regular polygon positioned in its core (or hub), creases on the flat membrane are tangential to the core polygon, and the fold behaviour transforms the flat membrane into a hollow cylinder having a regular polygon in its core. Along the transition process from flat to cylinder configuration, zigzag patterns are generated in the cylindrical shape which is due to the shape and configuration of the mountain and valleys of the membrane of Fig. \ref{foldscn}. An example of the layout structure of the spiral fold pattern is shown by Fig. \ref{flatspiral}(a), and an example of a wrapped configuration using paper is shown by Fig. \ref{flatspiral}(b).

\begin{figure}[ht]
\begin{center}
{\includegraphics[width=0.98\textwidth]{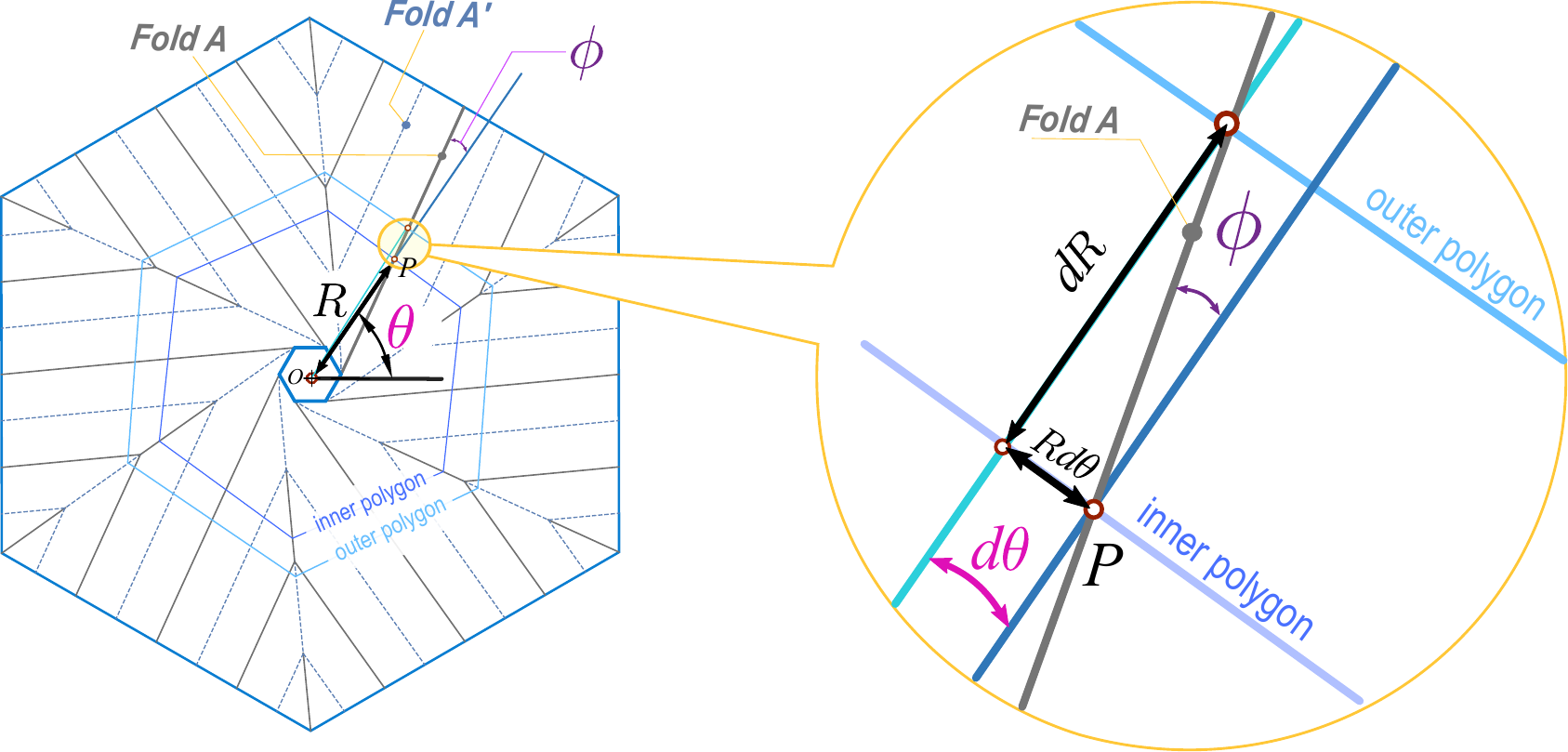}}
\end{center}
\caption{Infinitesimal region in polar coordinates. Relationship between $R$, $d \theta$, $dR$ and $\phi$. Variations in the rotation angle by $d\theta$ generates the arc length $Rd\theta$ and radius $R+dR$ between the \emph{inner} and \emph{outer polygon}.}
\label{figtanphi}
\end{figure}

In order to portray the governing equations of the spiral folding pattern, we first refer to Fig. \ref{flatelem}, which describes the key elements involved in the rendering of the crease geometry and the overall folding structure. By observing Fig. \ref{flatelem}, given a regular polygon with center at origin $O$, the crease \emph{Fold A} is tangential to the edge of the core polygon. Considering a regular polygon with $N$ sides, the spiral folding pattern is expected to have $N$ creases of the type \emph{Fold} $A$. Then, let $P$ be an arbitrary point in \emph{Fold A}, it is possible to describe an infinitesimal region bounded by an \emph{inner polygon} and an \emph{outer polygon} in polar coordinates, as shown by Fig. \ref{figtanphi}, which will aid in outlining the governing equations to render the geometry of \emph{Fold} $A$. Here, by looking at the relation among the infinitesimal distances $dR$ and $Rd\theta$, and the angle $\phi$ in Fig. \ref{figtanphi}, the following holds

\begin{equation}\label{tanphiplan}
\tan(\phi) = \frac{R d\theta}{dR}.
\end{equation}

The above expression is useful to find the angle $\theta$ by numerical integration of $\frac{d\theta}{dR}$ for a defined interval in $R \in [R_0, R_f]$. However, since the angle $\phi$ is expected to vary with $R$ as well, it becomes necessary to find further relationships between $\phi$ and $R$. In the following we find such relationships.

\begin{figure}[t]
%\hfill
\begin{center}
{\includegraphics[width=1\textwidth]{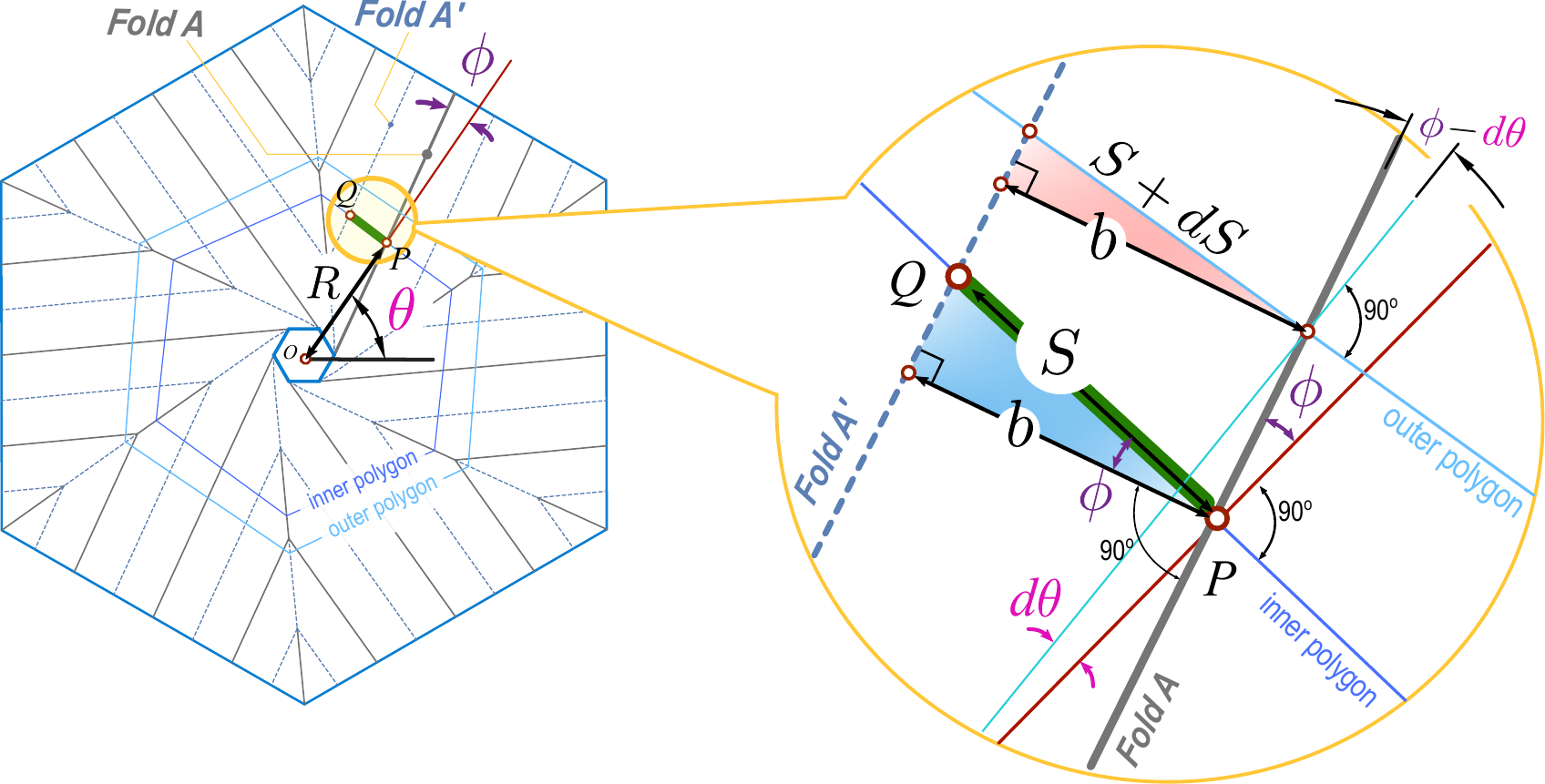}}
\end{center}
\caption{Main elements in the relationship among $R$, $S$, $b$, $\phi$ and $\theta$. Considering that \emph{Fold} $A'$ is consecutive to \emph{Fold} $A$, the segment $\overline{PQ}$ is prescribed in the periphery of the \emph{inner polygon}, where $\overline{PQ} \perp \overline{OP}$ and $Q$ is the intersection between \emph{Fold} $A'$ and the edge of the \emph{inner polygon}. For simplicity of further descriptions, $S$ is the length of the segment $\overline{PQ}$. $b$ is the distance between the creases \emph{Fold} $A$ and \emph{Fold} $A'$ and determines the height of the membrane in packed cylindrical configuration. The relationships between $b$, $S$, $\phi$ and $\theta$ are defined from the colored triangular regions.}
\label{mainflat}
\end{figure}

Let \emph{Fold} $A'$ be the adjacent crease to \emph{Fold} $A$ (as shown by Fig. \ref{flatelem}), it is possible to define an infinitesimal region between the \emph{inner polygon} and the \emph{outer polygon}, as shown by Fig. \ref{mainflat}, which will help in outlining the governing equations of the \emph{Fold} $A'$. In Fig. \ref{mainflat}, let $S$ be the length of the segment $\overline{PQ}$ and let $b$ be the breadth (distance) between \emph{Fold} $A$ and \emph{Fold} $A'$. From Fig. \ref{mainflat}, we can define the breadth $b$ by the triangles colored by blue and red. One of the relation is derived from the triangle in blue, as follows

\begin{equation}\label{bcos}
b = S\cos\phi.
\end{equation}

We can also compute the breadth $b$ from the red triangle in Fig. \ref{mainflat} as

\begin{equation}\label{bSdS}
b = (S + dS)\cos(\phi - d\theta).
\end{equation}

We can then combine Eq. \ref{bcos} and Eq. \ref{bSdS} to obtain the following estimation:

\begin{equation}\label{ds}
\frac{dS}{S} = -\tan(\phi) d\theta.
\end{equation}

The above expression is useful to find $S$ by numerical integration of $\frac{dS}{dR}$ in a defined interval of $R \in [R_0, R_f]$, and along Eq. \ref{tanphiplan} represents the system of differential equations being useful to render \emph{Fold} $A$ and \emph{Fold} $A'$.

Next, we study the relationship between the flat membrane and the folded configuration. A regular polygon with $N$ edges and apothem $R$ has area $kR^2$ and perimeter $2kR$, in which

\begin{equation}\label{kk}
k = N\tan \Big (\frac{\pi}{N} \Big ).
\end{equation}

\begin{figure}[ht]
%\hfill
\begin{center}
{\includegraphics[width=1\textwidth]{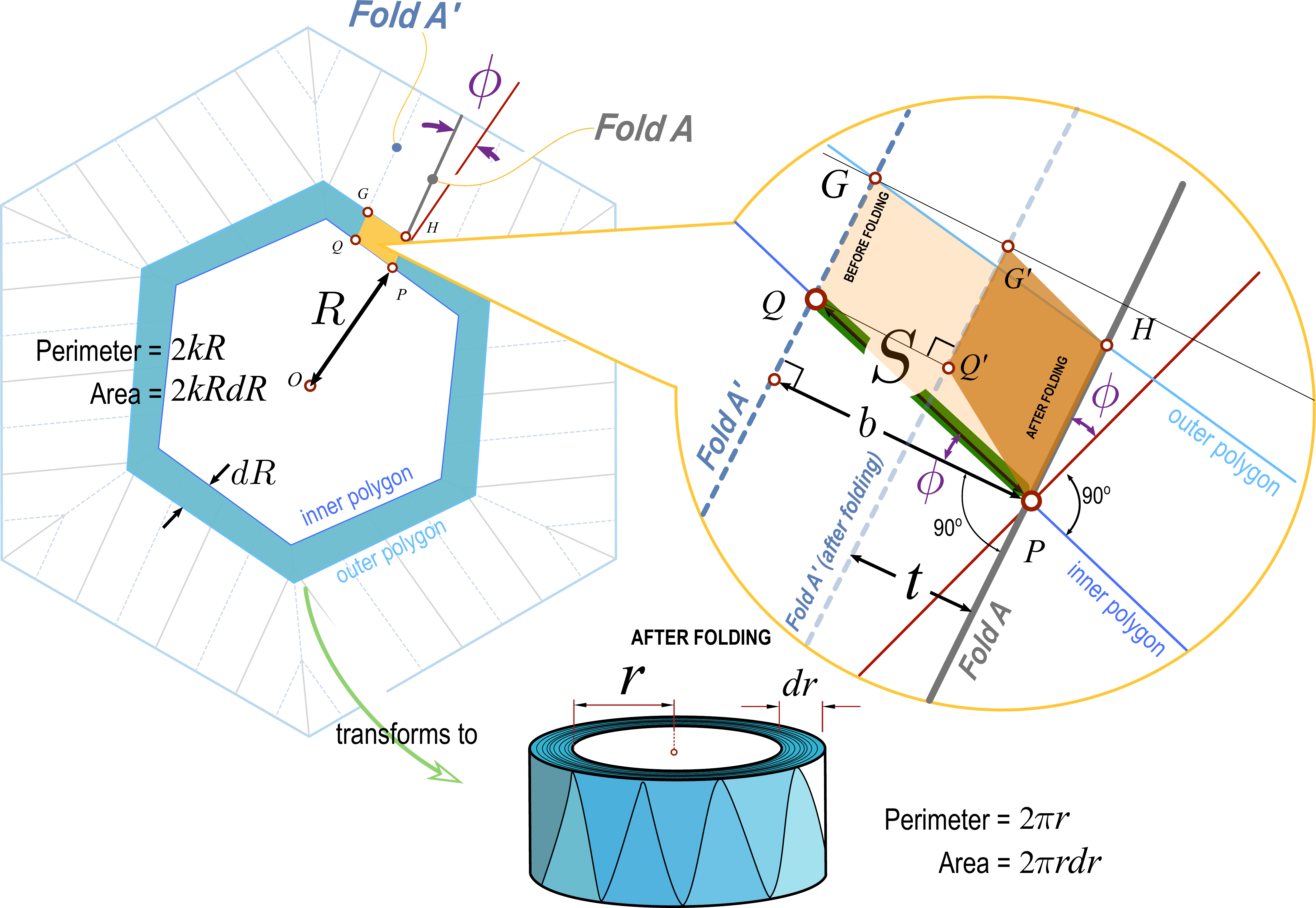}}
\end{center}
\caption{Crease geometry showing the relationship between the flat and folded configurations of the membrane. The infinitesimal region between the \emph{inner polygon} and the \emph{outer polygon} folds into a cylindrical shell with radius $r$ and thickness $dr$. The infinitesimal region $PQGH$ transforms into $PQ'G'H$ during the folding process, bounded by the thickness $t$ between folded membranes, in which $Q$ ($G$) transforms into $Q'$ ($G'$).}
\label{baflat}
\end{figure}

The above-mentioned constant $k$ is useful to describe the folding behaviour in terms of change of perimeter and area. To portray a glimpse of the folding behaviour between planar and cylindrical forms, Fig. \ref{baflat} shows the relations between flat and folded configurations of an infinitesimal polygonal region. By observing Fig. \ref{baflat}, it is possible to outline the principles relating change of lengths and areas between the flat and cylindrical configurations of the membrane. As such, assuming membrane thickness $t$, the perimeter of the cylindrical shell is $2\pi r$ whereas the perimeter of its corresponding flat polygonal region is $2kR$. Thus, from Fig. \ref{baflat}, since the infinitesimal region $PQGH$ transforms into $PQ'G'H$ during the folding process, in which $Q$ ($G$) transforms into $Q'$ ($G'$, the segment $\overline{PQ'}$ is related to the segment $\overline{PQ}$ by the similarity ratio of change of perimeters

\begin{equation}\label{pqratio}
\frac{{PQ'}}{{PQ}} = \frac{2\pi r}{2kR},
\end{equation}
such that the relation holds from Fig. \ref{baflat}

\begin{equation}\label{pqdash}
PQ' = \sqrt{(S \sin \phi)^2 + t^2}.
\end{equation}

Given that $S = PQ$, the above Eq. \ref{pqratio} and Eq. \ref{pqdash} can be arranged to be

\begin{equation}\label{phiplanar}
\frac{2\pi r}{2kR} = \sqrt{\sin^2\phi + \Big (\frac{t}{S} \Big )^2}.
\end{equation}

Similarly, the area of the upper face of the cylindrical shell in Fig. \ref{baflat} is $2\pi r dr$ whereas the area of its corresponding polygonal region is $2kRdR$. Thus, the corresponding areas are related by the similarity relation

\begin{equation}\label{pir2v1}
\frac{A_{PQ'G'H}}{A_{PQGH}} = \frac{2\pi rdr}{2kRdR},
\end{equation}
where
\begin{equation}\label{areafol}
A_{PQ'G'H} = (PH)t
\end{equation}
denotes the area of the region $PQ'G'H$ in Fig. \ref{baflat}, and

\begin{equation}\label{areaflat}
A_{PQGH} = (PH)S\cos \phi,
\end{equation}
denotes the area of the region $PQGH$ in Fig. \ref{baflat}.

The above relations (\ref{pir2v1}) to (\ref{areaflat}) can be arranged to be

\begin{equation}\label{phiplan2}
\frac{2\pi rdr}{2kRdR} = \frac{t}{S\cos \phi}.
\end{equation}

Finally, the governing equations outlining the geometry of \emph{Fold} $A$ are mainly determined by Eq. (\ref{tanphiplan}), Eq. (\ref{ds}), Eq. (\ref{kk}), Eq. (\ref{phiplanar}) and Eq. (\ref{phiplan2}). It is also possible to define the system of differential equations able to render \emph{Fold} $A$. As such, we can isolate $\phi$ from Eq. (\ref{phiplanar}) by

\begin{equation}\label{phifunR}
\phi = \mathrm{asin}\left( \frac{\sqrt{( \pi rS + kRt) ( \pi rS-kRt)}}{kRS} \right ),
\end{equation}
which can be used as a constraint to find the differential equations of $\theta$, $S$ and $r$ with respect to $R$. By replacing Eq. \ref{phifunR} into Eq. (\ref{ds}), Eq. (\ref{phiplanar}) and Eq. (\ref{phiplan2}) and accommodating terms we obtain

\begin{equation}\label{dqdr}
\frac{d\theta}{dR} = \displaystyle \frac{\sqrt{\left(\pi  rS  + kR t\right) \left(\pi  rS -kR t\right)}}{kSR^2 \sqrt{ \Big ( \frac{t}{S} \Big )^2 - \Big ( \frac{\pi r}{kR} \Big )^2 +1} },
\end{equation}

\begin{equation}\label{dSdR}
\frac{dS}{dR} = \displaystyle \frac{S \Big (  (kRt)^2 - (\pi rS)^2 \Big)}{ R \Big ( (kRt)^2 + (kRS)^2 - (\pi rS)^2 \Big) },
\end{equation}

\begin{equation}\label{drdR}
\frac{dr}{dR} = \displaystyle \frac{R k t}{\pi  rS  \sqrt{ \Big ( \frac{t}{S} \Big )^2 - \Big ( \frac{\pi r}{kR} \Big )^2 +1}  }.
\end{equation}

The above-mentioned system of differential equations Eq. \ref{dqdr} to Eq. \ref{drdR} can be solved numerically for given membrane parameters $k$ (Eq. \ref{kk}), thickness $t$, boundaries $[R_0, R_f]$, and initial conditions $\theta_0$, distance $S_0$ and $r_0 = R_0$. Here, $R_0$ denotes the circumradius of the polygon at the core of the flat membrane. Conversely, $R_f$ denotes the circumradius of the polygon at the outer boundary of the flat membrane, thus it represents the user-defined upper bound on the value of $R$.

\begin{figure*}[ht]
\centering
%\hfill
\subfigure[$S$]{\includegraphics[width=0.48\textwidth]{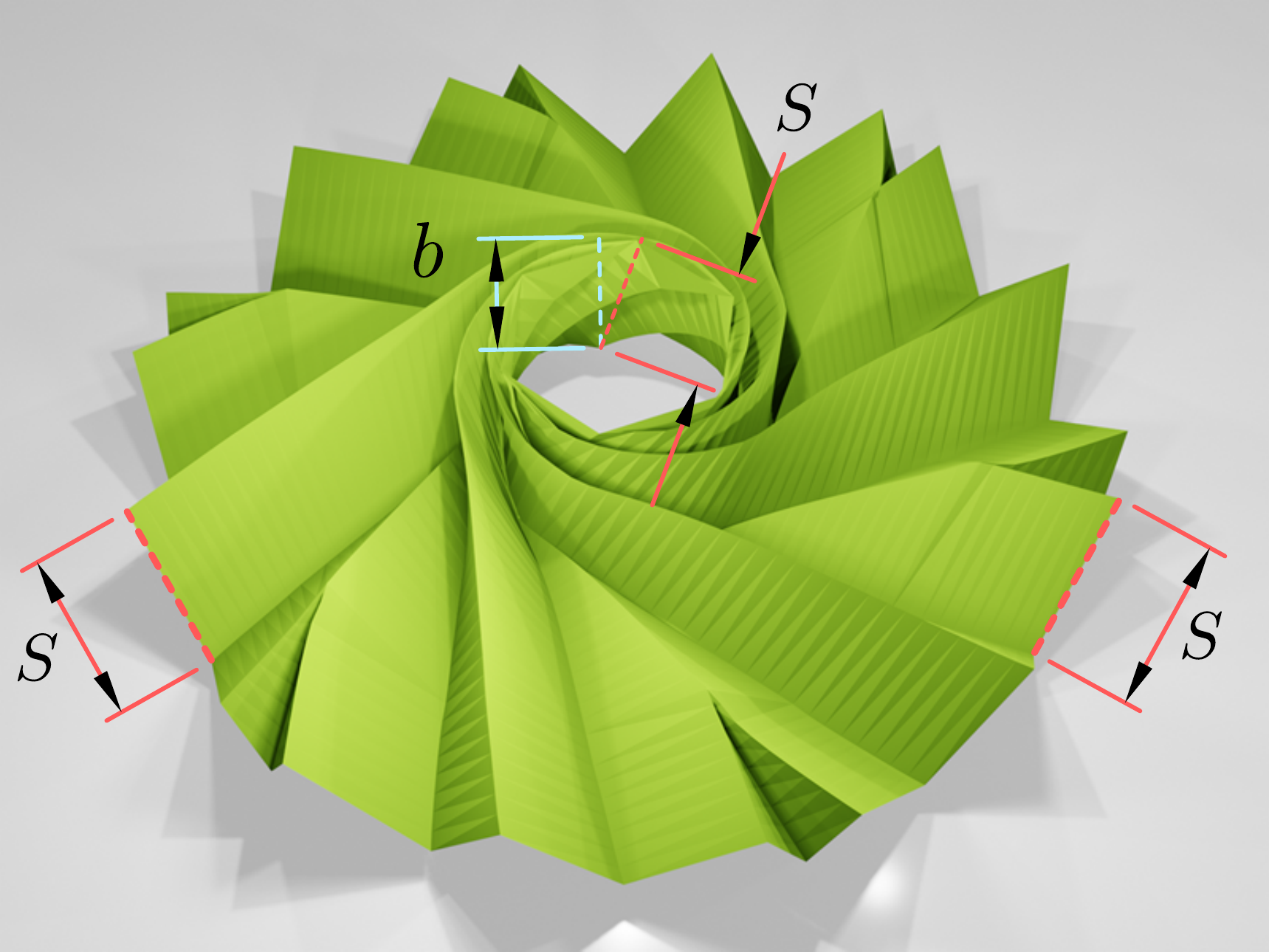}}
%\hfill
\subfigure[$b$]{\includegraphics[width=0.48\textwidth]{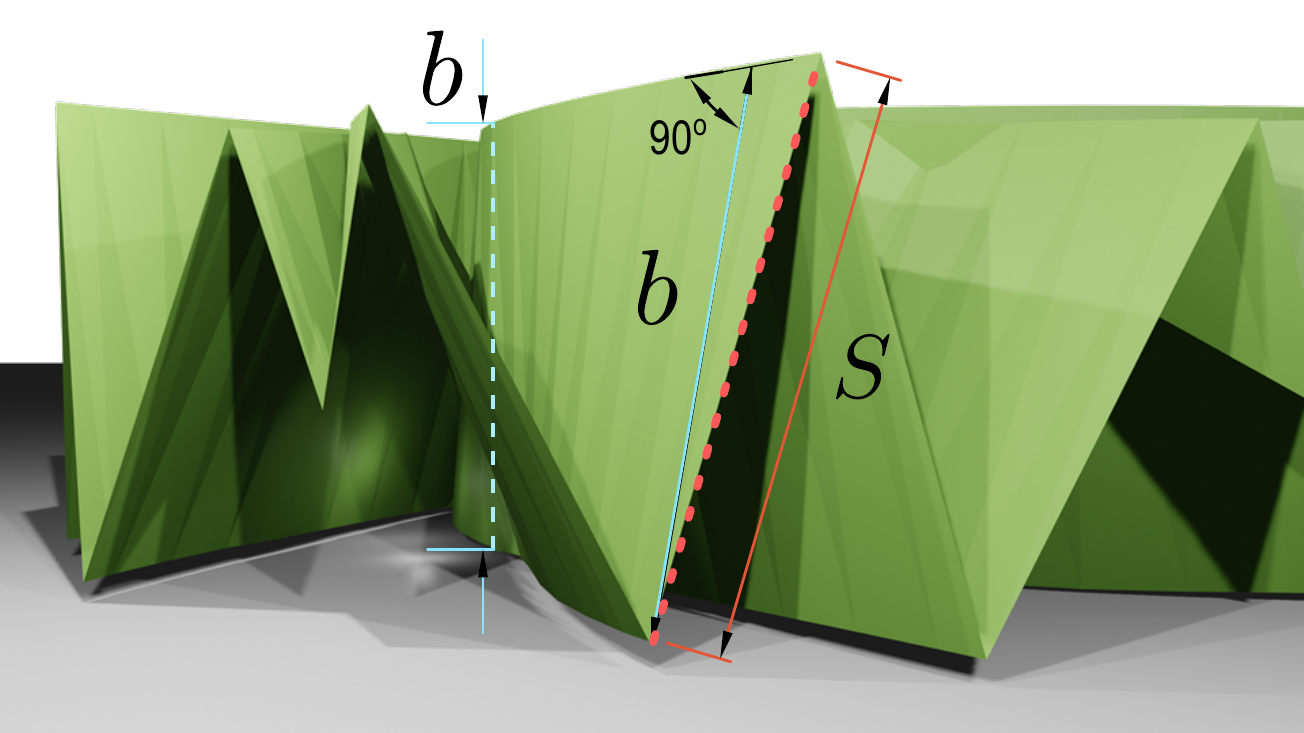}}
%\hfill
\caption{Physical significance of the quantities $S$ and $b$. Whereas the variable $S$ denotes the length of the zigzag within the same plane of a membrane when packed in cylindrical form, the variable $b$ denotes the height of the membrane in packed cylindrical configuration. }
\label{showSb}
\end{figure*}

\begin{figure*}[ht]
\centering
%\hfill
\subfigure[$S$]{\includegraphics[width=0.48\textwidth]{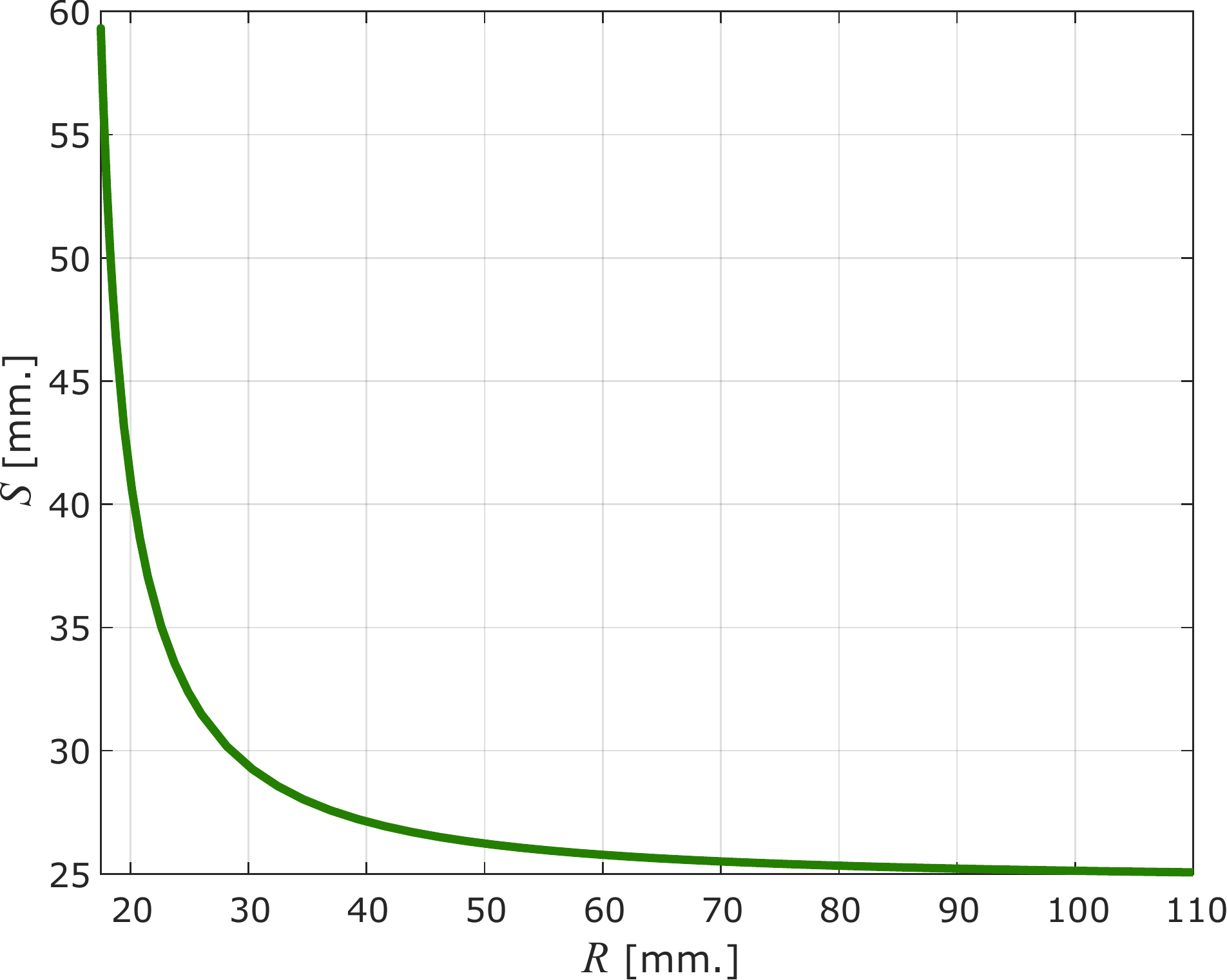}}
%\hfill
\subfigure[$b$]{\includegraphics[width=0.48\textwidth]{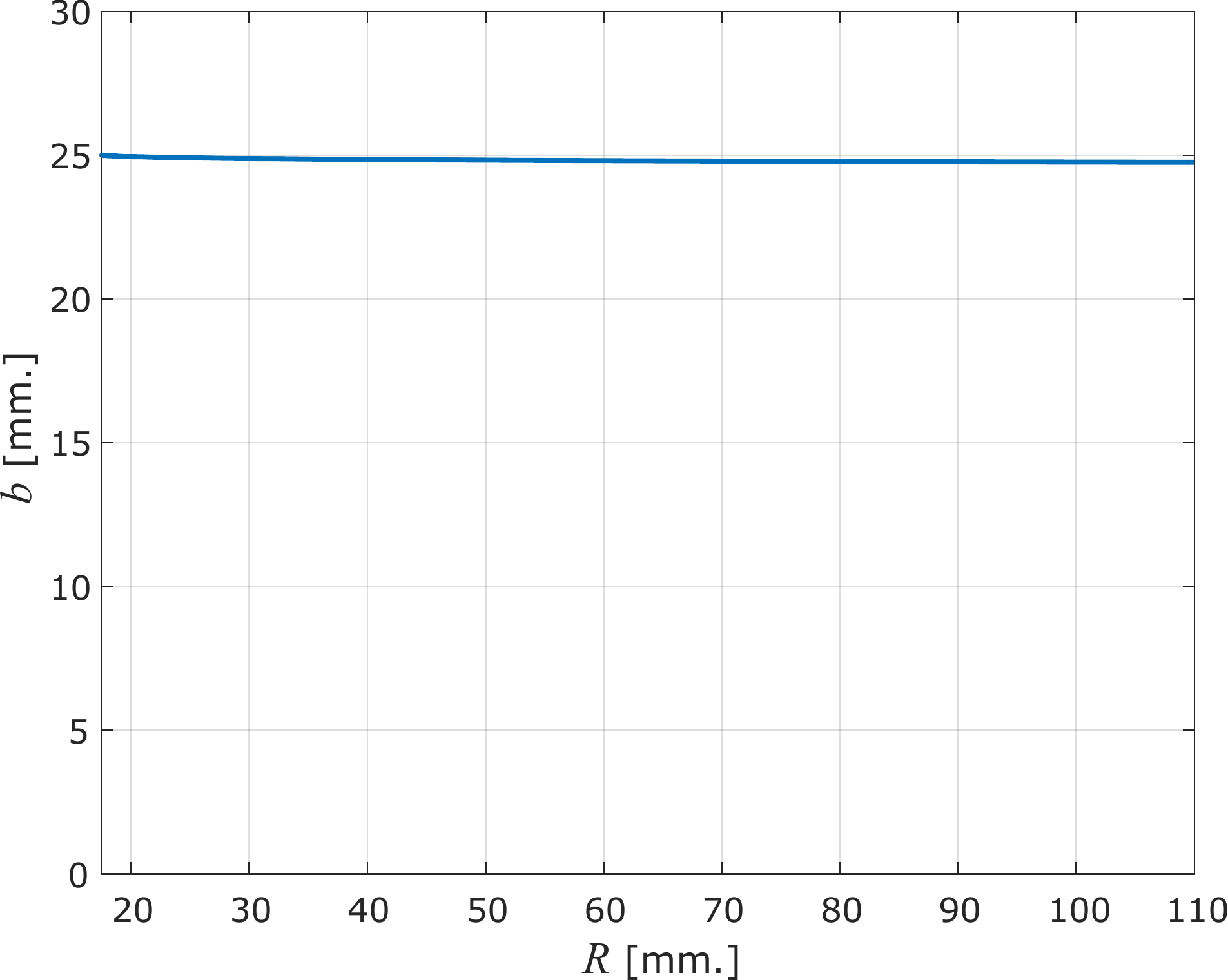}}
%\hfill
\caption{Solutions of $S$ and $b$ after numerical solution of the system of differential equations. Here, membrane parameters consider $N = 6$ (hexagon), $\theta_0 = 0$, $R_0 = 17.5 \si{mm},~ R_f = 110\si{mm}, ~ t = 0.1 \si{mm}, ~ S_0 = 59.33\si{mm}$.}
\label{solSb}
\end{figure*}

From a physical point of view, the variable $S$ denotes the length of the segment of the planar membrane when it is wrapped diagonally in a region between the two bases of the surface of the cylindrical shell, as shown by Fig. \ref{showSb}, and thus the variable $b$ determines the height of the membrane in packed cylindrical configuration, as shown by Fig. \ref{showSb}. From the solution of the system of differential equations in Eq. \ref{dqdr} to Eq. \ref{drdR}, one can find the values of $S$, and $b$ from Eq. \ref{bcos}. An example of the solution of variables $S$ and $b$ is portrayed by Fig. \ref{solSb}. Here, the initial values of $S$ are large and converge towards 25.06 mm., whereas the initial value of $b$ was 25 mm and decreased in minute proportions to 24.75 mm. The large values of $S$ at the beginning are due to the inner layers of the membrane being wrapped in longer diagonal segments of the cylinder. The final values of $S$ and $b$ differ slightly when $R$ is close to 110 mm. This is due to the outer segments of the membrane are wrapped in almost vertical configurations in the cylinder.

Given the above-mentioned membrane parameters, it is possible to compute the initial value of angle $\phi$ from Eq. \ref{phifunR}, as follows

\begin{equation}\label{phifunR0}
\phi_0 = \mathrm{asin}\left( \frac{\sqrt{( \pi r_0S_0 + kR_0t) ( \pi r_0S_0-kR_0t)}}{kR_0S_0} \right ),
\end{equation}
thus the height of the cylinder can be estimated from Eq. \ref{bcos} by

\begin{equation}\label{bcos0}
b_0 = S\cos \phi_0.
\end{equation}

Also, considering that we use regular polygons with $N$ edges, we use the following to compute initial values of angle $\theta_0$, as follows
\begin{equation}\label{theta0}
\theta_0 = \frac{2i\pi}{N},
\end{equation}
where $i \in [0, N-1]$ is an integer allowing to define the initial conditions to render $N$ creases of type \emph{Fold} $A$.

\begin{figure*}[h]
\centering
%\hfill
\subfigure[\emph{Fold} $A$]{\includegraphics[width=0.4\textwidth]{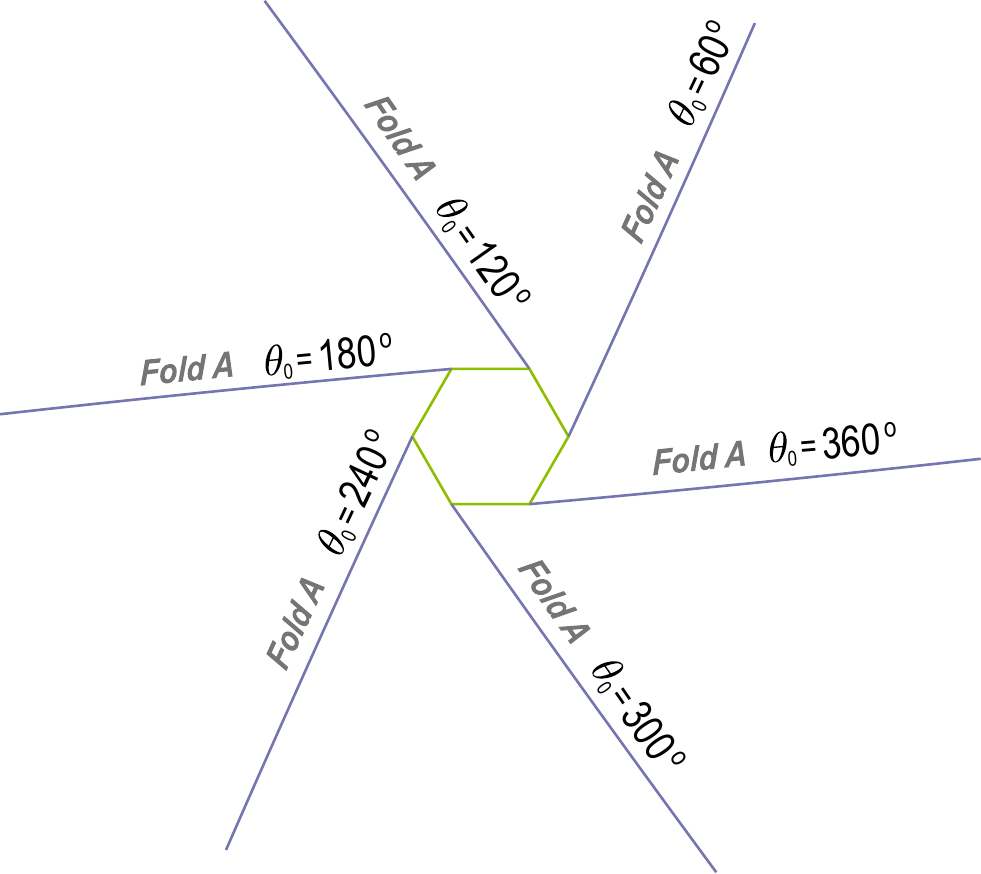}}
%\hfill
\subfigure[Concentric polygons]{\includegraphics[width=0.4\textwidth]{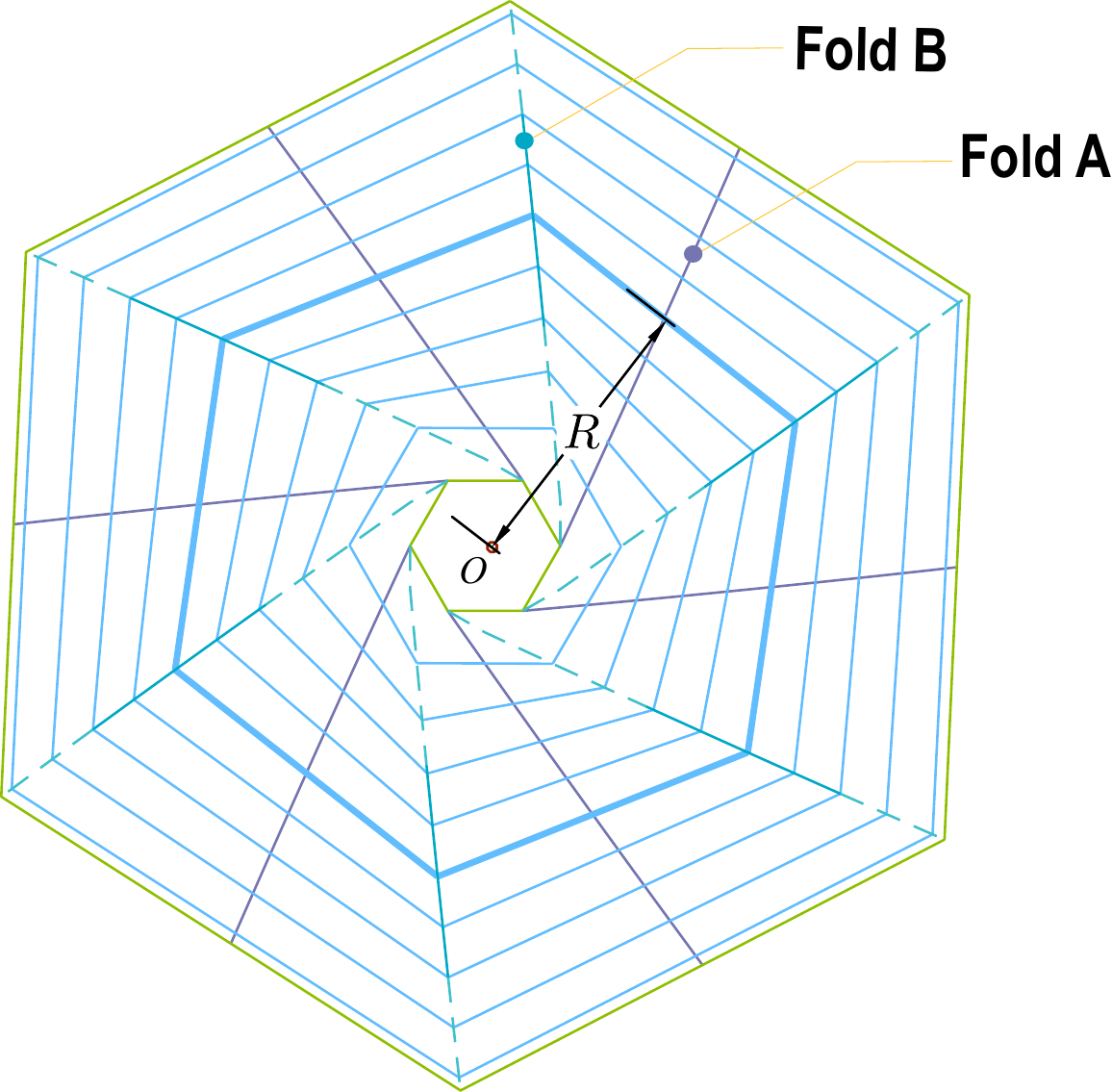}}
%\hfill
\subfigure[\emph{Fold} $B$ ]{\includegraphics[width=0.35\textwidth]{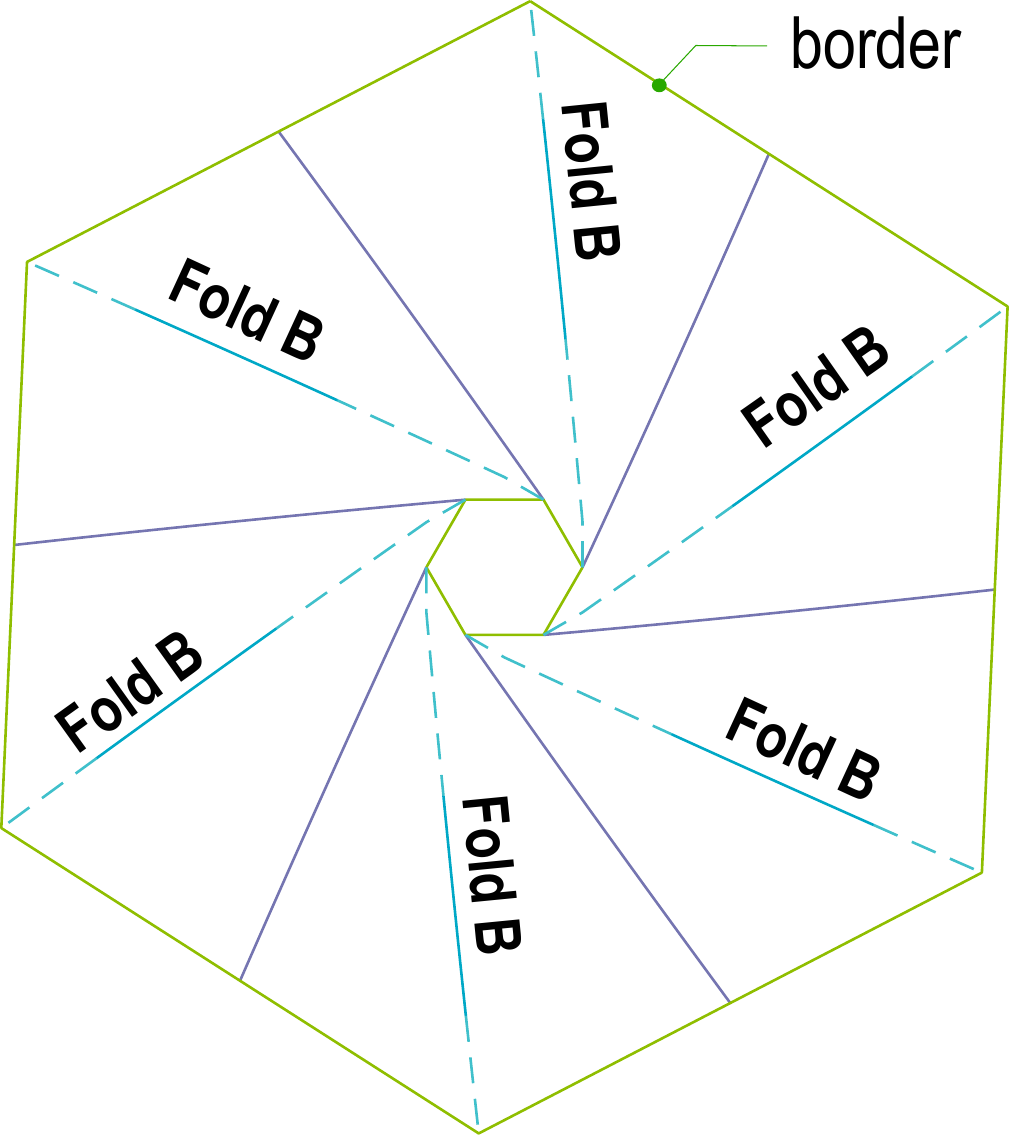}}
%\hfill
\subfigure[\emph{Fold} $A'$ ]{\includegraphics[width=0.35\textwidth]{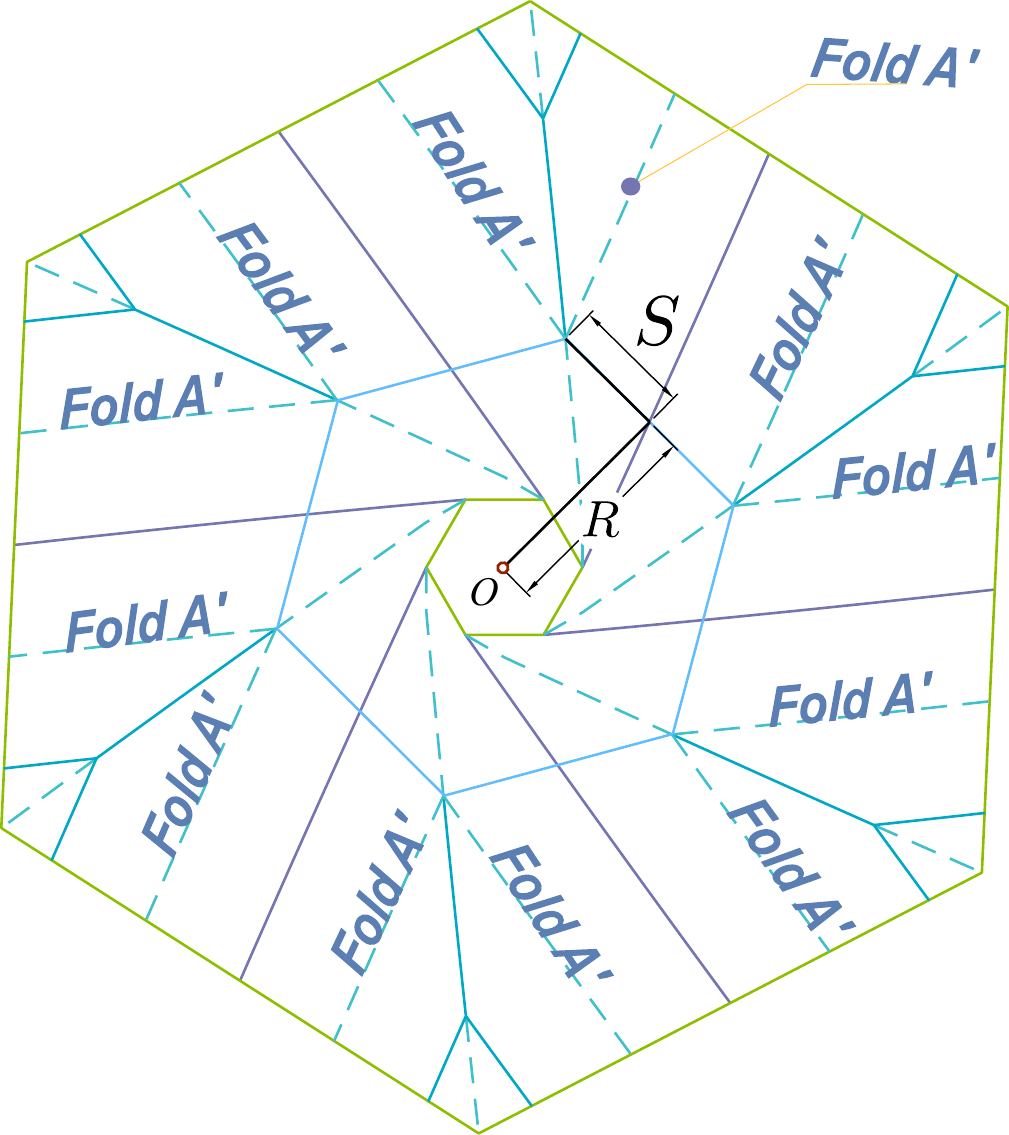}}
%\hfill
\caption{General steps to render the spiral folding pattern. Here, membrane parameters consider $N = 6$ (hexagon), $R_0 = 17.5 \si{mm},~ R_f = 110\si{mm}, ~ t = 0.1 \si{mm}, ~ S_0 = 59.33\si{mm}$.}
\label{planarsteps}
\end{figure*}

\begin{figure*}[h]
\centering
%\hfill
\subfigure[$N = 3$]{\includegraphics[width=0.32\textwidth]{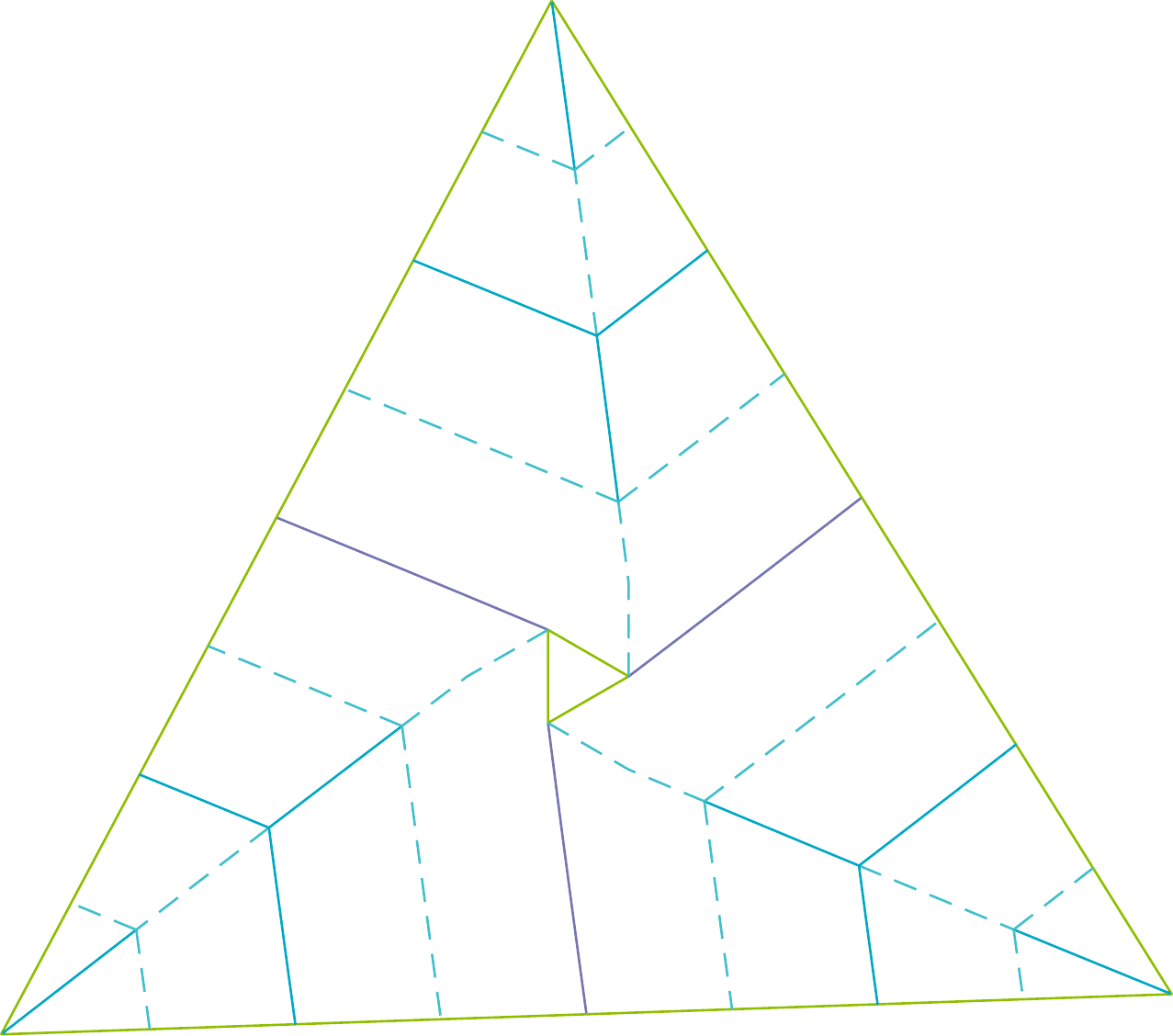}}
%\hfill
\subfigure[$N = 4$]{\includegraphics[width=0.32\textwidth]{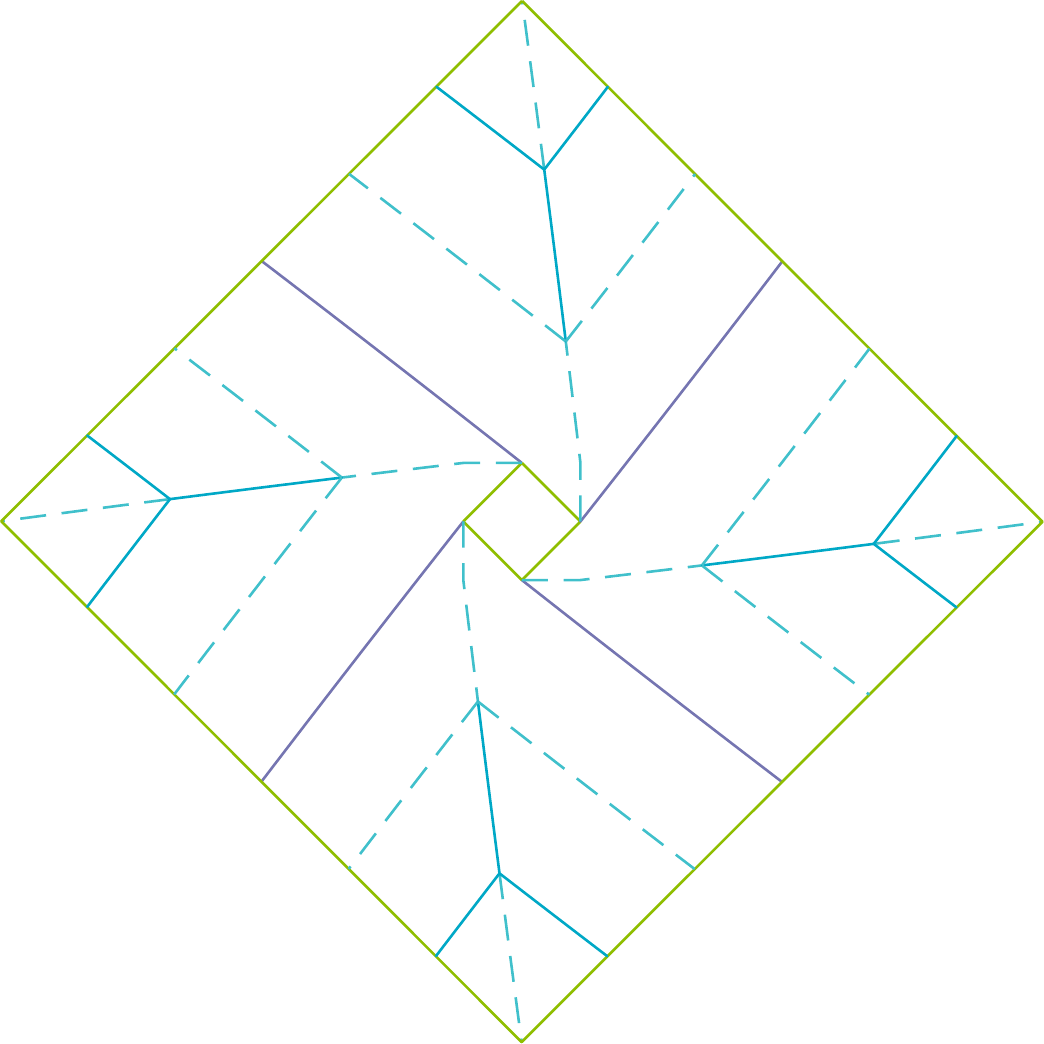}}
%\hfill
\subfigure[$N = 5$]{\includegraphics[width=0.32\textwidth]{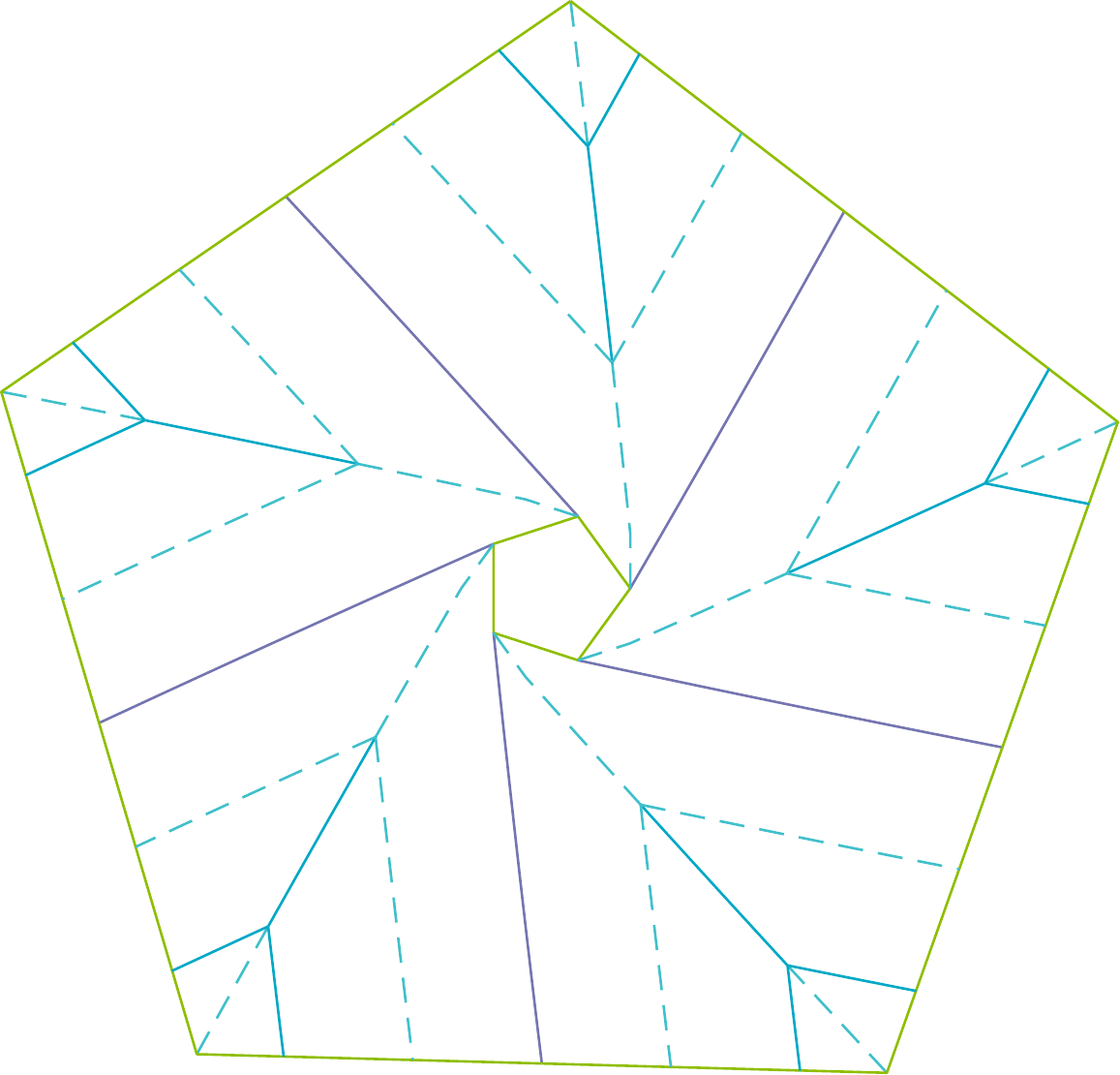}}
%\hfill
\subfigure[$N = 6$]{\includegraphics[width=0.29\textwidth]{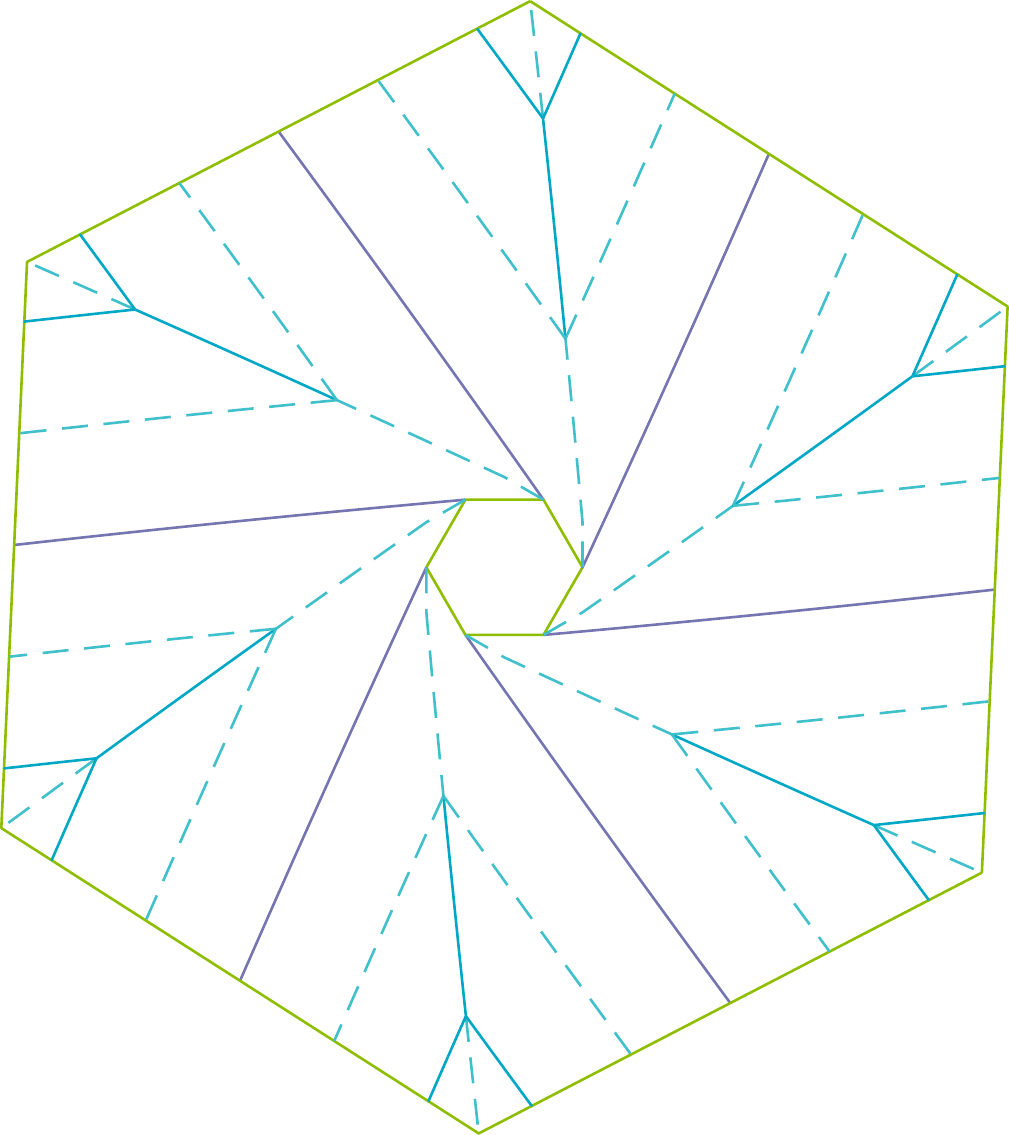}}
%\hfill
\subfigure[$N = 8$]{\includegraphics[width=0.32\textwidth]{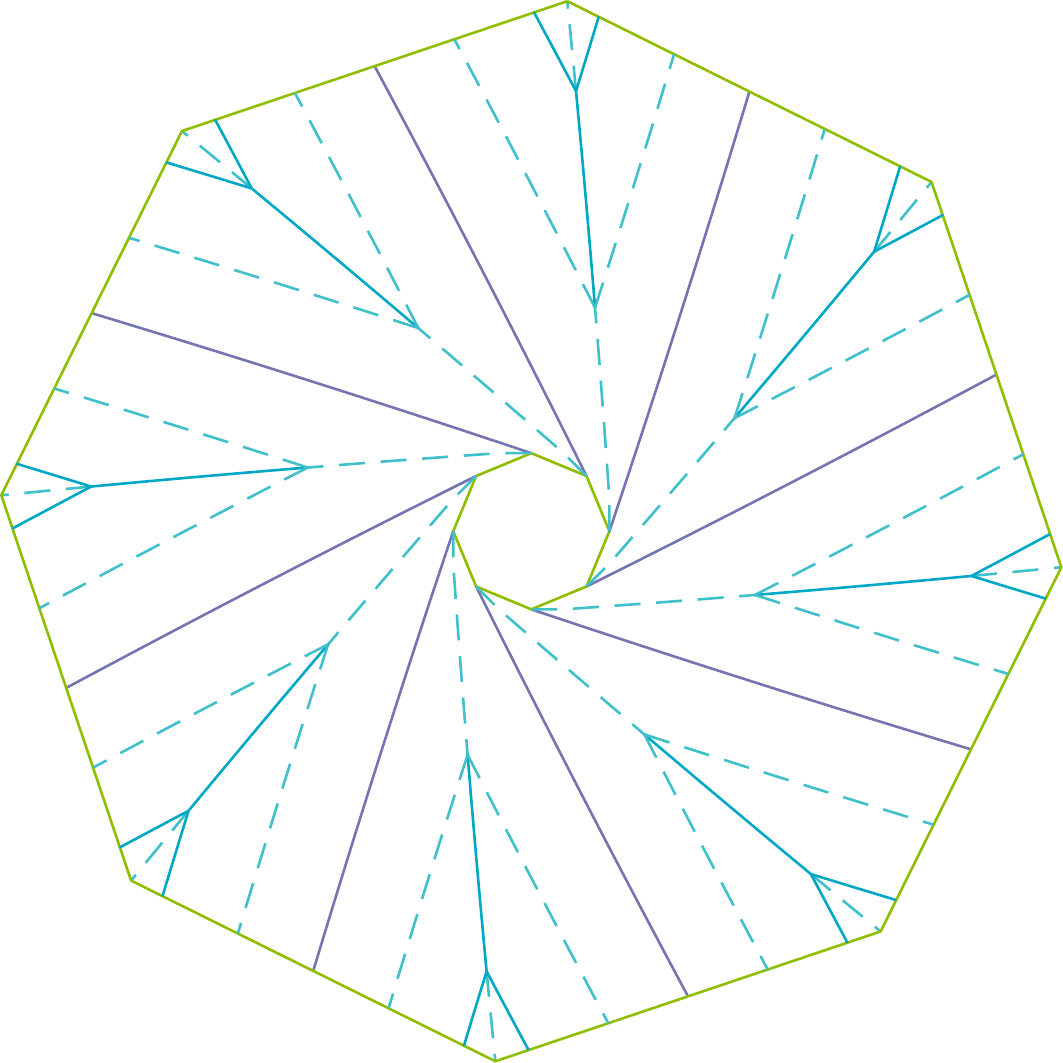}}
%\hfill
\subfigure[$N = 15$]{\includegraphics[width=0.32\textwidth]{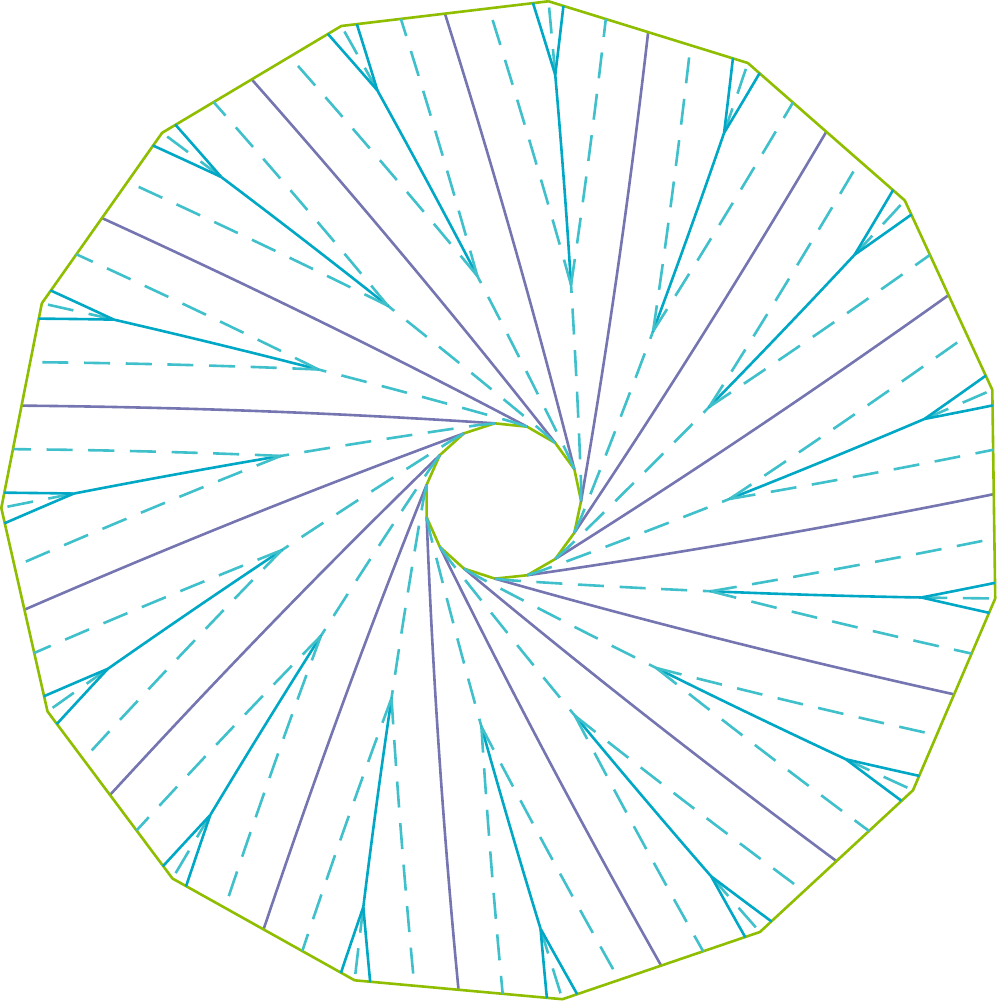}}
%\hfill
\caption{Basic layouts of spiral folding pattern for membrane parameters $R_0 = 17.5 \si{mm},~ R_f = 110\si{mm}, ~ t = 0.1 \si{mm}, ~ S_0 = 59.33 \si{mm}$.}
\label{planarcases}
\end{figure*}

\begin{figure*}[ht]
\centering
%\hfill
\subfigure[$R_f = 150$ \si{mm}]{\includegraphics[width=0.48\textwidth]{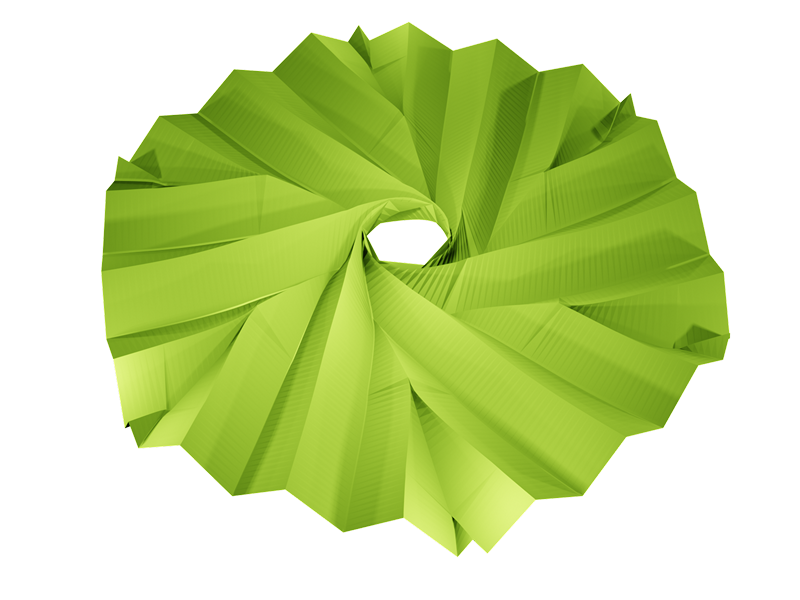}}
%\hfill
\subfigure[$R_f = 300$ \si{mm}]{\includegraphics[width=0.48\textwidth]{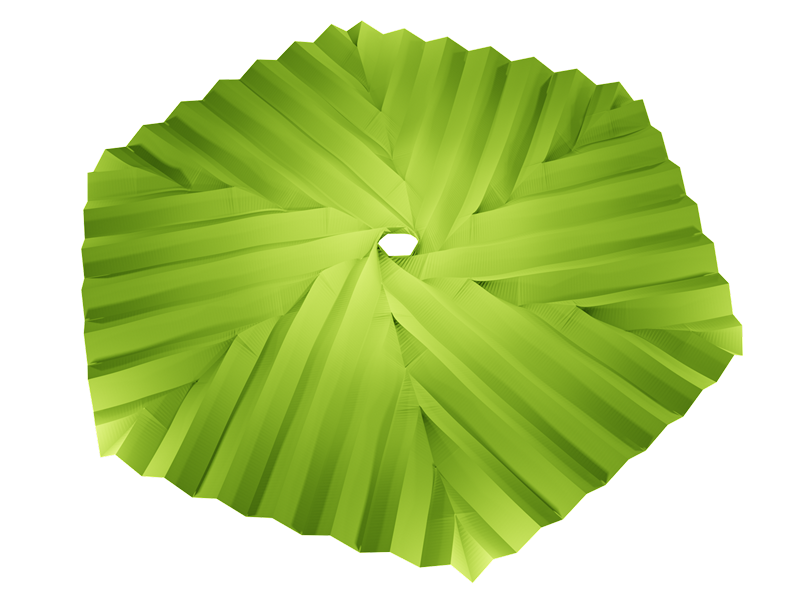}}
%\hfill
\caption{Basic layouts of spiral folding pattern for membrane parameters consider $N = 6$, $R_0 = 17.5 \si{mm}, ~ t = 0.1 \si{mm}, ~ S_0 = 59.33 \si{mm}$. }
\label{planar150300}
\end{figure*}

The procedure to compute the geometry of the crease layout in the spiral folding pattern follows a three-step approach, as Fig. \ref{planarsteps} shows.
\begin{itemize}
  \item First, as shown by Fig. \ref{planarsteps}-(a), $N$ creases of type \emph{Fold} $A$ are computed by numerically solving the system of differential equations presented in Eq. \ref{dqdr} - Eq. \ref{drdR} with membrane parameters $k$ (Eq. \ref{kk}), thickness $t$, boundaries $[R_0, R_f]$, and initial conditions $\theta_0$ (Eq. \ref{theta0}), length $S_0$ and radius $r_0 = R_0$.
  \item Second, given that $R \in [R_0, R_f]$ is part of the solution in the system of differential equations of Eq. \ref{dqdr} - Eq. \ref{drdR}, $R$ is the apothem of polygons concentric at origin $O$, as Fig. \ref{planarsteps}-(b) shows. Then, it is possible to concatenate the vertices of such concentric polygons to compute \emph{Fold} $B$, as Fig. \ref{planarsteps}-(c) shows. Thus, considering a regular polygon with $N$ sides, the layout is expected to have $N$ creases of \emph{Fold} $B$. For simplicity of visualization, Fig. \ref{planarsteps}-(b) shows a limited number of polygons, yet the more granular case should be expected across the whole interval of $R \in [R_0, R_f]$.
  \item Third, \emph{Fold} $A'$ is traced by points from \emph{Fold} $A$ which are displaced by length $S$ along the edges of the concentric polygons with apothem $R$ (see previous step). Basically, \emph{Fold} $A'$ is parallel to \emph{Fold} $A$ and intersects \emph{Fold} $B$ where $S$ is half of the edge of the concentric polygon with apothem $R$, as Fig. \ref{planarsteps}-(d) shows. Also, due to the symmetric configuration of \emph{Fold} $A'$, there will be two units of fold \emph{Fold} $A'$ for each \emph{Fold} $A$.
\end{itemize}

The reader may note that \emph{Fold} $A$ is a \emph{mountain}-type fold, whereas crease segments of \emph{Fold} $A'$ and \emph{Fold} $B$ are either \emph{mountain} or \emph{valley} folds. We use the following expression to classify the fold type in \emph{Fold} $A'$ and \emph{Fold} $B$:

\begin{equation}\label{foltype}
ft =  \Bigg \lceil \frac{\lambda}{S}  \Bigg \rceil \text{ mod } 2 ,
\end{equation}
where $ft$ is a binary variable denoting the type of fold (\emph{mountain} or \emph{valley} fold), $\lceil  ... \rceil$ denotes the ceil function, the term $\lambda$ denotes the half of the edge of the concentric polygon with apothem $R \in [R_0, R_f]$. For a regular polygon with $N$ edges

\begin{equation}\label{lambda}
\lambda = R \tan \Big (\frac{\pi}{N} \Big )
\end{equation}

The modulo operator in Eq. \ref{foltype} renders either 0 or 1, thus for apothem $R \in [R_0, R_f]$, segments of \emph{Fold} $A'$ and \emph{Fold} $B$ are outlined as \emph{mountain} ($ft = 0$) or \emph{valley} fold ($ft = 1$).

The solution of the system of differential equations in Eq. \ref{dqdr} to Eq. \ref{drdR} enable to render spiral folding patterns whose generalization to other regular geometries and lengths is straightforward, as Fig. \ref{planarcases} and Fig. \ref{planar150300} show.

\begin{figure}[t]
%\hfill
\begin{center}
{\includegraphics[width=1\textwidth]{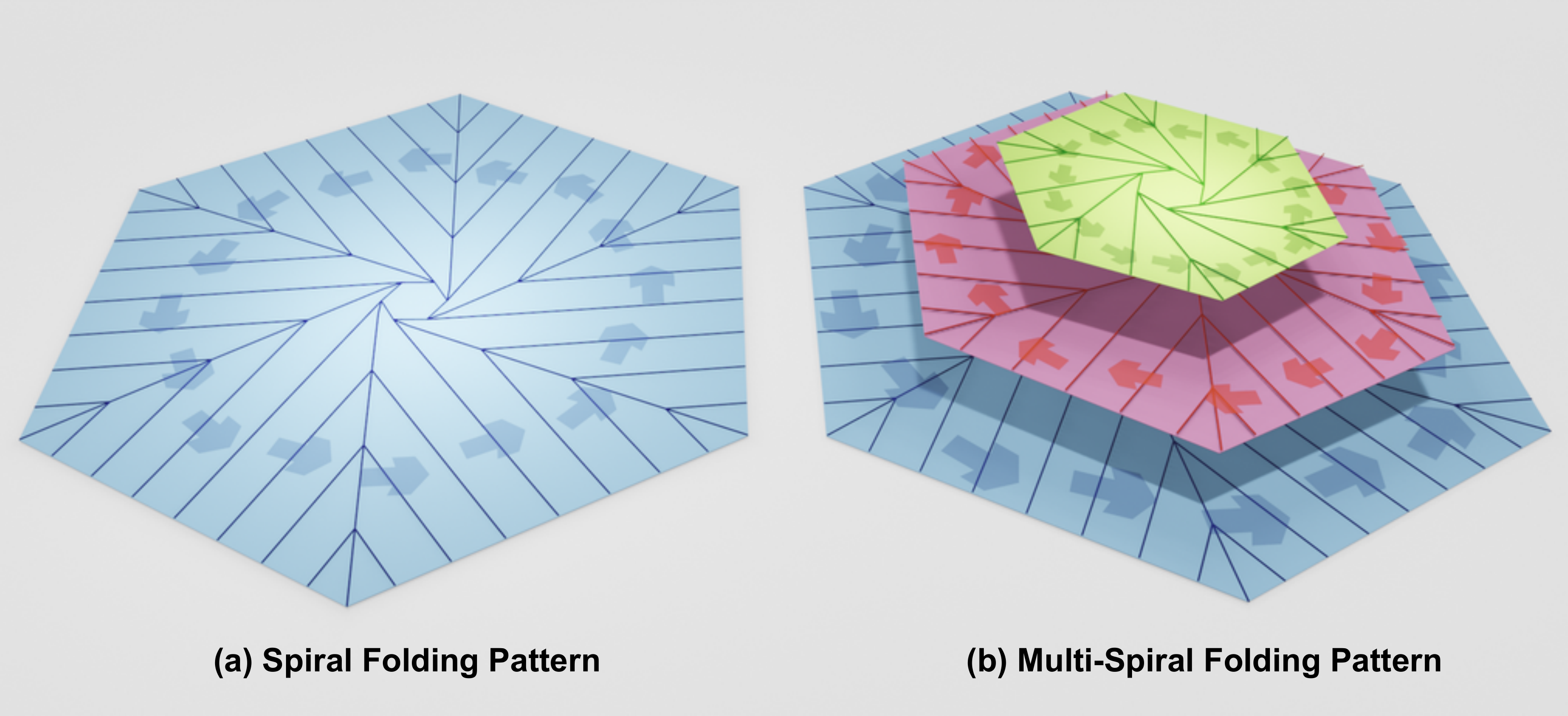}}
\end{center}
\caption{Basic concept of the proposed idea. In the left, the spiral folding pattern of a membrane. In the right, the basic concept of the superimposition of clockwise and counterclockwise folding patterns to enable the multi-spiral folding pattern.}
\label{concept}
\end{figure}

\begin{figure}[t]
%\hfill
\begin{center}
{\includegraphics[width=0.9\textwidth]{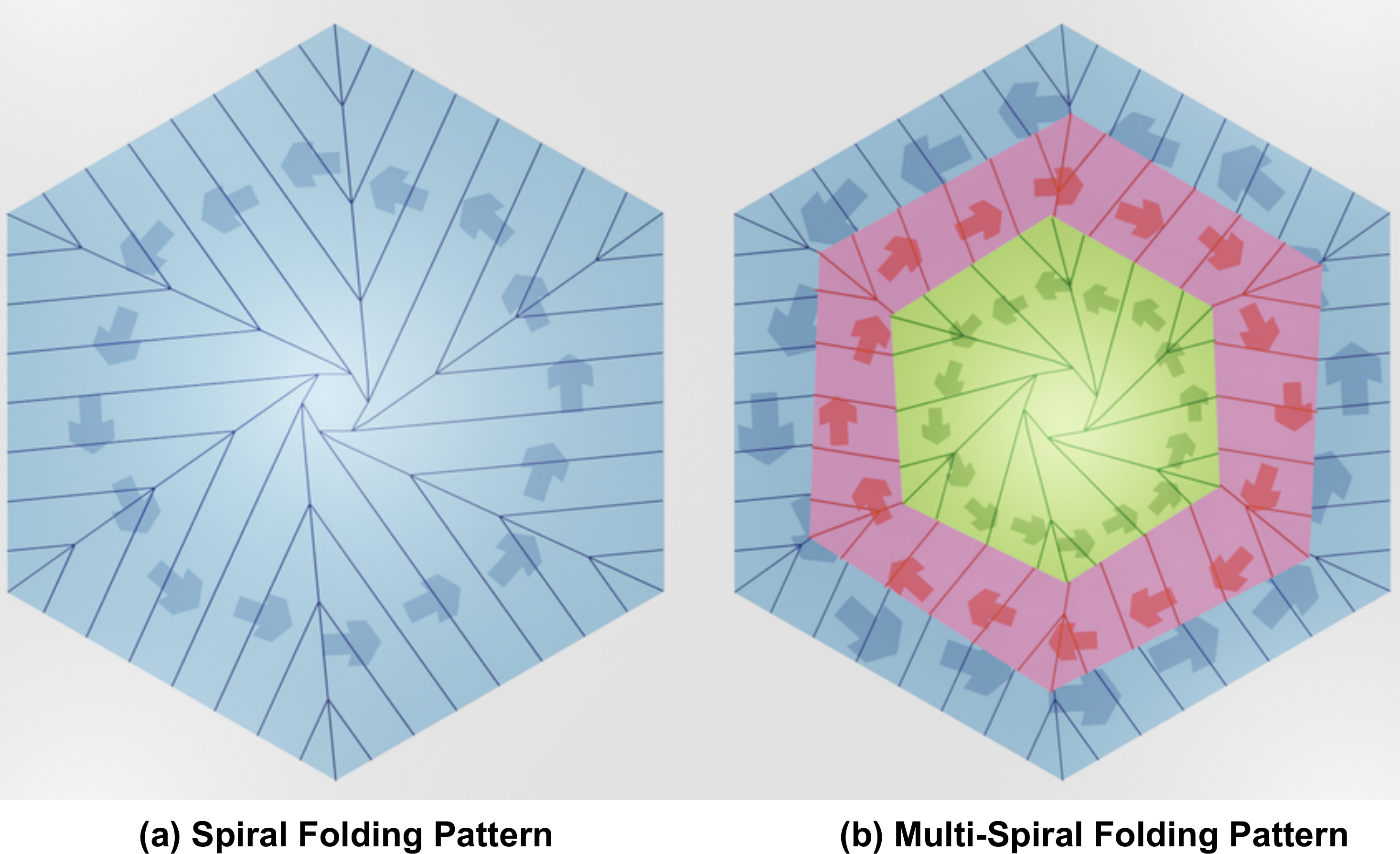}}
\end{center}
\caption{Layout of the membranes in our proposed approach. In the left, the spiral folding pattern of a membrane. In the right, the basic concept of the layout of the multi-spiral folding pattern.}
\label{conceptb}
\end{figure}

\begin{figure*}[h]
\begin{center}
     \subfigure[Rotation direction of spiral]{\includegraphics[width=0.45\textwidth]{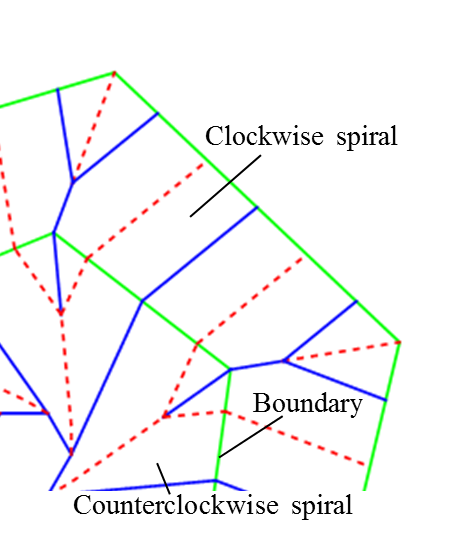}}
     \hfill
     \subfigure[Coordinate System]{\includegraphics[width=0.45\textwidth]{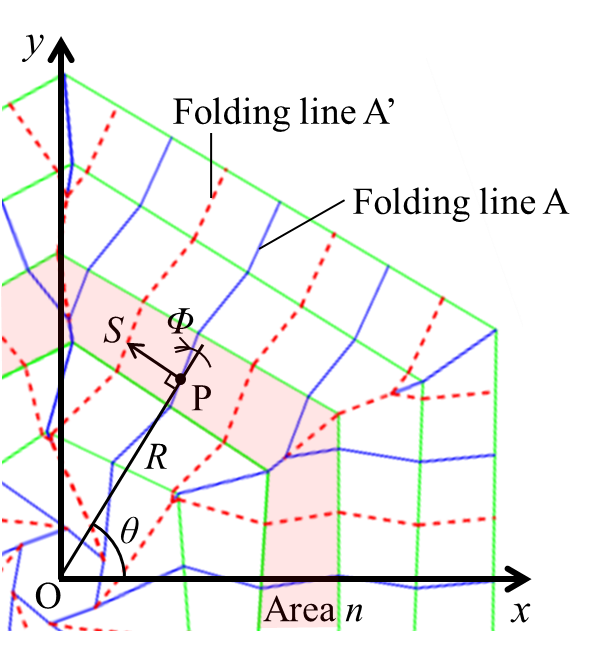}}
\end{center}
  \caption{Basic concept of the rotation direction of the spiral and coordinate system.}
  \label{fig4}
\end{figure*}

\begin{figure*}[h]
\centering
%\hfill
\subfigure[Folded state]{\includegraphics[width=0.25\textwidth]{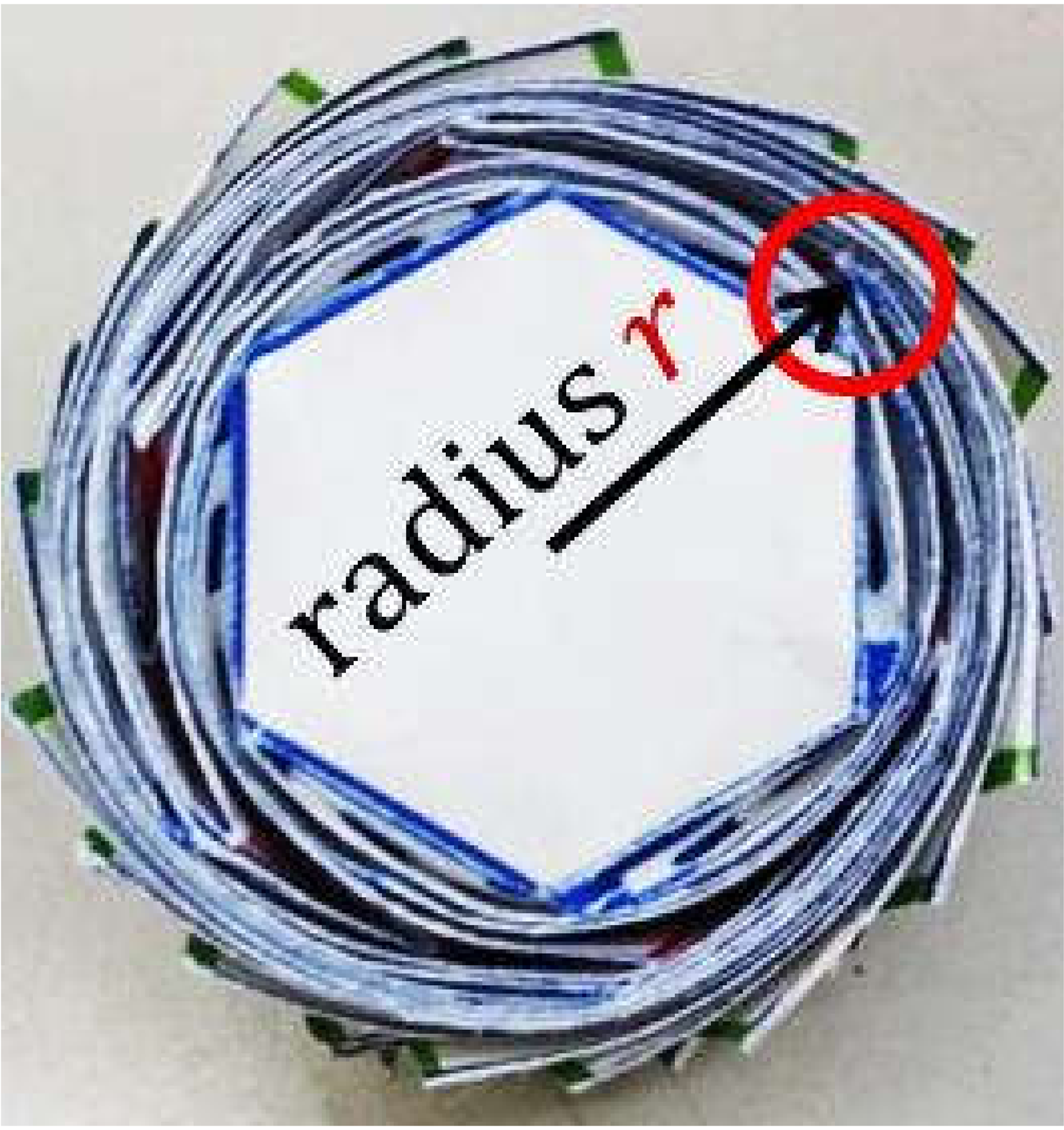}}
%\hfill
\subfigure[Thickness]{\includegraphics[width=0.25\textwidth]{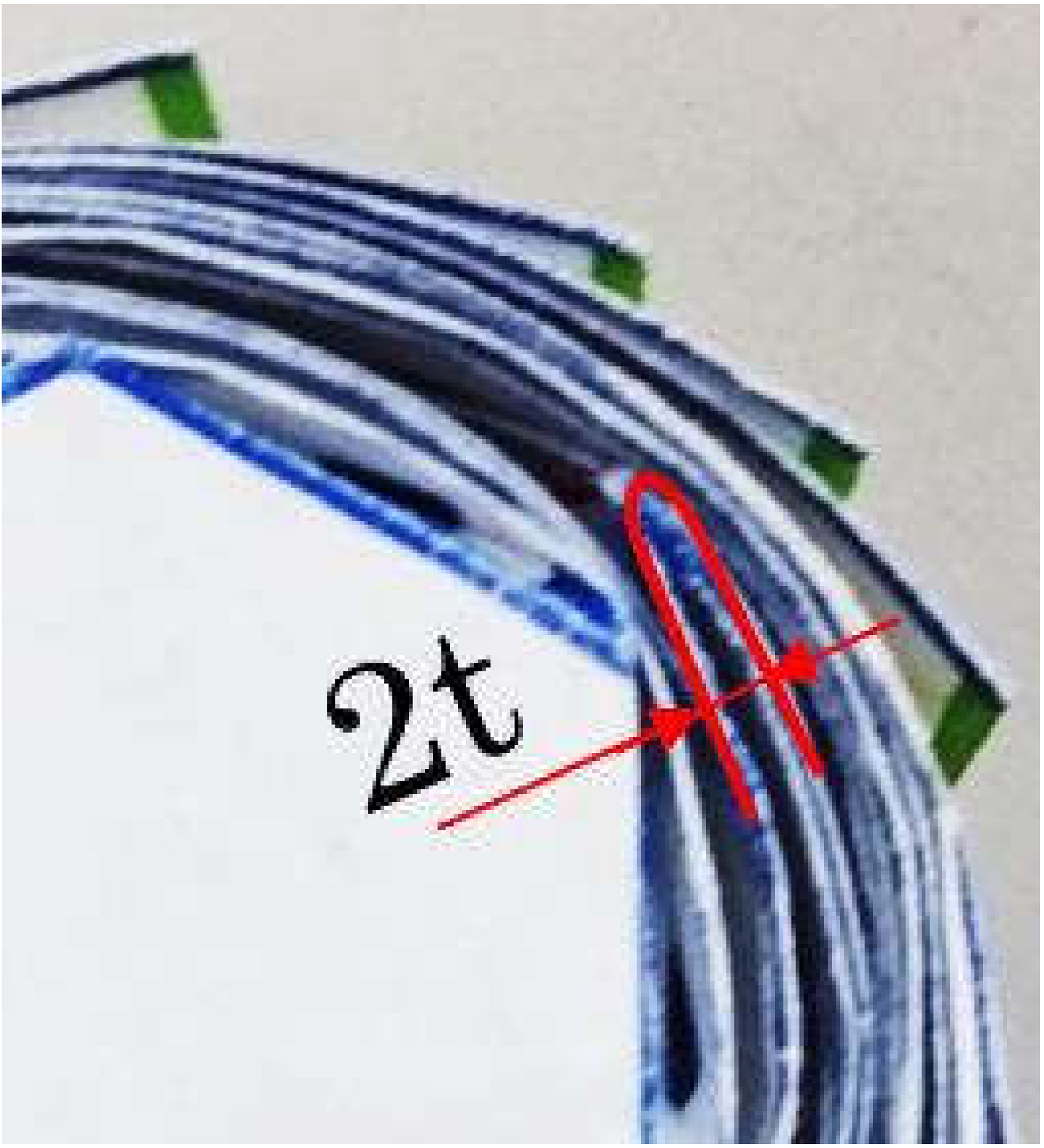}}
%\hfill
\caption{Basic concept to update of the radius.}
\label{radius}
\end{figure*}

\begin{figure}[ht!]
%\hfill
\begin{center}
{\includegraphics[width=0.98\textwidth]{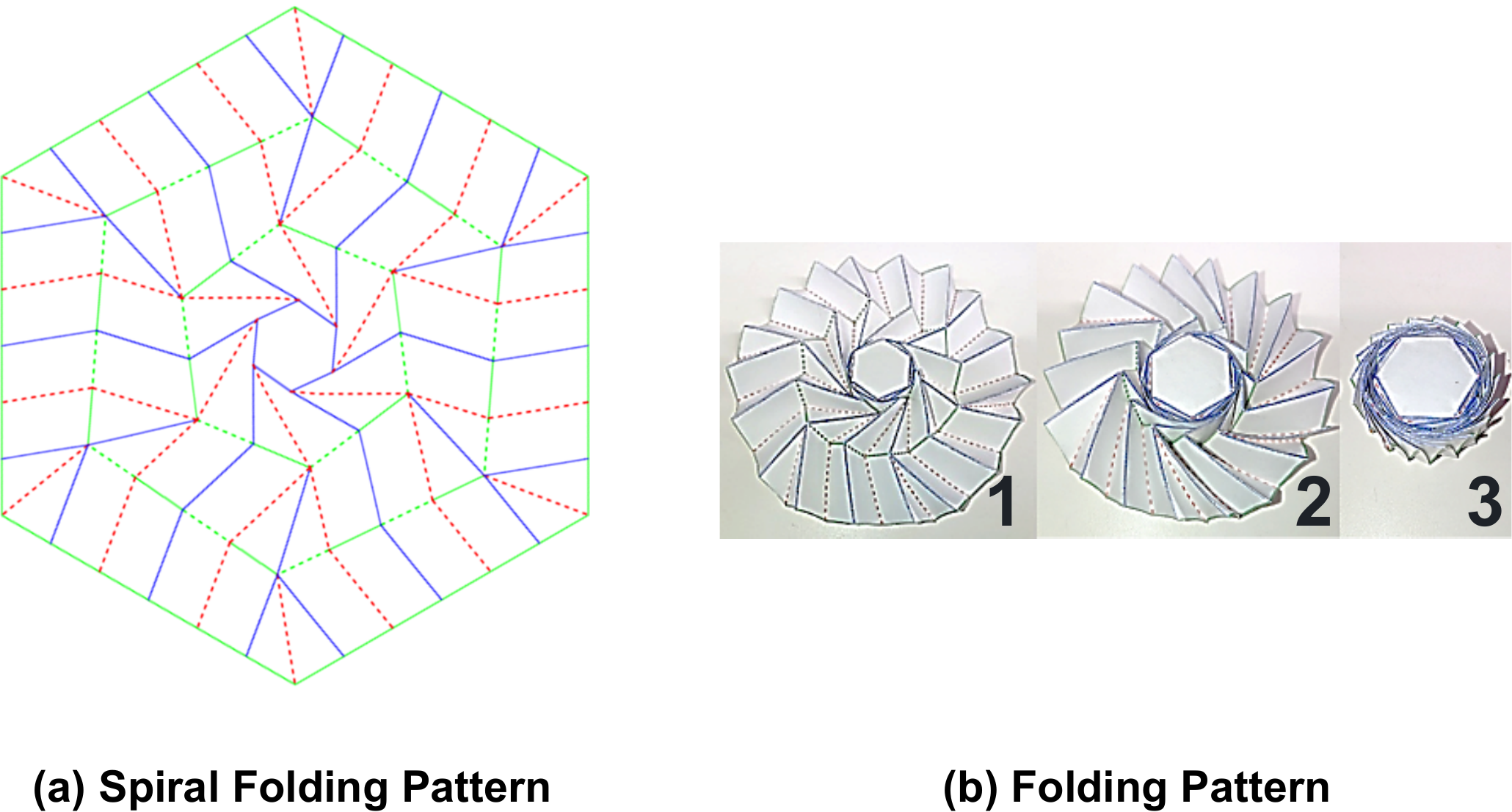}}
\end{center}
\caption{Layout of the multi-spiral folding pattern on a \emph{membrane} and its folding process.}
\label{samplespiral}
\end{figure}

\subsection{Multi-Spiral Folding Pattern}

We extend the above-mentioned principles to develop the multi-spiral folding patterns. The basic idea of our approach is portrayed by Fig. \ref{concept}. Here, in the left side of Fig. \ref{concept}, we portray the spiral folding on a single membrane, in which the orientation is deemed to be counterclockwise; and in the right side of Fig. \ref{concept}, we show the basic concept of our proposed approach combining two types of spiral folding patterns: (1) the counterclockwise spiral pattern as shown by the membranes with blue and green color in Fig. \ref{concept}, and (2) the clockwise spiral patterns as shown by the membrane with red color in Fig. \ref{concept}. The basic idea in the multi-spiral folding pattern is to superimpose membranes folding in clockwise and counterclockwise orientations, whose result renders a single membrane with multi-spiral and concentric folding patterns as portrayed by Fig. \ref{conceptb}. The reader may note that the membrane of Fig. \ref{conceptb} shows the concentric multiple spiral by different colors (blue, red and green), in which consecutive spirals have opposing orientations. In line of the above, the proposed multi-spiral folding pattern consist of (1) bellows folding on the circumferential direction, and (2) the spiral folding on the radial direction. The folding lines that form bellows are applied to the boundaries dividing the two spiral patterns. Thus, we extend the governing equations described in Eq. \ref{ds}- Eq. \ref{kk} to allow the rendering of  multi-spiral folding into a single membrane.

In order to portray the basic idea of our approach, the coordinate system is shown by Fig. \ref{fig4}. Here, the boundaries of the spirals with clockwise and the counterclockwise orientations are shown by green color, and the arbitrary point $P$ along the \emph{Folding Line A} within the $n$th area is shown by Fig. \ref{fig4}(b).

From Fig. \ref{fig4}, the direction of rotation of the spiral depends on whether the area $n$ is an odd or an even number. When the area $n$ is an odd number, the orientation pattern of its associated spiral becomes counterclockwise. On the other hand, when the area $n$ is an even number, the orientation pattern of the spiral becomes clockwise. Thus, It becomes necessary to arrange the above-mentioned Eq. \ref{ds}- Eq. \ref{kk} to allow both the counterclockwise and the clockwise orientation patterns. We add a negative sign to the angle $\phi$ the Eq. \ref{ds}- Eq. \ref{kk} in order to reverse the rotation of the spiral. Also, the folding by bellows depends on whether the area $n$ is an odd or an even number.

Based on the above motivations, the governing equations of the multi-spiral folding pattern is formulated as follows:

\begin{equation}\label{eq:ds}
\frac{dS_n}{S_n} = -\tan((-1)^{n-1}\phi_n) \cdot d\theta_n,
\end{equation}

\begin{equation}\label{eq:tanphi}
\tan((-1)^{n-1}\phi_n) = \frac{R  d\theta_n}{dR},
\end{equation}

\begin{equation}\label{pir}
\frac{2\pi r_n}{2kR} = \sqrt{\sin^2((-1)^{n-1}\phi_n) + \Big (\frac{t}{S_n} \Big )^2},
\end{equation}

\begin{equation}\label{pir2}
\frac{2\pi r_n}{2kR}  \frac{dr_n}{dR} = \frac{t}{S_n  \cos ((-1)^{n-1}\phi_n)}.
\end{equation}

In the above formulations, the subscript $n$ of the angles $\theta$, $\phi$, the radius $r$ and the fold interval $S$ indicates the corresponding value of the $n$th area. Additionally, the angle $\theta_n$, the radius $r_n$ and the fold interval $S_n$ are defined as  $\theta_{n,n}$, $r_{n,n}$, and $S_{n,n}$, respectively,  when the distance $R$ is equal to the radius $R_n$.

%The second numerical subscript n indicates the value when R is equal to $R_n$.

The boundary conditions on the $n$th boundary folded like bellows are defined as follows:

\begin{equation}\label{thetann}
\theta_{n+1,n} = \theta_{n,n}
\end{equation}

\begin{equation}\label{rnn}
r_{n+1,n} = r_{n,n} + 2t
\end{equation}

\begin{equation}\label{snn}
S_{n+1,n} = S_{n,n}
\end{equation}

The above formulations are motivated by our aim to consider the thickness effects of the membranes when folding, thus the boundary condition of the radius $r_n$  (due to of bellows folding), and the radius are updated between the consecutive boundaries of spirals. Thus, the double of the thickness of the membrane is added to the radius just before the boundary $r_{n,n}$ in order to allow the radius of the boundary $r_{n+1,n}$ to become thicker.

In order to show the an example of the multi-spiral folding pattern, Fig. \ref{samplespiral} shows the folding pattern and its folding process. Here, blue continuous lines denote the mountain folding lines, whereas, red and green dashed lines denote the valley folding lines. Also, lines colored in green denote the concentric spiral regions in which the folding of bellows in the circumferential direction occurs.

\section{Multi-Spiral Folding Pattern of Curved Surfaces}

In this section, we describe the governing equations of the multi-spiral folding pattern on a curved membrane.

\subsection{Preliminaries}

\begin{figure}[t]
%\hfill
\begin{center}
{\includegraphics[width=0.98\textwidth]{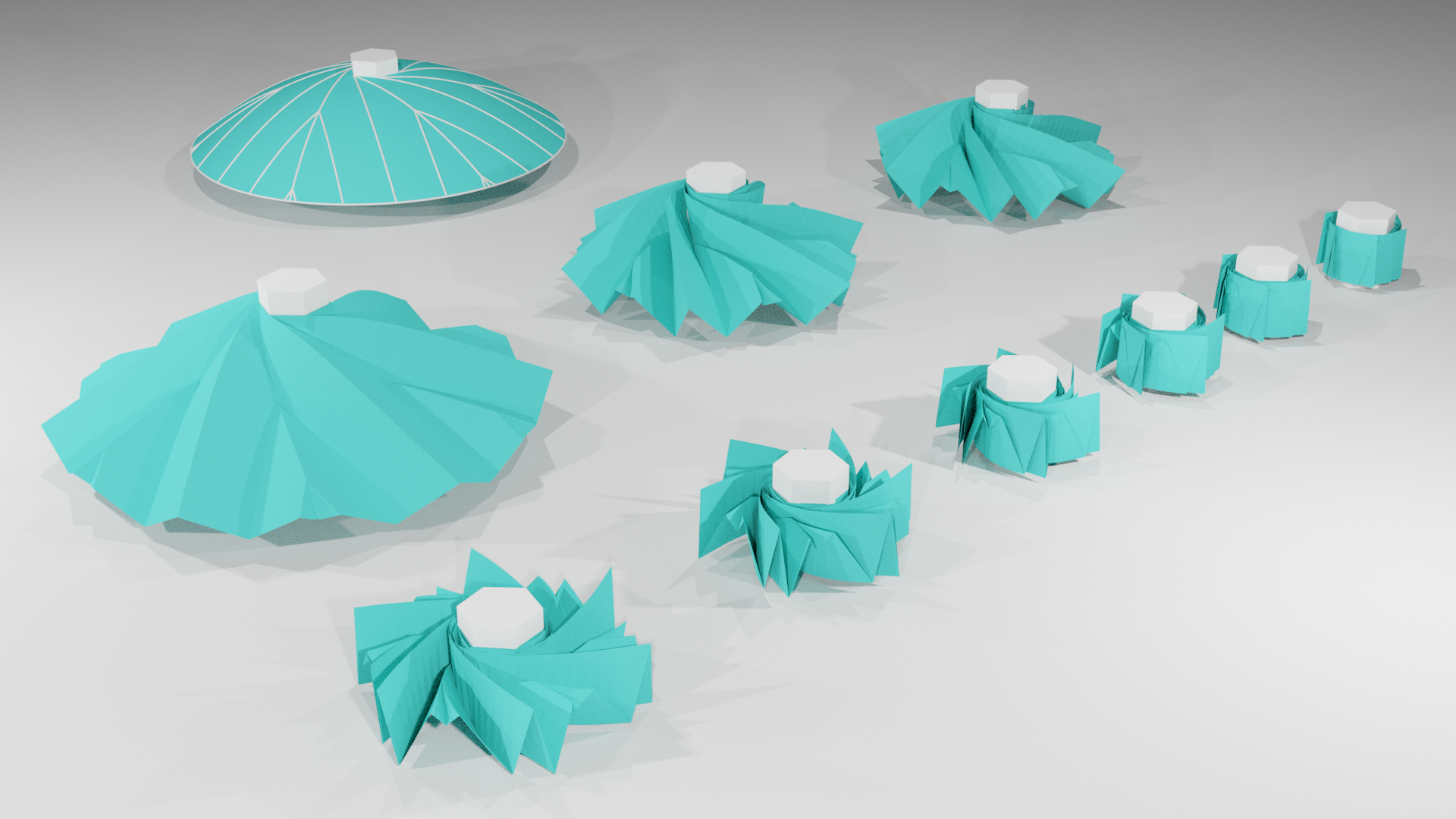}}
\end{center}
\caption{Basic idea of the folding of a curved surface by using the spiral folding pattern.}
\label{foldcuscn}
\end{figure}

\begin{figure}[t]
%\hfill
\begin{center}
{\includegraphics[width=1\textwidth]{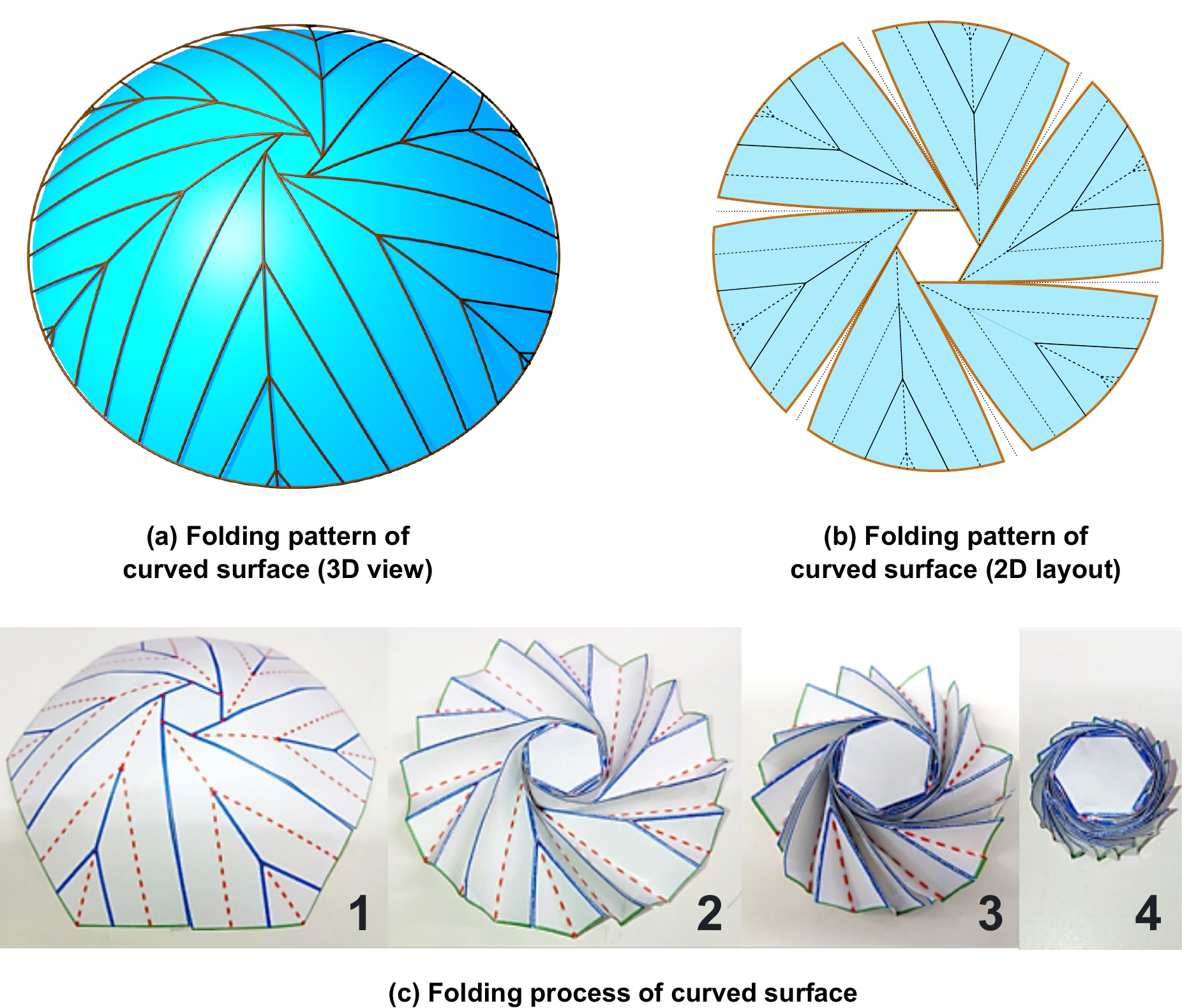}}
\end{center}
\caption{ Basic idea for folding into a curved surface.}
\label{domepaper}
\end{figure}

In contrast to the planar membranes, the folding of a curved surface (\emph{paraboloid} or \emph{dome-shaped} membrane) by the spiral folding method aims at transforming a curved membrane into a cylinder shell, as shown by Fig. \ref{foldcuscn} and Fig. \ref{domepaper}. As such, an early approach in this direction was proposed by Nojima by using circular membranes extended by the archimedean spiral\citep{pelle92} with folds in the radial/circumferential directions\citep{nojima01a} and by conical and parabolic wrapping models with slits\citep{nojimavipsi}. In this paper, we consider a curved membrane whose geometry is derived from juxtaposing the polynomial equations of a parabola and those of the spiral folding pattern described in the previous section.

\begin{figure}[t]
%\hfill
\begin{center}
{\includegraphics[width=0.98\textwidth]{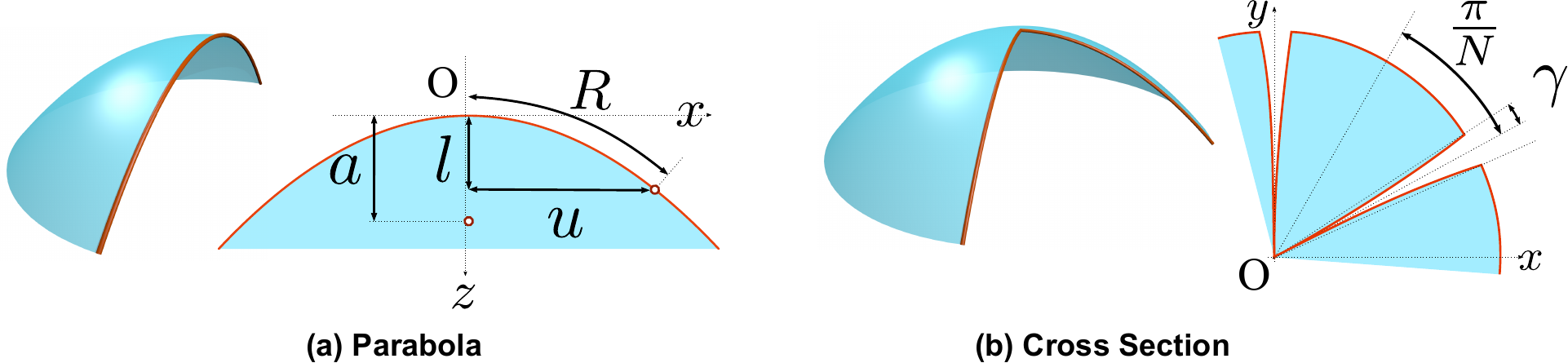}}
\end{center}
\caption{Basic idea of (a) cross-section views of a parabolic surface. (b) slits which allow folding into curved surface.}
\label{curvesection}
\end{figure}

As such, in order to portray the governing equations of the spiral folding pattern, we refer first to Fig. \ref{curvesection} which portrays  the key elements that allow to describe the geometry of the curved layout structure. Given a parabola with features as shown by Fig. \ref{curvesection}-(a), its governing equation is

\begin{equation}\label{eq:para}
x^2 = 4az,
\end{equation}
where $a$ is the focal length of the 2-dimensional parabola, and $x, z$ are the coordinates of the parabola in the cross-section. In Fig. \ref{curvesection}-(a), $R$ denotes the arc length of the parabola, thus

\begin{equation}\label{eq:Rarc}
dR^2 = du^2 + dl^2,
\end{equation}
and due to Eq. \ref{eq:para}, $u^2 = 4al$, then

\begin{equation}\label{eq:dlcurve}
dl = \frac{udu}{2a}.
\end{equation}
Then by replacing Eq. \ref{eq:dlcurve} into Eq. \ref{eq:Rarc} and accommodating terms

\begin{equation}\label{eq:dudR}
\frac{du}{dR} = \frac{2a}{\sqrt{u^2 + 4a^2}}.
\end{equation}

An alternative way to express the relationship between $R$ and $u$ is by the arc-length formulation from Fig. \ref{curvesection}-(a), as follows
\begin{equation}\label{RInt}
R = \int_{0}^{u} \sqrt{1 + (z')^2}dx,
\end{equation}
whose closed form expression translates into

\begin{equation}\label{eq:Rcu}
R = \sqrt{al + l^2} + a\sinh^{-1} \Bigg ( \sqrt{\frac{l}{a}} \Bigg).
\end{equation}

Furthermore, for a slit angle $\gamma$ in Fig. \ref{curvesection}-(b), the ratio of the length $R$ to the horizontal distance $u$ of the parabola is to the ratio of the angle $\pi/N$ to the angle of a half petal out of $N$ petals

\begin{equation}\label{eq:ru}
\frac{R}{u} = \frac{\frac{\pi}{N}}{\Big( \displaystyle \frac{\pi}{N} - \gamma\Big)},
\end{equation}
and accommodating terms we obtain

\begin{equation}\label{eq:gammacu}
\gamma = \frac{\pi}{RN}\Big (R-u \Big ).
\end{equation}

\begin{figure}[t]
%\hfill
\begin{center}
{\includegraphics[width=0.5\textwidth]{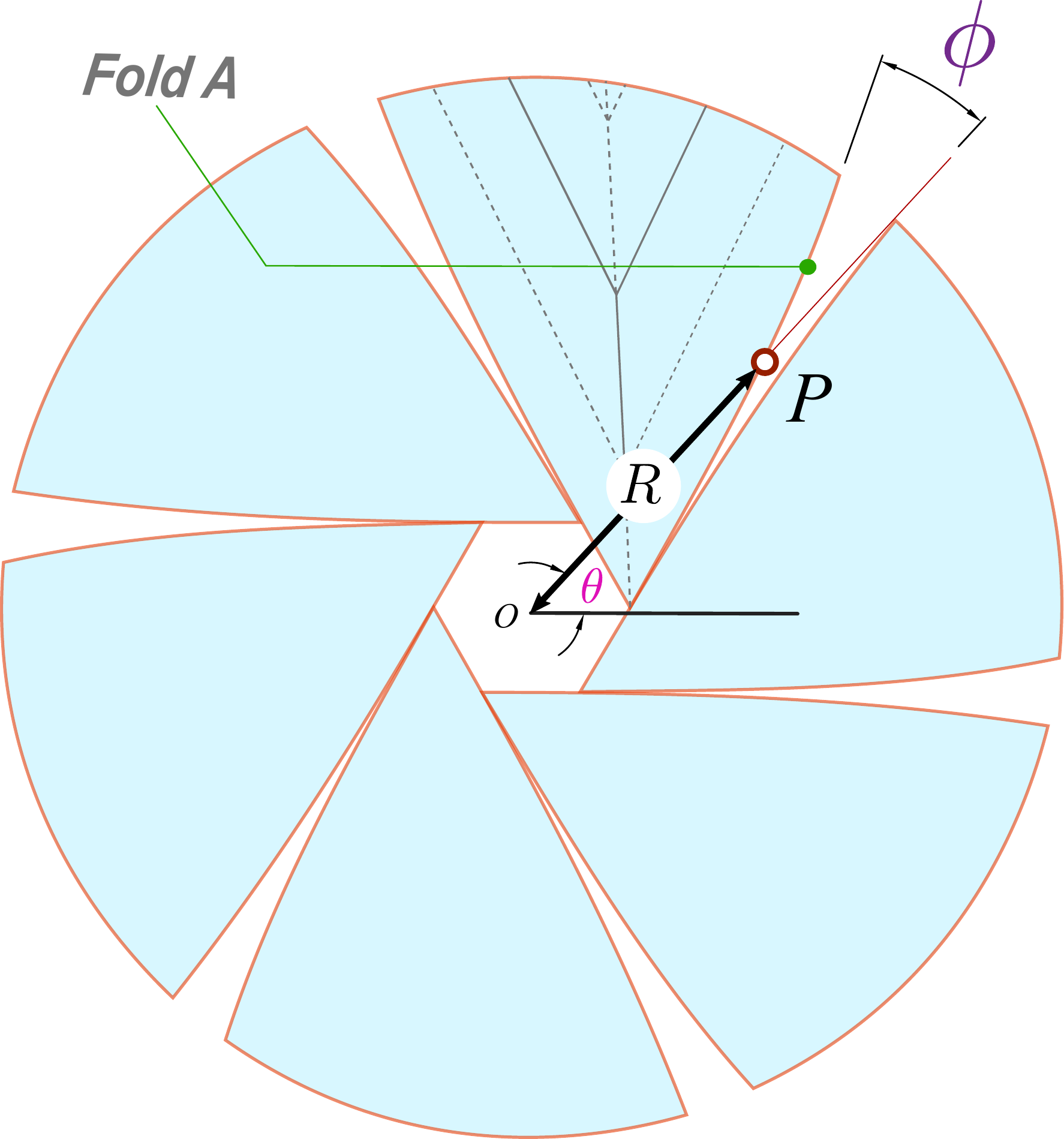}}
\end{center}
\caption{Main elements in the spiral folding pattern of a curved surface. The center of the polygon hub is located at $O$,  \emph{Fold A} is tangential to edges of the polygon hub, $P$ is a point in \emph{Fold A}, $R$ is the length of segment $\overline{OP}$, $\theta$ is the angle between the $x$-axis and the segment $\overline{OP}$, and $\phi$ is the acute angle between the segment $\overline{OP}$ and \emph{Fold} $A$ at point $P$.}
\label{samplecurve}
\end{figure}

In the above definition, the slits in the curved surface enable to fold and unfold according to the line segments extending the vertices and the edges of a polygon with $N$ sides. For simplicity, and without loss of generality, we use a hexagon as portrayed by Fig. (\ref{samplecurve}), thus it is possible to concatenate $N$ petals being singly curved, in which petal edges match the characteristics of the above-described paraboloid. In a more general case, it is expected that the core polygon converges to a circle with fold segments being tangential to it, and creases being curved due to the thickness effect in the curved membrane.

Then, we refer to Fig. \ref{samplecurve} to define the main elements in the spiral folding with curved surface. Given a regular polygon with center at $O$, the \emph{Fold A} is tangential to the edge of the polygon, and $P$ is a point in \emph{Fold A}.

\begin{figure}[t]
%\hfill
\begin{center}
{\includegraphics[width=0.98\textwidth]{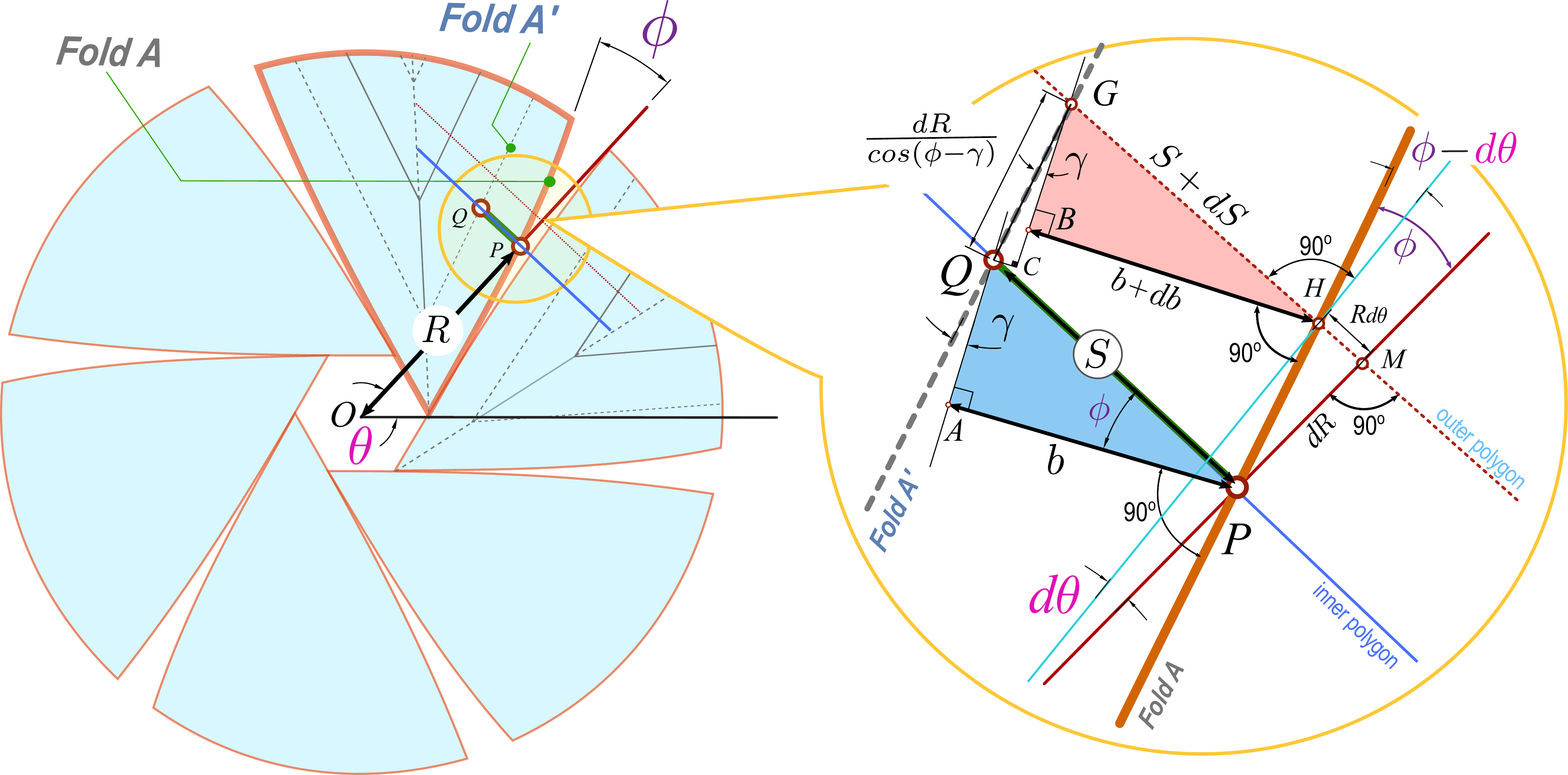}}
\end{center}
\caption{Main elements in the relationship among $R$, $S$, $b$, $\phi$ and $\theta$. Considering that \emph{Fold} $A'$ is consecutive to \emph{Fold} $A$, the segment $\overline{PQ}$ is prescribed in the side of an \emph{inner polygon}, where $\overline{PQ} \perp \overline{OP}$ and $Q$ is the intersection between \emph{Fold} $A'$ and the edge of the \emph{inner polygon}. For simplicity of further descriptions, $S$ is the length of the segment $\overline{PQ}$. Also, $b$ is the distance from $P$ to \emph{Fold} $A$ when it is subject to a slit $\gamma$. The relationships between $b$, $S$, $\phi$ and $\theta$ are defined from the colored triangular regions $\triangle$\emph{PAQ} and $\triangle$\emph{HBG}.}
\label{curvecoord}
\end{figure}

 It is possible study the infinitesimal region bounded by an \emph{inner polygon} and an \emph{outer polygon}, as shown by Fig. \ref{curvecoord}, to find the governing equations that render the geometry of \emph{Fold} $A$. Thus, by observing the relation among $dR$, $Rd\theta$, and $\phi$ in the triangle $\triangle$\emph{PMH}, the following holds

\begin{equation}\label{eq:tanphicu}
\tan(\phi) = \frac{R d\theta}{dR},
\end{equation}
then, from $\triangle$\emph{PAQ} in Fig. \ref{curvecoord}

\begin{equation}\label{bcoscu}
b = S\cos\phi,
\end{equation}
and from $\triangle$\emph{HBG} in Fig. \ref{curvecoord}

\begin{equation}\label{bSdScu}
b + db = (S + dS)\cos(\phi - d\theta);
\end{equation}
in which by combining Eq. \ref{bcoscu} and Eq. \ref{bSdScu} we obtain the following estimation:

\begin{equation}\label{dscu}
\frac{dS}{S} = \frac{db}{S\cos \phi}  -\tan(\phi) d\theta.
\end{equation}

Also, from $\triangle$\emph{GCQ} in Fig. \ref{curvecoord}, we find

\begin{equation}\label{singamma}
\sin \gamma =  \cfrac{-db}{   \mfrac{dR}{\cos (\phi - \gamma)}}.
\end{equation}

\begin{figure}[t]
%\hfill
\begin{center}
{\includegraphics[width=1\textwidth]{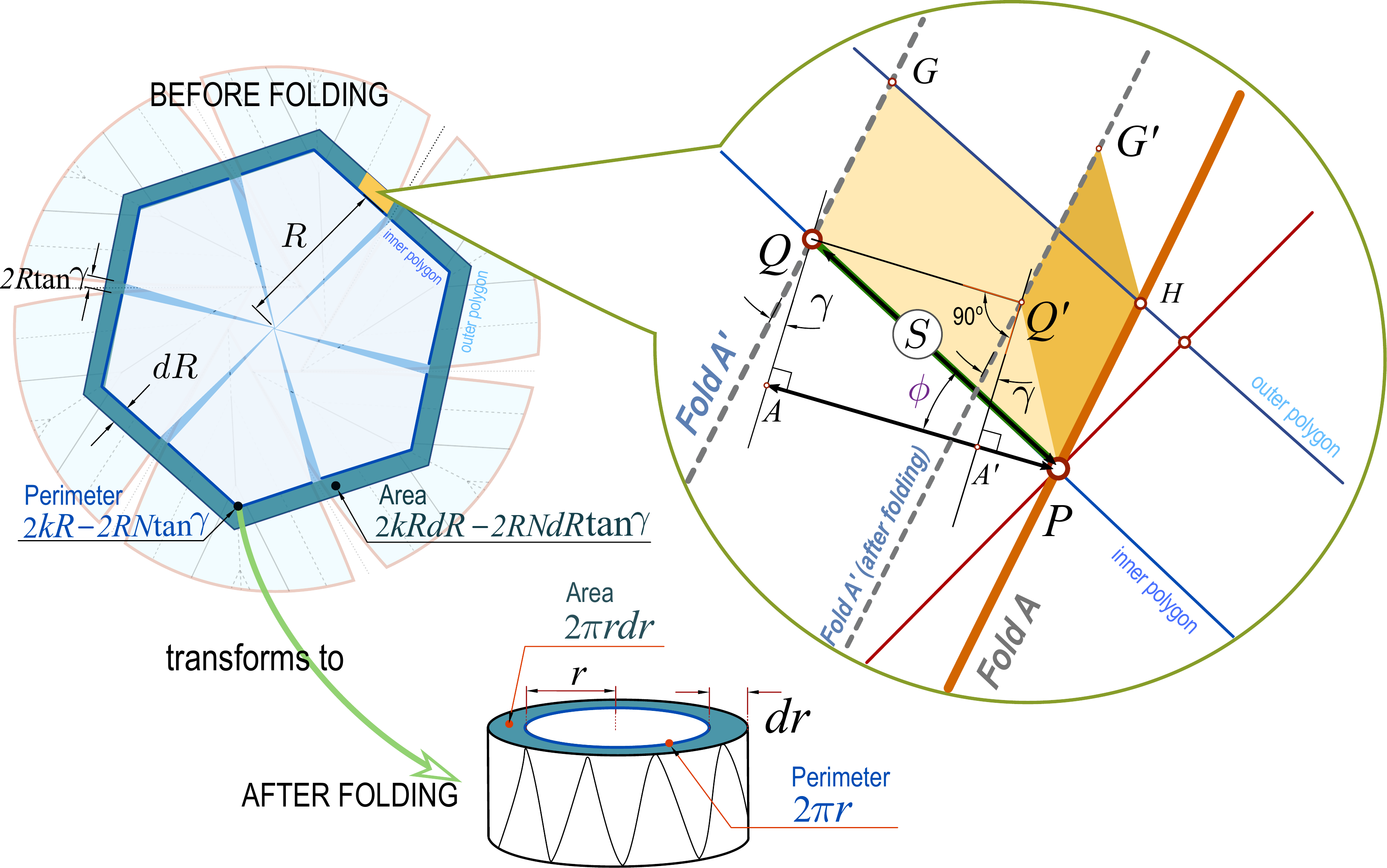}}
\end{center}
\caption{Crease geometry showing the relationship between the flat and folded configurations of the membrane. The infinitesimal region between the \emph{inner polygon} and the \emph{outer polygon} folds into a cylindrical shell with radius $r$ and thickness $dr$. The infinitesimal region $PQGH$ changes into $PQ'G'H$ during the folding process, bounded by the thickness $t$ between folded membranes, in which $Q$ ($G$) transforms into $Q'$ ($G'$).}
\label{curvebeforeafter}
\end{figure}

Furthermore, it is possible to study the changes during the folding behaviour between planar and cylindrical configurations, Fig. \ref{curvebeforeafter} shows the relations between an infinitesimal part of the flat membrane and a cylindrical shell. Here, we outline the changes relating lengths and areas between the flat and the cylindrical configurations. The perimeter of the base of the cylindrical shell is $2\pi r$ whereas the perimeter of its corresponding flat polygonal region is $2kR - 2RN\tan \gamma$, in which $k$ is obtained by Eq. \ref{kk}. Thus, from Fig. \ref{curvebeforeafter}, the segment $\overline{PQ'}$ is related to the segment $\overline{PQ}$ by the ratio

\begin{equation}\label{pqratiocu}
\frac{{PQ'}}{{PQ}} = \frac{2\pi r}{2kR - 2RN\tan \gamma},
\end{equation}
such that for membrane thickness $t$,

\begin{equation}\label{pqdashcu}
PQ' = \sqrt{(S \sin \phi)^2 + t^2}.
\end{equation}

Given that $S = PQ$, the above can be arranged to be

\begin{equation}\label{phiplanarcu}
\frac{\pi r}{kR - RN\tan \gamma} = \sqrt{\sin^2\phi + \Big (\frac{t}{S} \Big )^2}.
\end{equation}

Similarly, the area of the upper face of the cylindrical shell is $2\pi r dr$ whereas the area of its corresponding polygonal region is $2kRdR - 2RNdR \tan \gamma$. Thus, the corresponding areas are related by

\begin{equation}\label{pir2v1cu}
\frac{A_{PQ'G'H}}{A_{PQGH}} = \frac{2\pi rdr}{2kRdR - 2RNdR \tan \gamma},
\end{equation}
where $A_{PQ'G'H}$ denotes the area of the region $PQ'G'H$, and $A_{PQGH}$  denotes the area of the region $PQGH$. Eq. \ref{pir2v1cu} can be approximated by

\begin{equation}\label{phiplan2cu}
\frac{t}{S\cos \phi} = \frac{\pi rdr}{(kR - RN \tan \gamma)dR}.
\end{equation}

Finally, the governing equations outlining the geometry of \emph{Fold} $A$ are defined by Eq. \ref{eq:Rarc} (or Eq. \ref{eq:Rcu}), Eq. \ref{eq:gammacu}, Eq. (\ref{eq:tanphicu}), Eq. (\ref{dscu}), Eq. \ref{singamma}, Eq. (\ref{phiplanarcu}), Eq. (\ref{phiplan2cu}) and Eq. (\ref{kk}). It is also possible to define the system of differential equations able to render \emph{Fold} $A$ with respect to $R$. As such, after accommodating terms in Eq. \ref{eq:Rarc}, Eq. (\ref{eq:tanphicu}), Eq. (\ref{dscu}), Eq. \ref{singamma} and Eq. (\ref{phiplan2cu}), respectively, we find

\begin{equation}\label{dudRcu}
\frac{du}{dR} = \frac{2a}{\sqrt{4a^2+{u}^2}},
\end{equation}

\begin{equation}\label{dthetadRcu}
\frac{d\theta}{dR} = \frac{\mathrm{tan}\left(\phi \right)}{R},
\end{equation}

\begin{equation}\label{dSdRcu}
\frac{dS}{dR} = -\frac{\sin(\gamma)}{\cos(\phi )\cos(\gamma-\phi )}-\frac{S {\mathrm{tan}^2(\phi )} }{R},
\end{equation}

\begin{equation}\label{dbdRcu}
\frac{db}{dR} =  -\frac{\sin(\gamma)}{\cos(\gamma-\phi )},
\end{equation}

\begin{equation}\label{drdRcu}
\frac{dr}{dR} = \frac{Rt( k-N\mathrm{tan}( \gamma) ) }{\pi S  r \cos( \phi ) },
\end{equation}
in which the slit angle $\gamma$ is defined by Eq. \ref{eq:gammacu}, the angle $\phi$ can be obtained from Eq. (\ref{phiplanarcu}), and the constant $k$ is defined by Eq. (\ref{kk}).

\begin{figure*}[ht]
\centering
%\hfill
\subfigure[$N = 3$]{\includegraphics[width=0.32\textwidth]{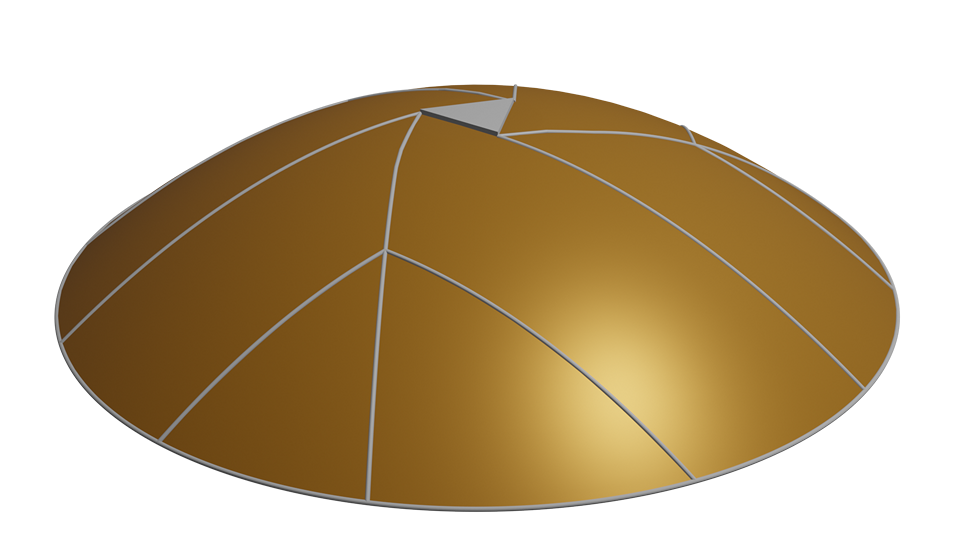}}
%\hfill
\subfigure[$N = 4$]{\includegraphics[width=0.32\textwidth]{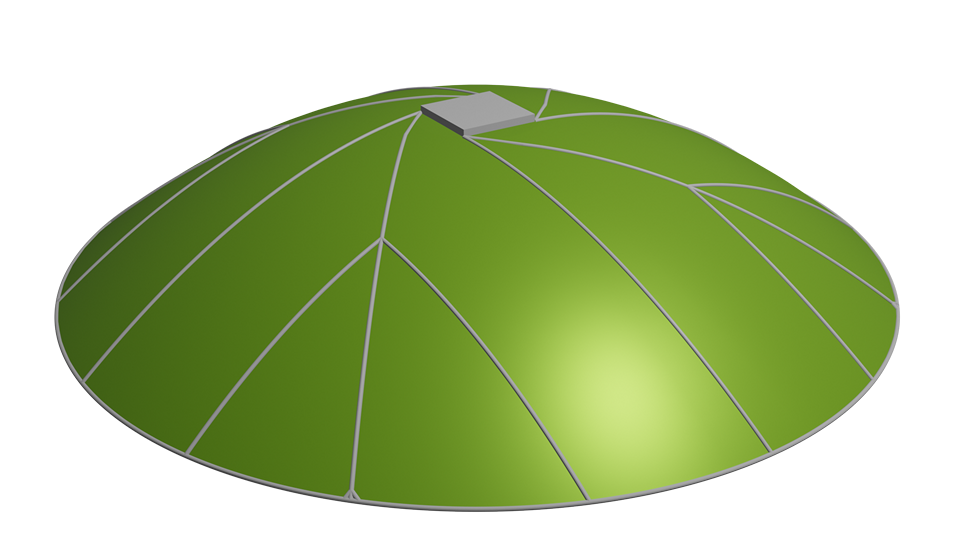}}
%\hfill
\subfigure[$N = 5$]{\includegraphics[width=0.32\textwidth]{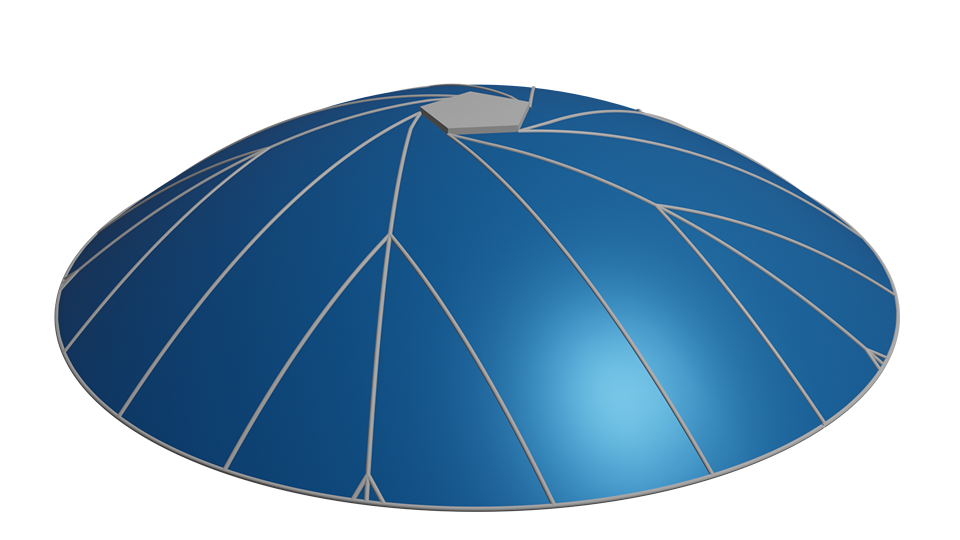}}
%\hfill
\subfigure[$N = 6$]{\includegraphics[width=0.32\textwidth]{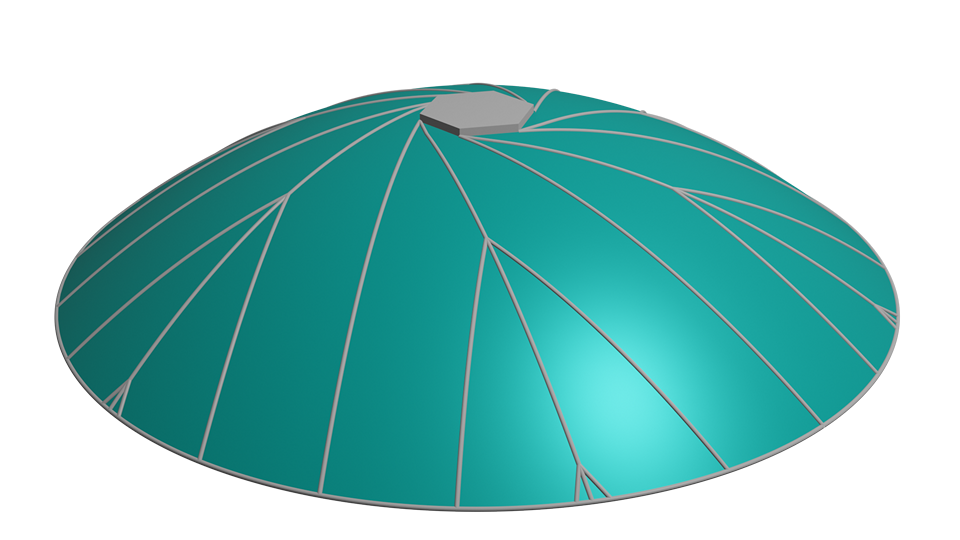}}
%\hfill
\subfigure[$N = 8$]{\includegraphics[width=0.32\textwidth]{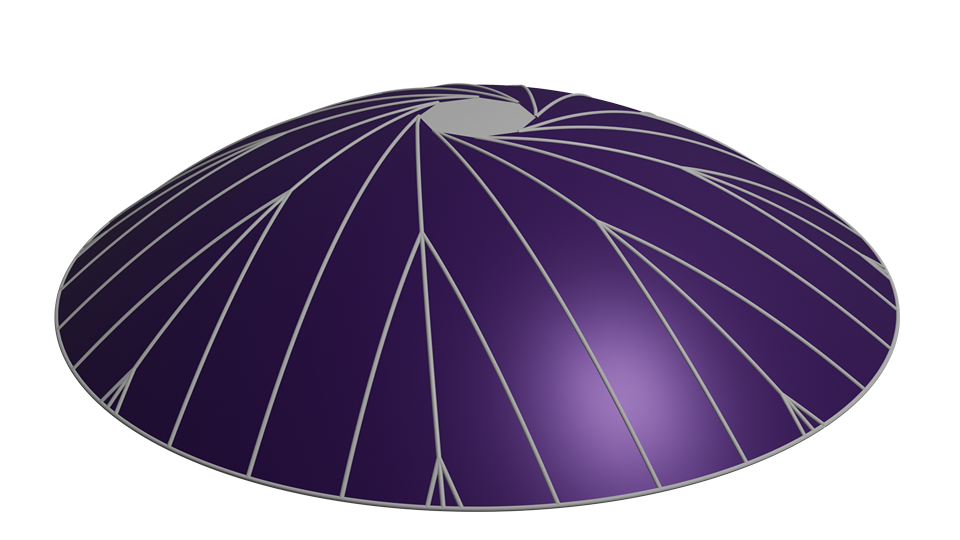}}
%\hfill
\subfigure[$N = 15$]{\includegraphics[width=0.32\textwidth]{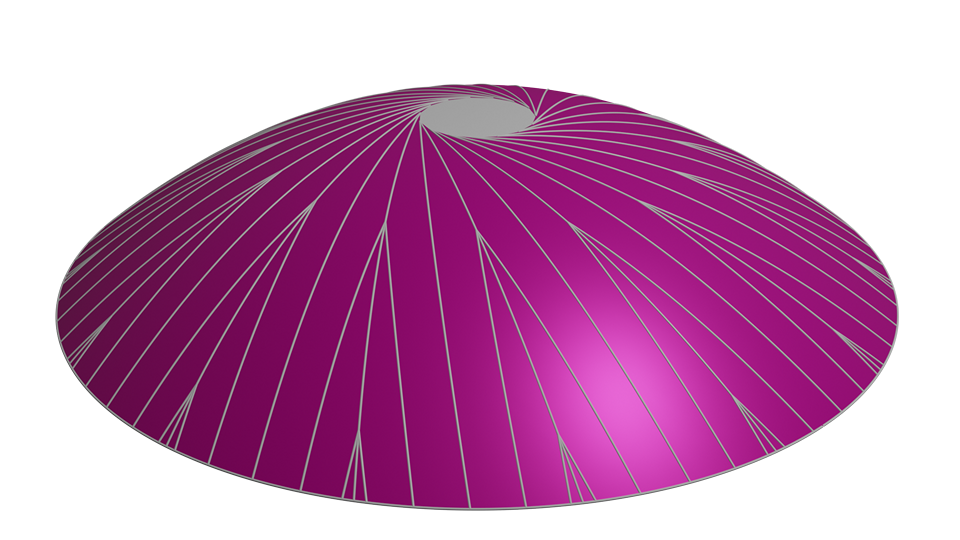}}
%\hfill
\caption{Basic layouts of spiral folding pattern in curved surface. Here, membrane parameters consider $R_0 = 15 \si{mm},~ R_f = 120\si{mm}, ~ t = 0.1 \si{mm}, ~ S_0 = 61 \si{mm}, ~ a = 70 \si{mm}$.}
\label{curvedcases}
\end{figure*}

The above-mentioned system of differential equations in Eq. \ref{dudRcu} - Eq. \ref{drdRcu} can be solved numerically for given focal length $a$, membrane parameters $k$, $t$, boundaries $[R_0, R_f]$, and initial conditions $u_0$, $\theta_0$, $S_0$, $b_0$ and $r_0$. As such, the solutions are generalizable to find folding patterns using arbitrary regular polygons in the core as shown by Fig. \ref{curvedcases}. Here, $u_0 = R_0= r_0$ denotes the circumradius of the core polygon of the flat membrane, $R_f$ denotes a user-defined upper bound on the value of $R$, and $b_0 = S_0\cos(2\pi/N)$ (due to Eq. \ref{bcoscu}). For a regular polygon with $N$ sides, $\theta_0$ is computed by Eq. \ref{theta0} and \emph{mountain}-\emph{valley} creases are assigned by Eq. \ref{foltype}.

%%%%%%%%%%%%%%%%%%%%%%%%%%%%%%%%%%%

\subsection{Multi-Spiral Folding Pattern of a Curved Surface}

We extend the governing equations described above to consider the multi-spiral folding patterns of a curved surface. Although the formulation of curved patterns is similar the case of non-curved surface, the consideration of curves and orientation becomes necessary. By following the same principle of superimposing the clockwise and the counterclockwise orientations (Fig. \ref{concept} and Fig. \ref{conceptb}), it becomes possible to model the multi-spiral folding patterns of a curved surface. In order to portray the basic idea of our approach, the coordinate system is shown by Fig. \ref{msection}. Here, the boundaries of the spirals with clockwise and the counterclockwise orientations are shown by green color, and the arbitrary point $P$ along within the $n$th area is shown by Fig. \ref{msection}-(b). When $n$-spiral is an odd (even) number, the orientation pattern is counterclockwise (clockwise). Then, the governing equations become as follows:

\begin{figure*}[t]
\begin{center}
     \subfigure[Rotation direction of spiral]{\includegraphics[width=0.48\textwidth]{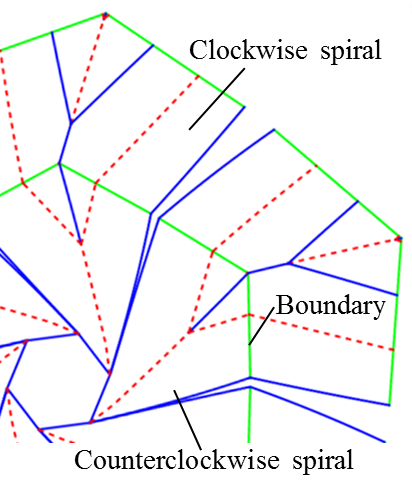}}
     \hfill
     \subfigure[Coordinate System]{\includegraphics[width=0.48\textwidth]{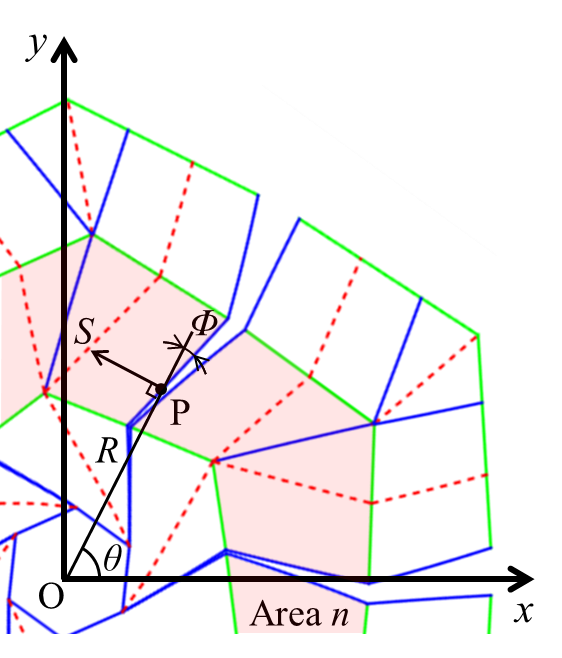}}
\end{center}
  \caption{Basic concept of the rotation direction of the spiral and coordinate system in curved surface.}
  \label{msection}
\end{figure*}

\begin{equation}\label{eq:tanphi}
\tan((-1)^{n-1}\phi_n) = \frac{R \cdot d\theta_n}{dR},
\end{equation}

\begin{equation}\label{eq:gamma}
  \gamma_n = \frac{\pi}{N\cdot R}\Big (R - 2\sqrt{a\cdot l_n} \Big),
\end{equation}

\begin{equation}\label{eq:R}
R = \sqrt{a \cdot l_n + l_n^2} + a.\sinh^{-1} \Bigg ( \sqrt{\frac{l_n}{a}} \Bigg),
\end{equation}

\begin{equation}\label{eq:ds}
\frac{dS_n}{{S_n}} = -\tan((-1)^{n-1}\phi_n).d\theta_n + \frac{db_n}{{S_n}.\cos((-1)^{n-1}\phi_n)},
\end{equation}

\begin{equation}\label{eq:slit}
\frac{db_n}{dR} = \frac{-\sin(\gamma_n)}{cos((-1)^{n-1}\phi_n-\gamma_n)},
\end{equation}

\begin{equation}\label{eq:len}
\frac{\pi r_n}{(kR - N\cdot R\tan\gamma_n)} = \sqrt{\sin^2 ( (-1)^{n-1} \phi_n ) + \frac{t^2}{S_n}},
\end{equation}

\begin{equation}\label{eq:area}
\frac{\pi r_n}{(k \cdot R - N \cdot R \tan\gamma_n)} \cdot \frac{dr_n}{dR} = \frac{t}{{S_n}\cdot \cos((-1)^{n-1} \phi_n)}.
\end{equation}

In the above formulations, the subscript $n$ indicates the ordinal number between boundaries with bellows folding (circumferential spirals). The folding lines that form the bellows are boundaries dividing clockwise and the counterclockwise patterns. And the boundary conditions on the $n$th spiral are defined as follows:

\begin{equation}\label{thetann}
\theta_{n+1,n} = \theta_{n,n}
\end{equation}

\begin{equation}\label{thetann}
\gamma_{n+1,n} = \gamma_{n,n}
\end{equation}

\begin{equation}\label{snn}
S_{n+1,n} = S_{n,n}
\end{equation}

\begin{equation}\label{snn}
b_{n+1,n} = b_{n,n}
\end{equation}

\begin{equation}\label{snn}
l_{n+1,n} = l_{n,n}
\end{equation}

\begin{equation}\label{rnn}
r_{n+1,n} = r_{n,n} + 2t
\end{equation}

\begin{figure*}[t]
\begin{center}
     \subfigure[Layout]{\includegraphics[width=0.5\textwidth]{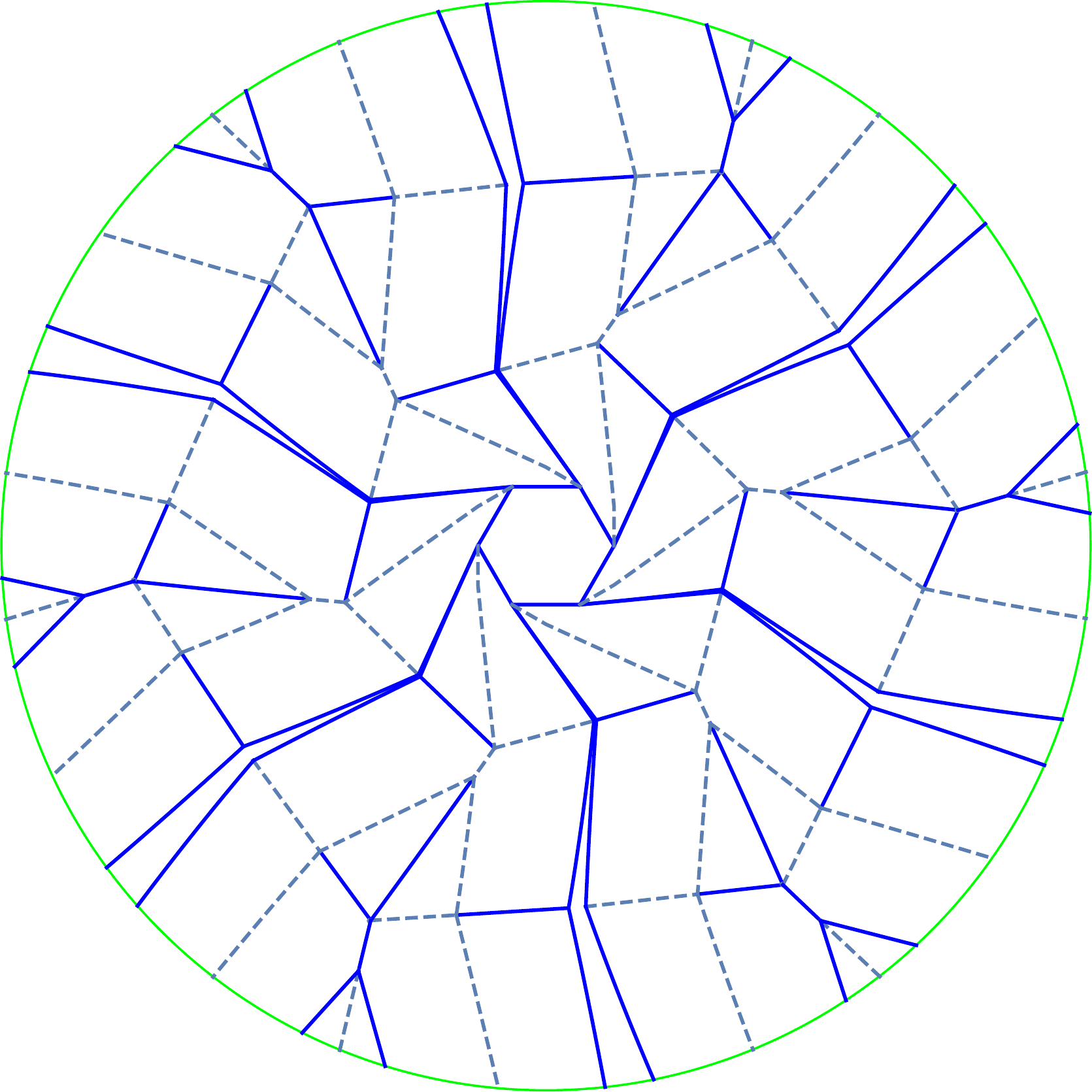}}
     \hfill
     \subfigure[Folding Process]{\includegraphics[width=0.98\textwidth]{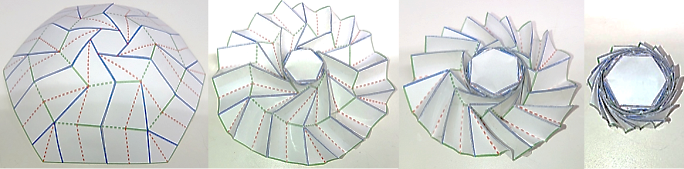}}
\end{center}
  \caption{Basic concept of the layout and folding of a curved surface with multi-spiral folds.}
  \label{mssample}
\end{figure*}

The above conditions are formulated due not only the superimposition of multiple spirals enabling to form bellows folding, but also due to our aim to consider the (double) thickness effect during the folding process. Thus, the angles $\theta_n$, $\gamma_n$, the radius $r_n$, the fold intervals $S_n$, $b_n$ and the height $l_n$ are defined as $\theta_{n,n}$, $\gamma_{n,n}$, $r_{n,n}$, $S_{n,n}$, $b_{n,n}$ and $l_{n,n}$ when the distance $R$ is equal to the radius $R_n$. The second numerical subscript $n$ in the above-mentioned formulations indicates the value of the variable when $R$ is equal to $R_n$.

In order to exemplify the multi-spiral folding pattern of curved sufaces, Fig. \ref{mssample} shows the layout of the folding pattern of a curved membrane and its folding process. Here, continuous lines denote the mountain folding lines, whereas, dashed lines denote the valley folding lines.

\section{Experiments}

In order to evaluate the deployment characteristics of the proposed folding approach, we performed experiments considering the folding and unfolding of planar and curved membranes. In this section we describe our experiment setup, analysis and obtained results.

\begin{figure}[t]
\begin{center}
{\includegraphics[width=0.98\textwidth]{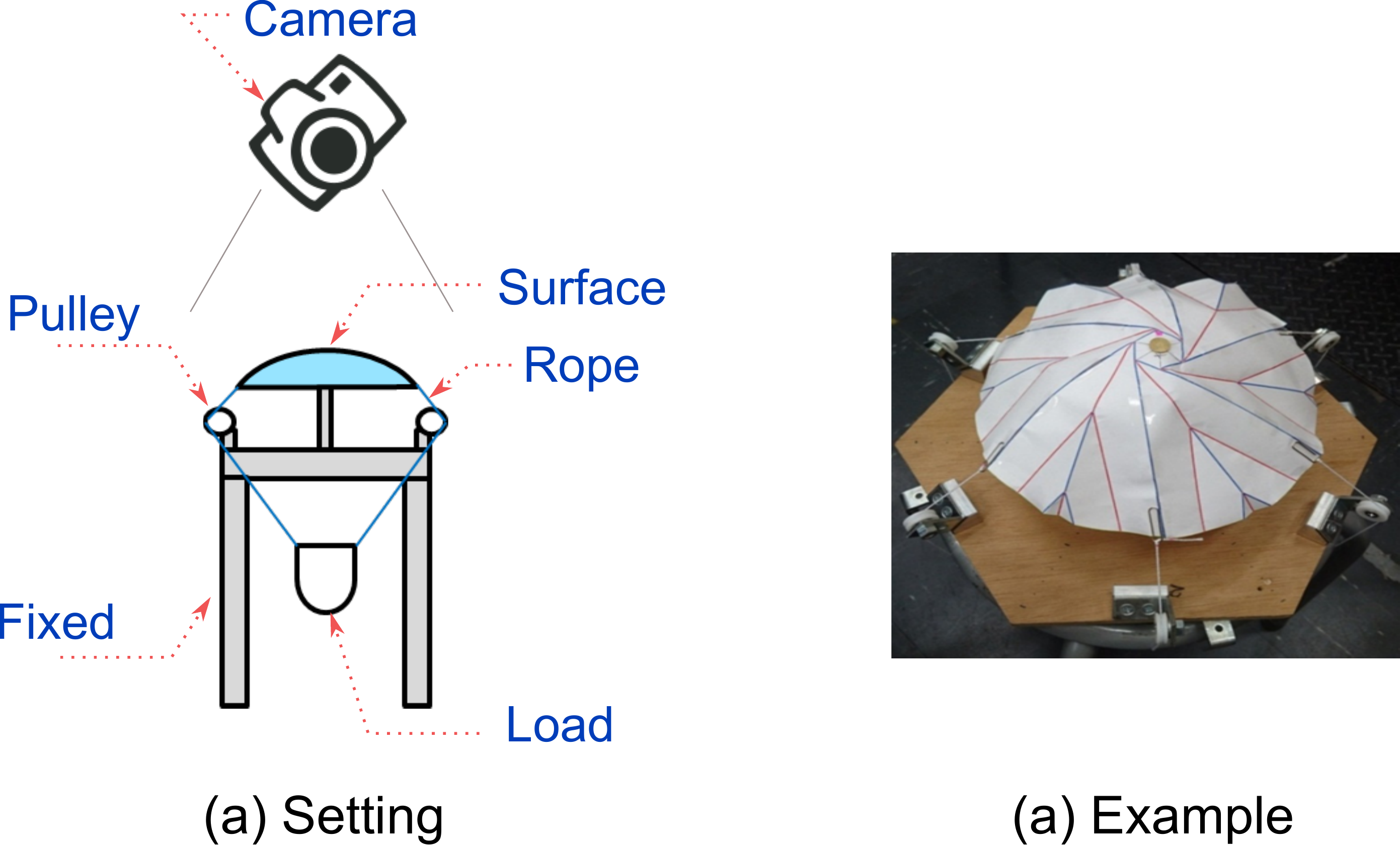}}
\end{center}
\caption{Basic idea of the experiment setup.}
\label{expe}
\end{figure}

\begin{figure*}[h]
\centering
%\hfill
\subfigure[$t*$]{\includegraphics[width=0.25\textwidth]{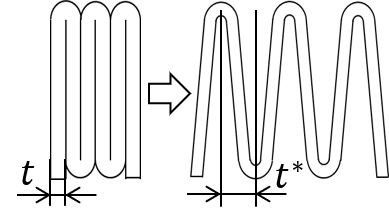}}
\hfill
\subfigure[Buckling]{\includegraphics[width=0.22\textwidth]{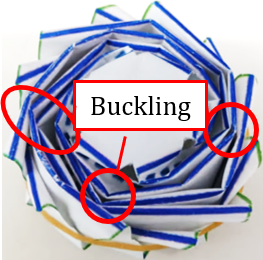}}
\hfill
\subfigure[Hole]{\includegraphics[width=0.25\textwidth]{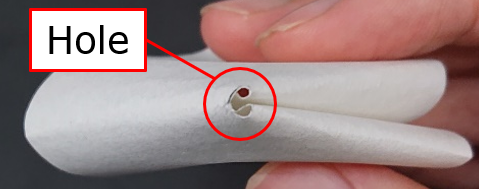}}
\hfill
\subfigure[Fold with holes]{\includegraphics[width=0.22\textwidth]{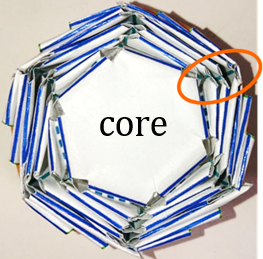}}
%\hfill
\caption{Basic concept of thickness $t$ and $t*$ and fold with holes.}
\label{tt}
\end{figure*}

\begin{figure*}[h]
\begin{center}
     \subfigure[Model 1]{\includegraphics[width=\mosize\textwidth]{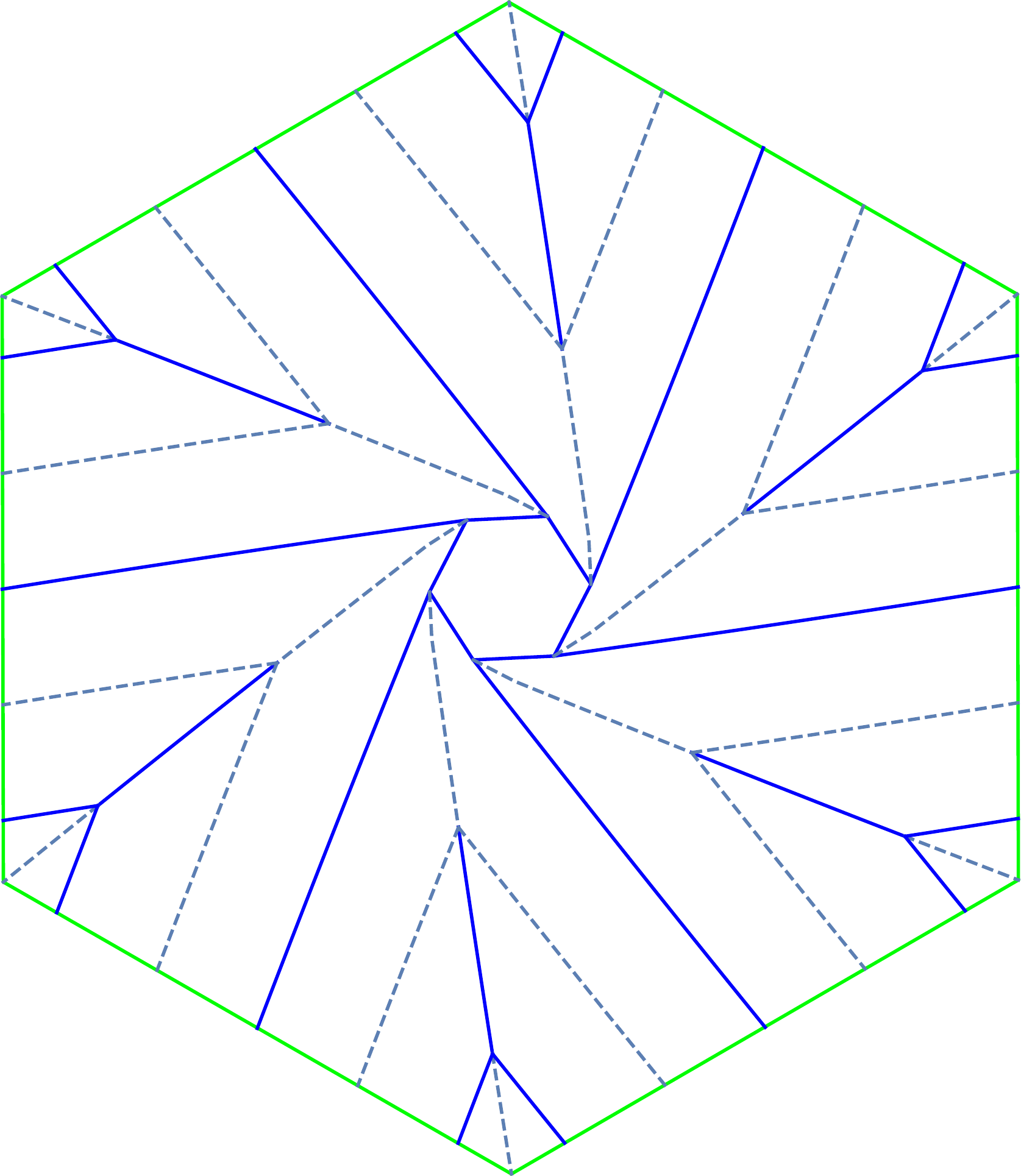}}
     \subfigure[Model 2]{\includegraphics[width=\mosize\textwidth]{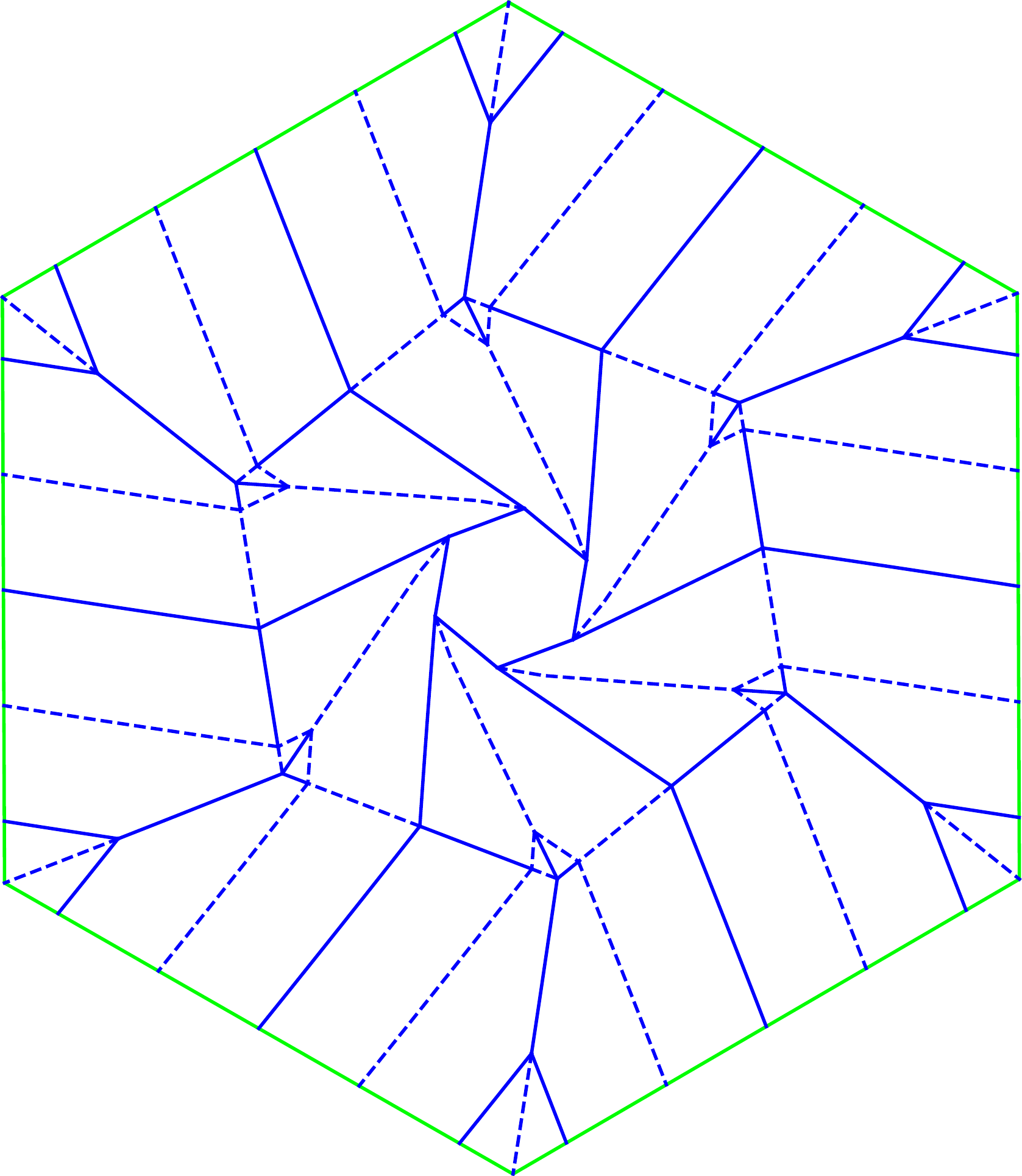}}
     \subfigure[Model 3]{\includegraphics[width=\mosize\textwidth]{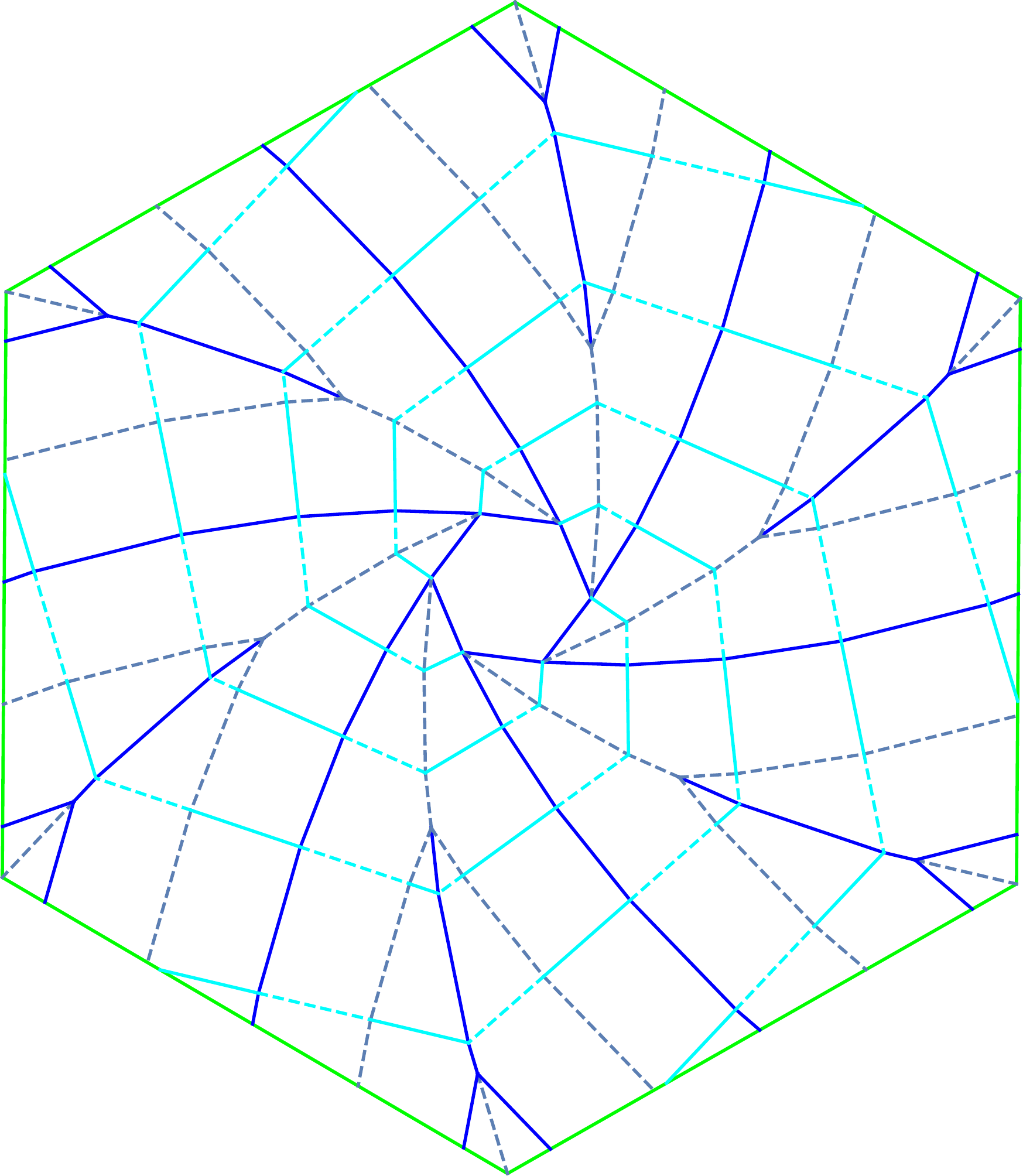}}
     \subfigure[Model 4]{\includegraphics[width=\mosize\textwidth]{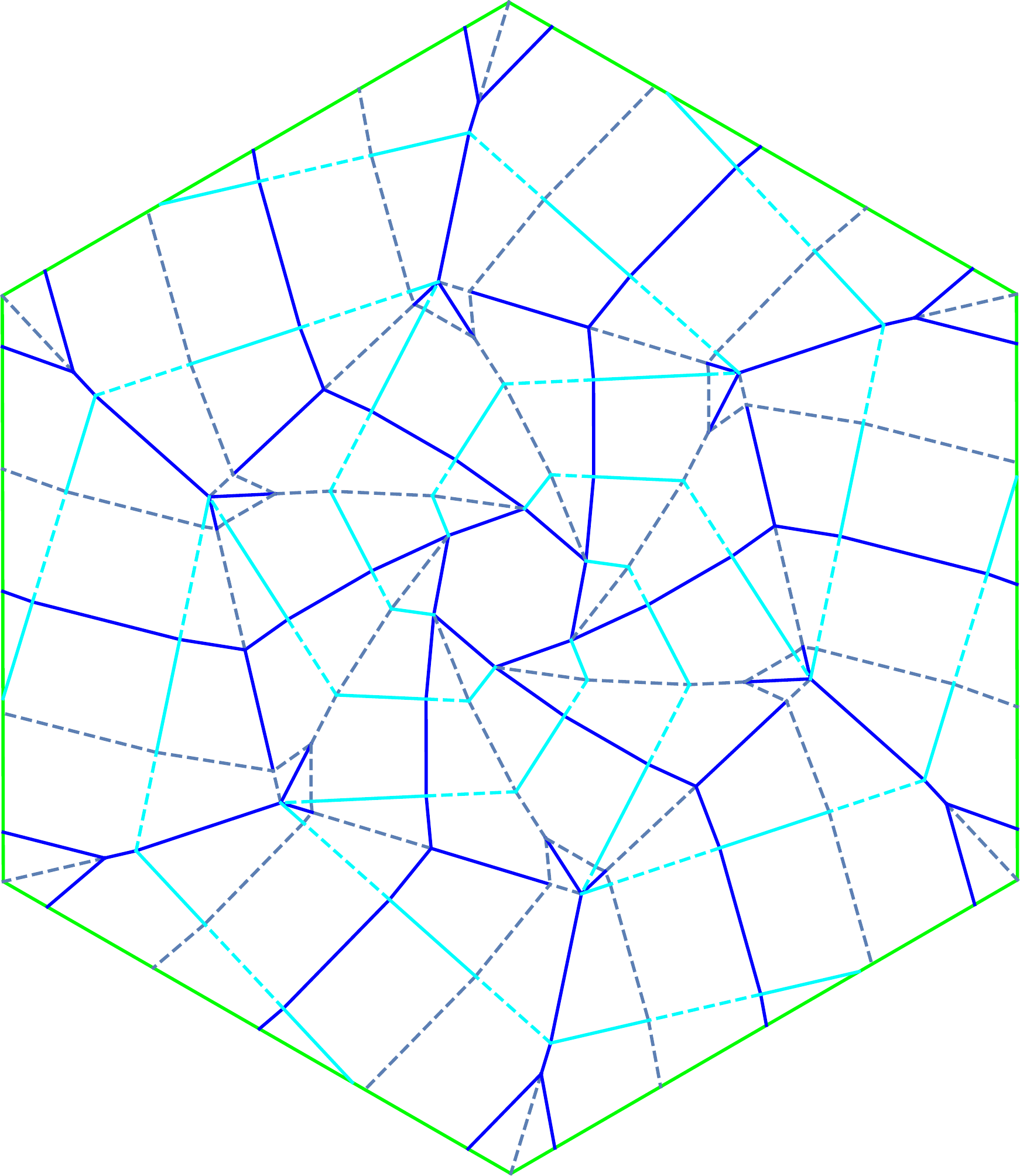}}
     \subfigure[Model 5]{\includegraphics[width=\mosize\textwidth]{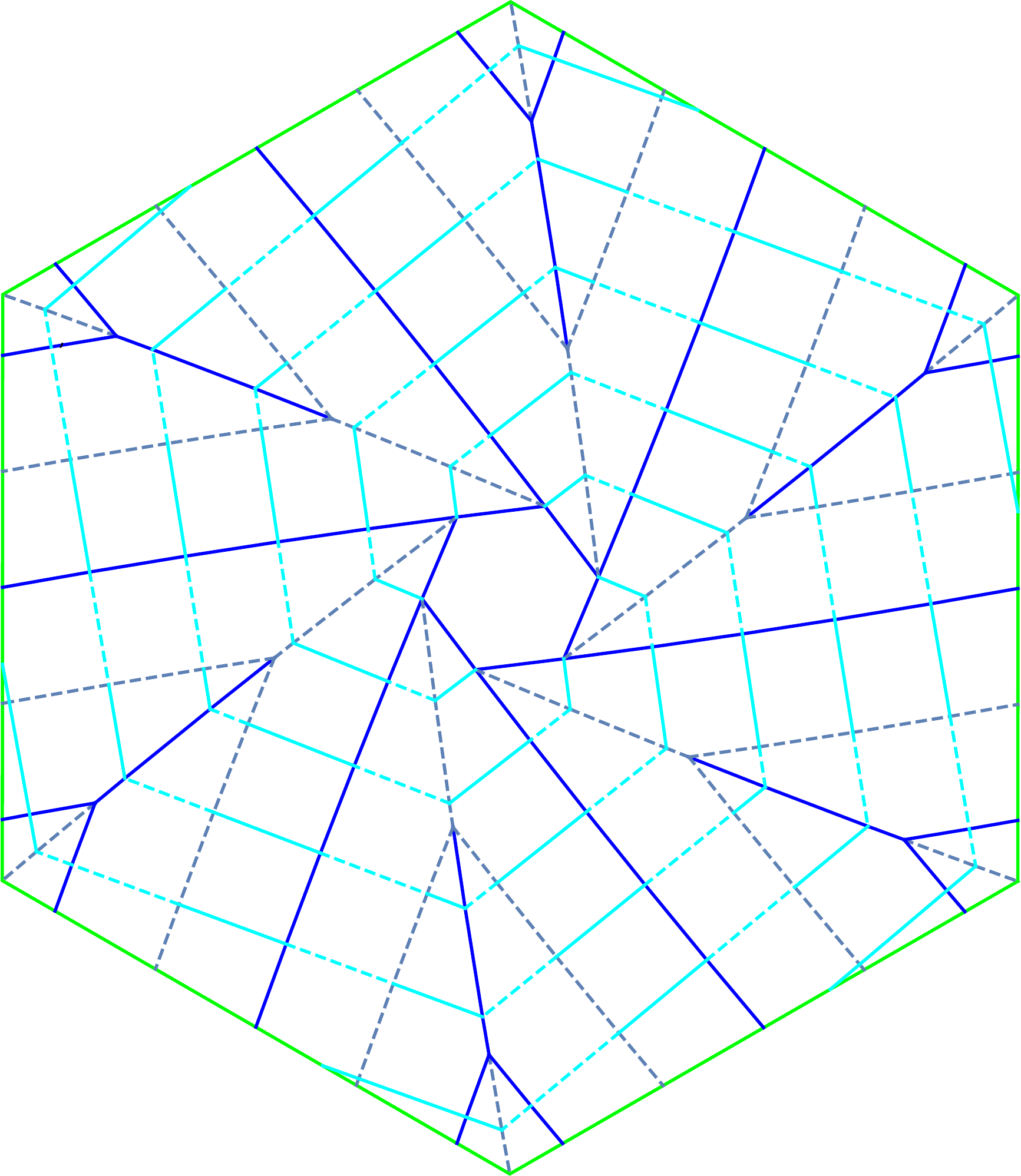}}
     \subfigure[Model 6]{\includegraphics[width=\mosize\textwidth]{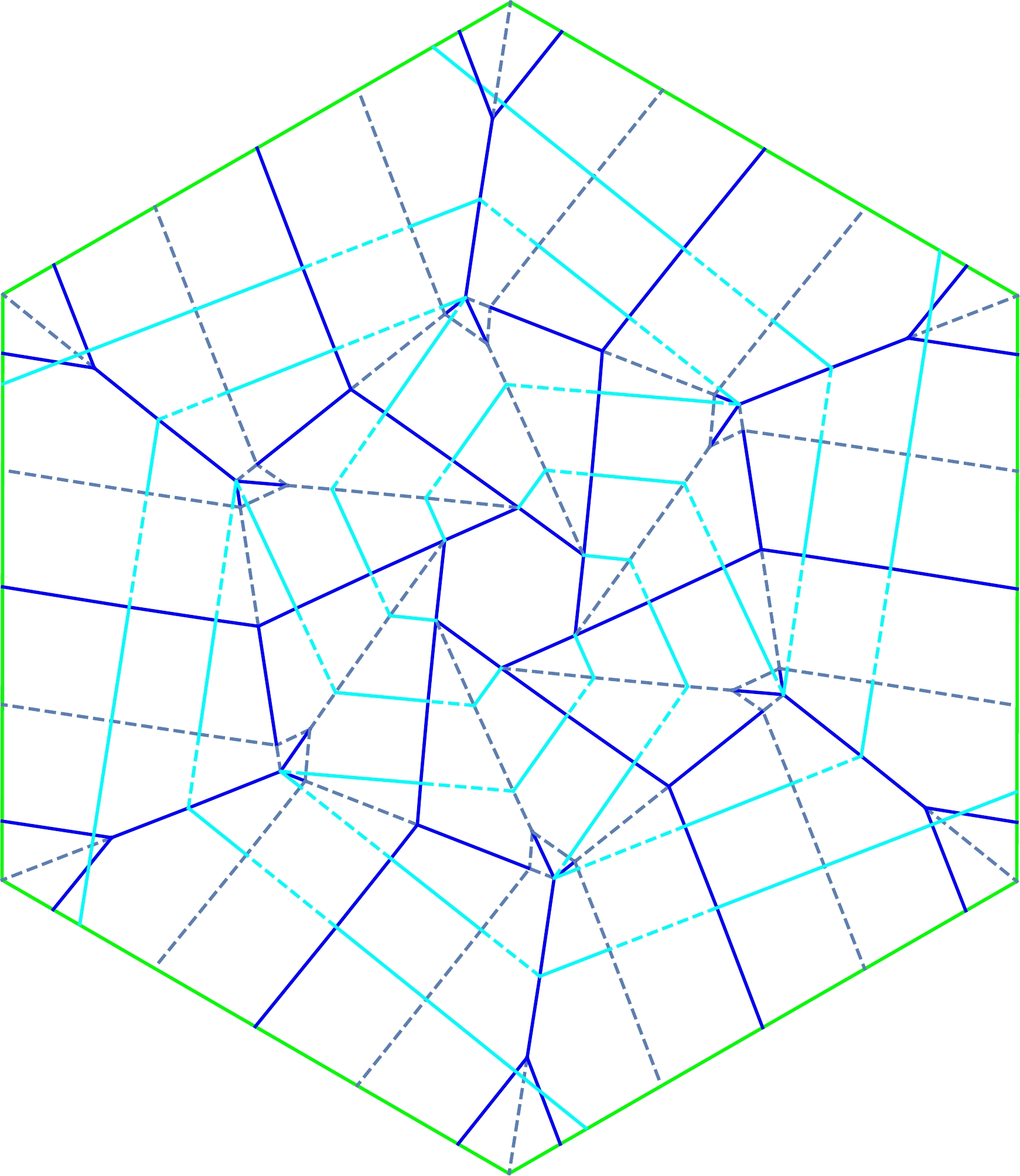}}
\end{center}
  \caption{Layouts of planar membranes with multi-spiral folding.}
  \label{planar}
\end{figure*}

\begin{figure*}[h]
\begin{center}
     \subfigure[Model 7]{\includegraphics[width=\mosize\textwidth]{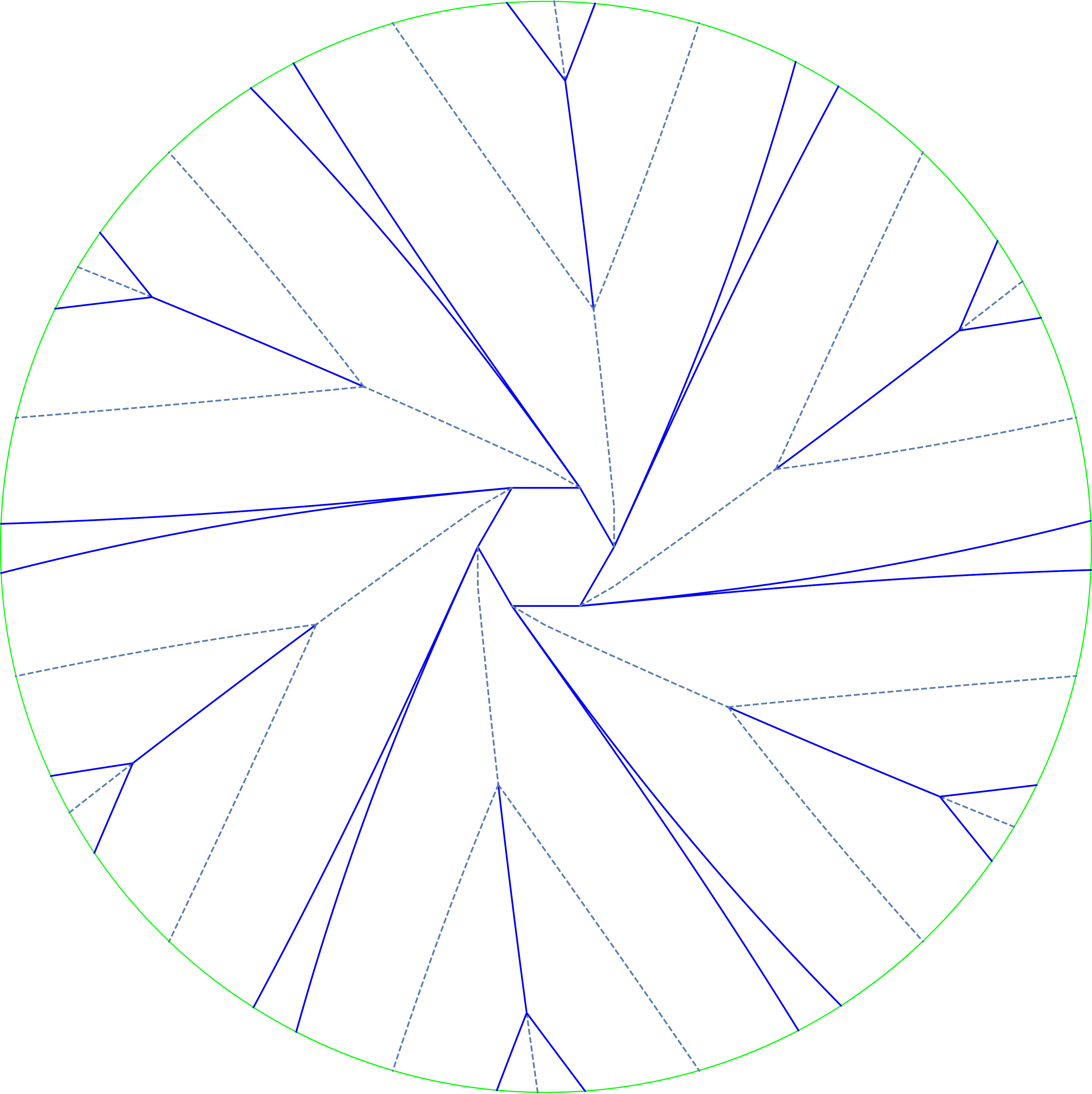}}
     \subfigure[Model 8]{\includegraphics[width=\mosize\textwidth]{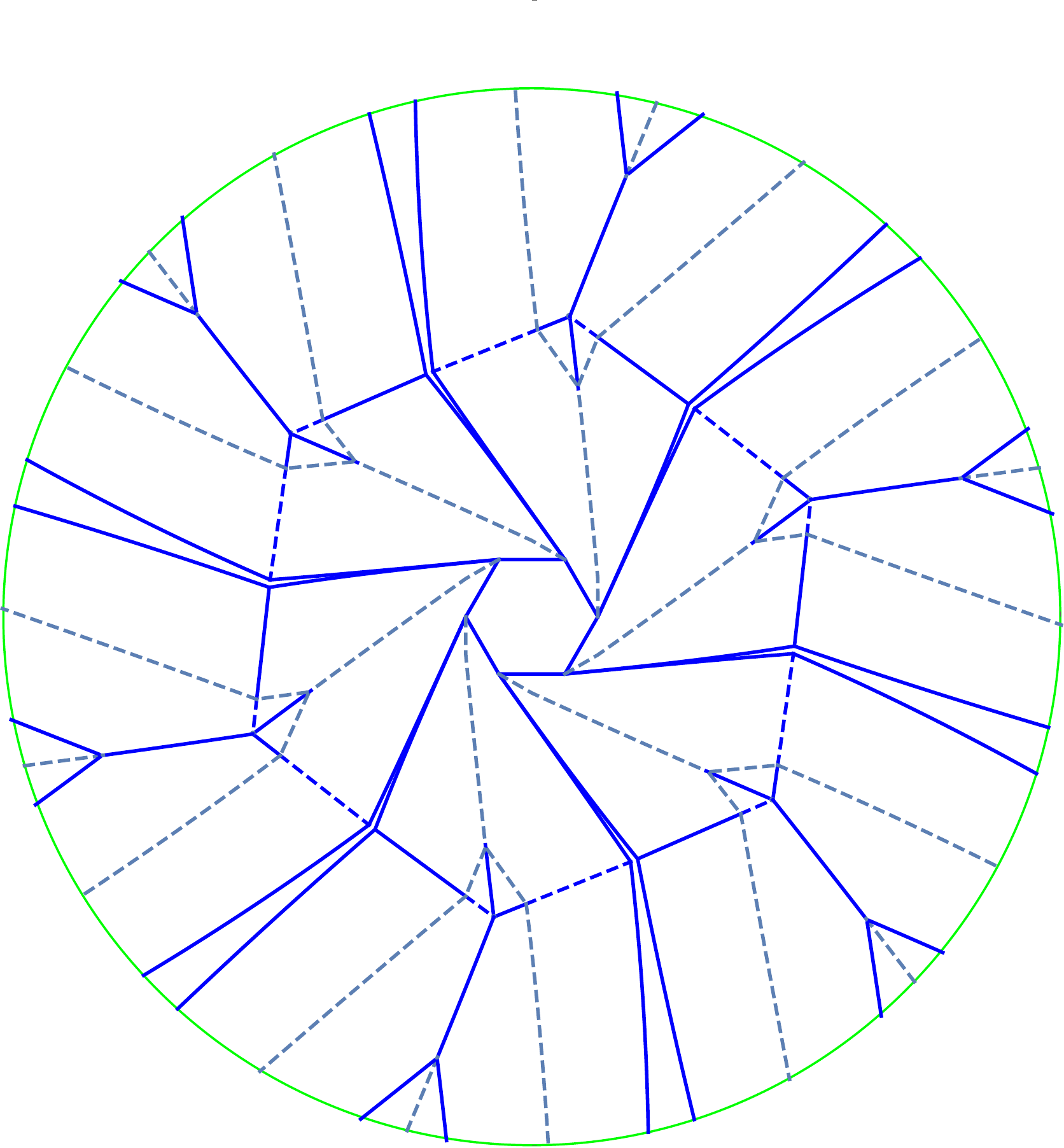}}
     \subfigure[Model 9]{\includegraphics[width=\mosize\textwidth]{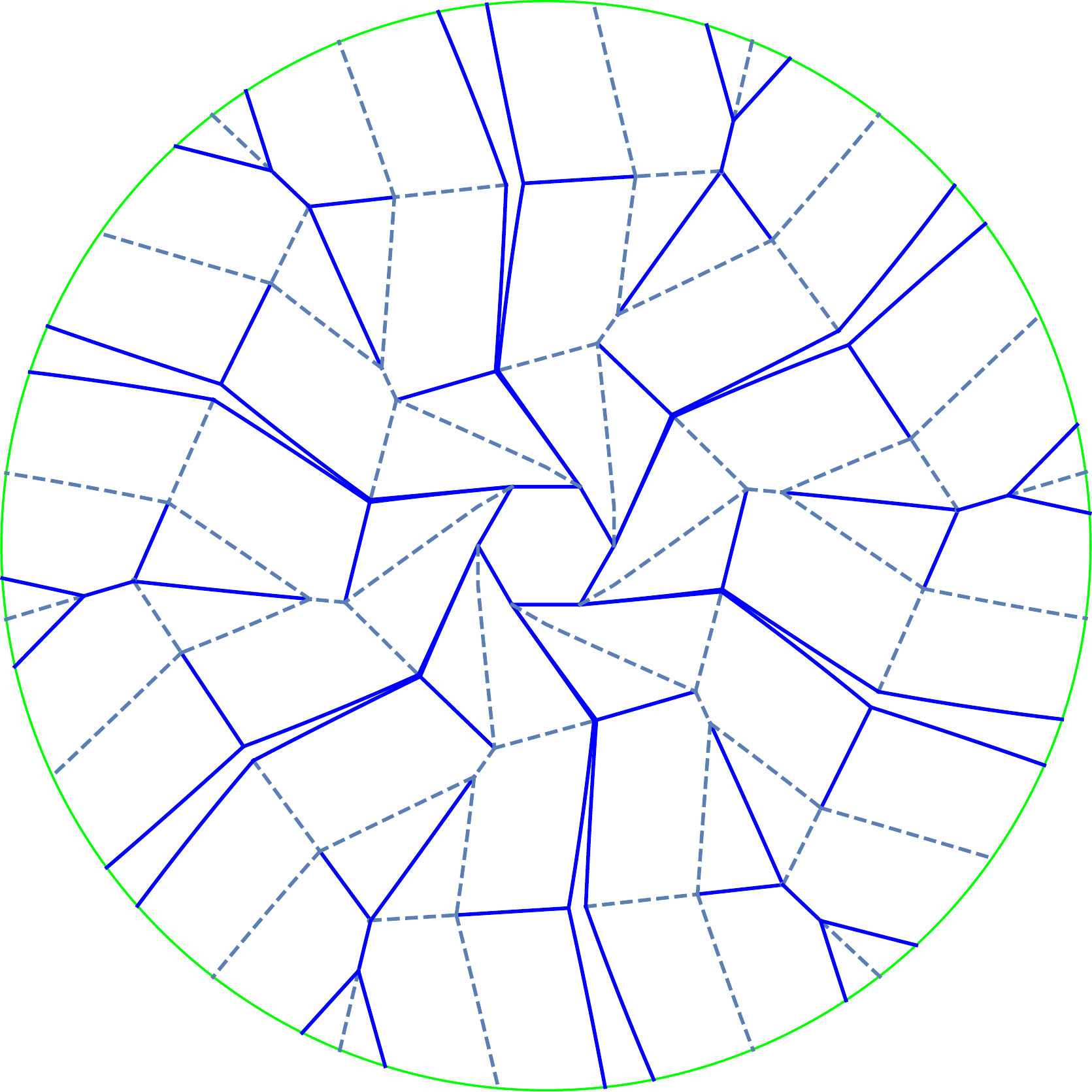}}
     \subfigure[Model 10]{\includegraphics[width=\mosize\textwidth]{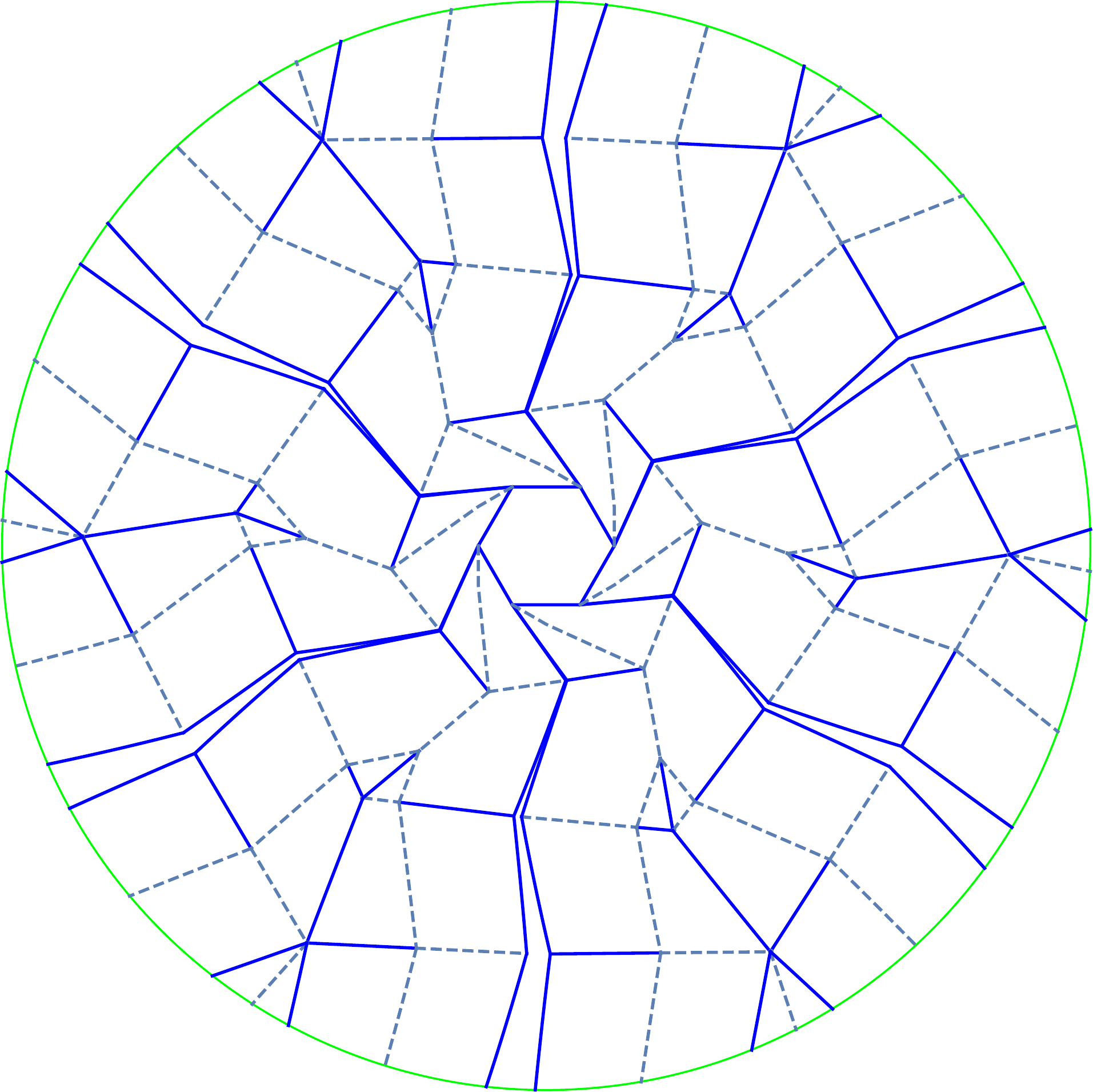}}
     \subfigure[Model 11]{\includegraphics[width=\mosize\textwidth]{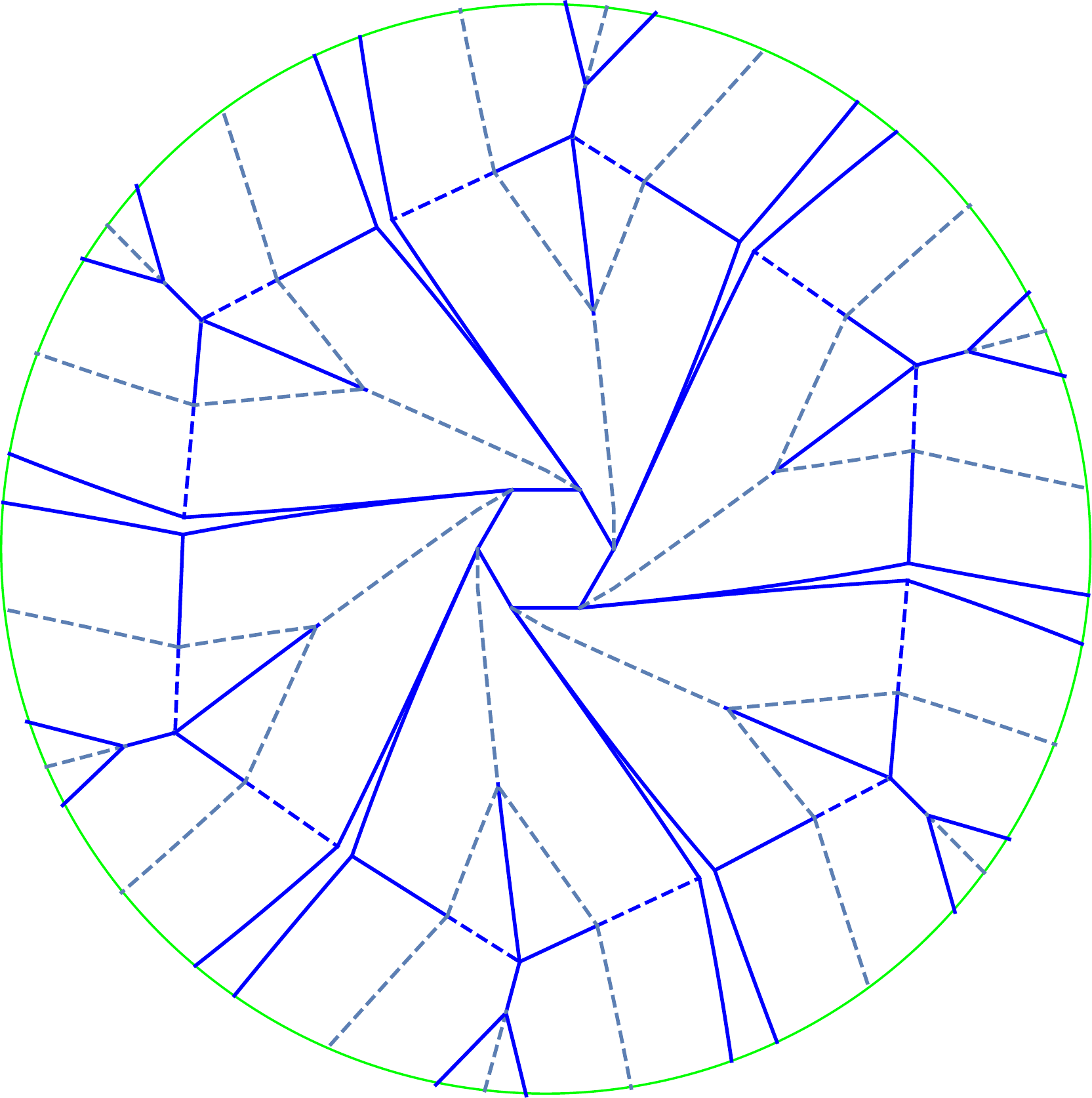}}
     \subfigure[Model 12]{\includegraphics[width=\mosize\textwidth]{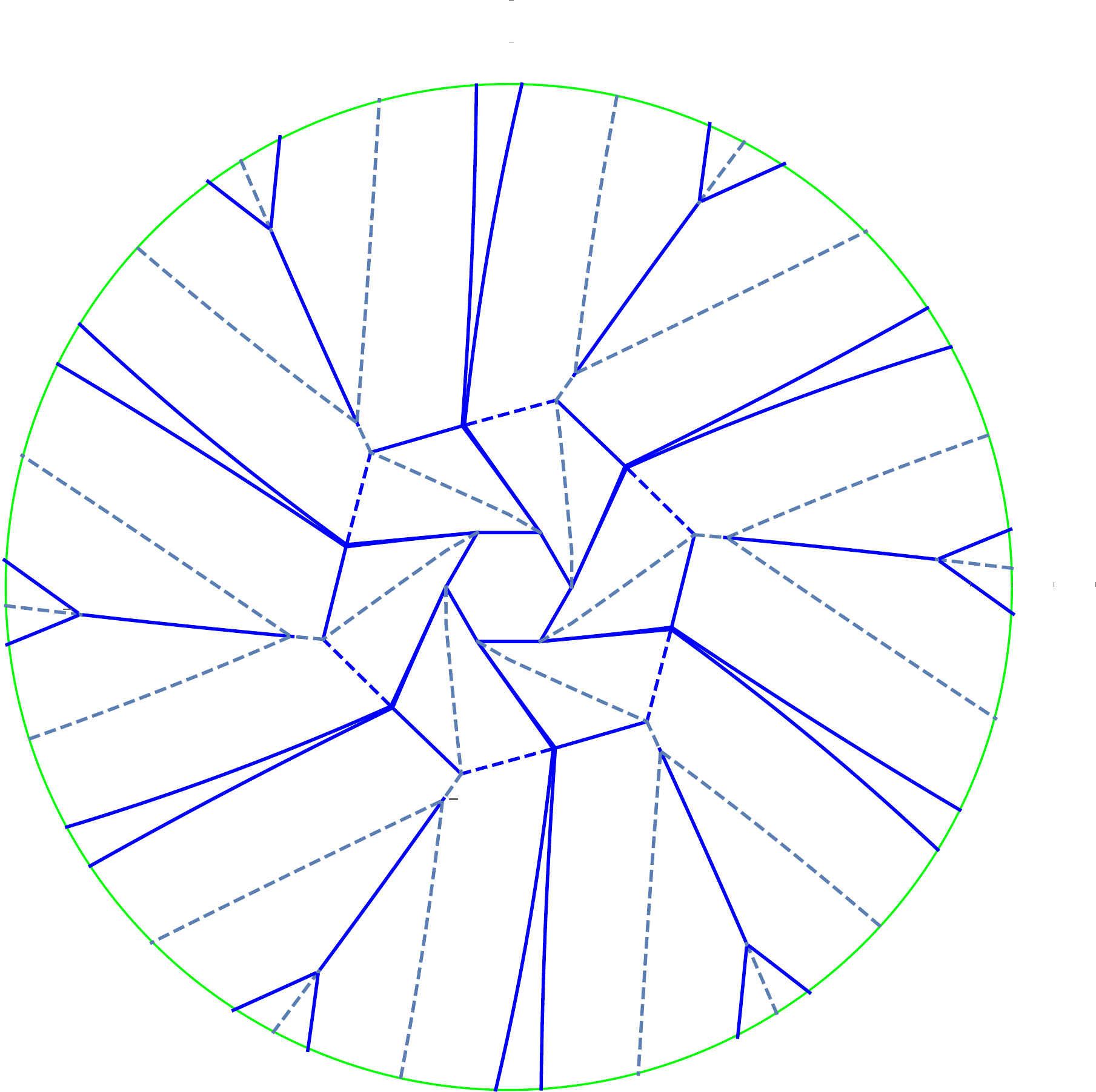}}
\end{center}
  \caption{Layouts of curved membranes with multi-spiral folding.}
  \label{curved}
\end{figure*}

\subsection{Settings}

The basic idea of our experimental setup is portrayed by Fig. \ref{expe}. Basically, we evaluated the following types of spiral folding methods: the planar and curved spiral folding with and without multiple spirals under thickness considerations (Fig. \ref{tt}). For deployment, a method that is less affected by the deployment intrinsics is desirable in order to contrast the deployment performance with respect to differences in folding methods. As such, inspired by the tensile deployment methods, we adopt a method of expanding the membrane by applying tension to the edges of the membrane model. We used threads with clips attached to the edges of the membranes, and loads inserted through a container below the membrane. Basically, we performed our experiments as follows:

\begin{itemize}
  \item Fold a new membrane model.
  \item Install and fixate the folded membrane in the center of the experimental device.
  \item Through the pulley, the membrane surface and the container for loading the weight are connected by a thread and a clip.
  \item Add 9 \si{g} of weight each time, and log the deployment performance with a camera from above.
  \item At attained equilibrium, log the load (\si{mN}) and the deployment radius (\si{mm}).
\end{itemize}

By inserting weights (with 9 \si{g} mass) into the container consecutively, tension is generated in the membranes, thus the deployment characteristics is obtained from the relationship between the applied tension (load), membrane expansion and the actual deployment radius.

Also, the influence of the film thickness during folding is larger than the actual film thickness $t$ when compounded over the spiral pattern. In other words, the thickness of compounding multiple layers is larger than the expected integer multiples of $t$. We rather consider a virtual thickness $t* >t$ (Fig. \ref{tt}-(a)) to model the kind of buckling effect observed due to the winding of several layers of thick membranes (Fig. \ref{tt}-(b)). Since it is difficult to predict where the buckling effect is to occur, for simplicity we use reasonably small holes at the corners of creases (Fig. \ref{tt}-(c)) to avoid the kind of buckling effect due to winding, thus enabling to make a compact folding while ensuring the virtual thickness to be within the upper bound $t*$ (Fig. \ref{tt}-(d)).

In line of the above-mentioned concepts, we evaluated the following types of planar layouts, as shown by Fig. \ref{planar}:

\begin{itemize}
\item Model 1. Spiral fold with $t = 0.1$ mm.; this model is inspired by the conventional formulation of the spiral folding with planar membrane as shown by Fig. \ref{flatspiral} - Fig. \ref{baflat}.
\item Model 2. Multiple spiral folding, with $t = 0.1$ mm.; this model takes into account the multi-spiral folding pattern of the above layout, in which superimposition of the clockwise and counterclockwise orientations is realized by two concentric spirals.
\item Model 3. Spiral fold, with $t* = 1$ mm.; this model considers the formulation of archimedean-type spiral fold segments originating at (being normal to) vertices (edges) of the polygon hub\citep{pelle92,nojima01a,nojimavipsi,thick13}.
\item Model 4. Multiple spiral folding, with $t* = 1$ mm.; this model considers the spiral configuration of Model 3 and the superimposition principle of multiple membrane layers. For simplicity and without loss of generality, we considered two membranes.
\item Model 5. Spiral fold, with $t = 0.1$ mm.; this model considers the spiral configuration of Model 3.
\item Model 6. Multiple spiral folding, with $t = 0.1$ mm.; this model considers the spiral configuration of Model 3 and straight folding lines under the principle of multiple spirals.
\end{itemize}

In the above configurations, we used the a regular polygonal hub with $N = 6$ and inscribed circle radius of the planar membrane $R_{f} = 110$ mm, the radius of the core circumcircle $r_0 = 17.5$ mm., the crease spacing $b = 25$ mm., and membrane thickness $t= 0.1$ mm. Fine tuning of the above parameters is out of the scope of this paper.

Furthermore, in addition to the above-mentioned planar membranes, we evaluated the deployment characteristics of the following curved surfaces, as shown by Fig. \ref{curved}:

\begin{itemize}
\item Model 7. Curved spiral fold with $n = 1$; this model is inspired by the spiral folding with curved surface as shown by Fig. \ref{samplecurve}.
\item Model 8. Multiple spiral folding, with $n = 2$; this model takes into account the multi-spiral folding pattern of the above layout, in which the boundary of an spiral is defined at $\frac{1}{2}$ of the circumscribed radius of the surface.
\item Model 9. Multiple spiral folding, with $n = 3$; this model takes into account the multi-spiral folding pattern with 3 spirals with boundaries at $\frac{1}{3}$ and $\frac{2}{3}$  of the circumscribed radius of the surface.
\item Model 10. Multiple spiral folding, with $n = 4$; this considers the multi-spiral folding pattern with 4 spirals with boundaries at $\frac{1}{4}$,  $\frac{1}{2}$ and $\frac{3}{4}$  of the circumscribed radius of the surface.
\item Model 11. Multiple spiral folding, with $n = 2$; this model is inspired by the model 8 in which the boundary of the mid spiral is defined at $\frac{2}{3}$ of the circumscribed radius of the surface. The boundary is located far from to the core polygon.
\item Model 12. Multiple spiral folding, with $n = 2$; this model is inspired by the above formulation in which the boundary of the mid spiral is defined at $\frac{1}{3}$ of the circumscribed radius of the surface. In this model, compared to the above formulation, the boundary of the spiral is located close to the core polygon.
\end{itemize}

In the above-mentioned configurations, we used the following parameters: $N = 6$, the inscribed circle radius of the planar membrane $R_{f} = 120$ mm, the radius of the core circumcircle $r_0 = 15$ mm., the focal length $a =70$ mm., and the membrane thickness $t= 0.1$ mm. Furthermore, we used paper as material for experimentations, and the above-mentioned dimensions were decided by judging the amenability of experimenting with paper at manoeuvrable scale. Also, we used 25 cases of feasible loads with masses in the range $[0, 215.24]$ \si{g}, which translates into $[0, 2110.78]$ \si{mN} by the standard gravity $9.80665$ \si{m/s^2}. For each type of the above-mentioned models, the unfolding behaviour by tensile deployment was repeated three times independently. Thus, taking into account the above configurations, we conducted 36 unfolding experiments for each type of applied load. Considering other types of materials relevant to solar sail operation and construction is left for future work in our agenda. Fine tuning of the above membrane parameters is out of the scope of this paper.

\subsection{Results}

In order to show the deployment characteristics of planar membranes, Fig. \ref{rplanar} shows the performance of the deployed radius after loads are applied (tensile deployment). In this figure, the $x$-axis denotes the magnitude of the applied load, and the $y$-axis denotes the magnitude of the deployed radius (the higher values are desirable). The vertical bars in Fig. \ref{rplanar} denote the lower and upper bounds over three independent unfolding experiments. Basically, since the effective deployment of the folded membrane with a small load is our aim, the folding patterns showing high deployed radius and small applied weights are preferable. By observing the overall trend of Fig. \ref{rplanar}, we can note that it is of a landscape with concave geometry, with positive gradient, as a function of the applied load. As such, reasonably small increments in load are shown to translate into proportional increments of deployed radius, specially at small and intermediate load intervals (around $(0, 1000]$ mN.), implying the smooth transition and deployment performance. Also, we observed that the final deployed radius of model 1 to model 6 lies around the range $[108, 110]$ mm. with mean 118.2 \si{mm}. Although the models are strain-free with equilibrium at a radius of 110 mm., we found that models were unable to reach such equilibrium as shown by the deployment behaviours portrayed by Fig. \ref{deploy1} - Fig. \ref{deploy6}. We believe the reason behind this phenomenon is due to the effect of plastic deformation of creased paper, which shifts the deployment equilibrium to a radius less than 110 mm. This observation implies that additional force is necessary to overcome the plastic deformation from paper.

Furthermore, to ease visualization of our results, Fig. \ref{rplanar}-(a) (Fig. \ref{rplanar}-(b)) shows the deployment of planar membranes with single (multi) spiral patterns. By observing Fig. \ref{rplanar}, we note the following facts:

\begin{figure*}[t]
\begin{center}
     \subfigure[Model 1, Model 3, Model 5]{\includegraphics[width=\plasize\textwidth]{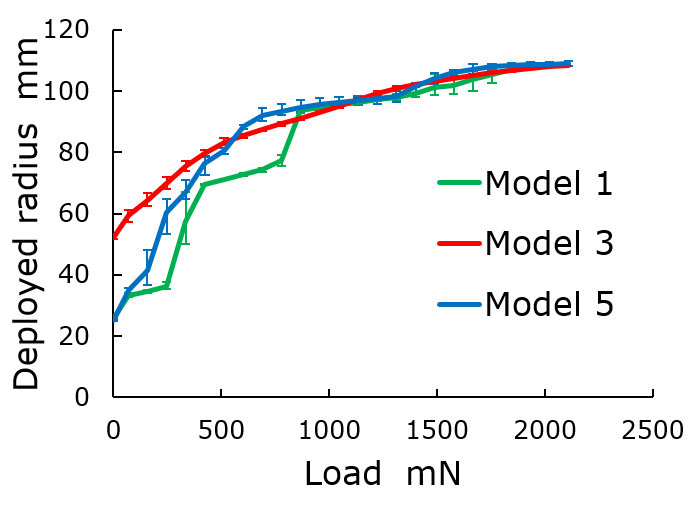}}
     \subfigure[Model 2, Model 4, Model 6]{\includegraphics[width=\plasize\textwidth]{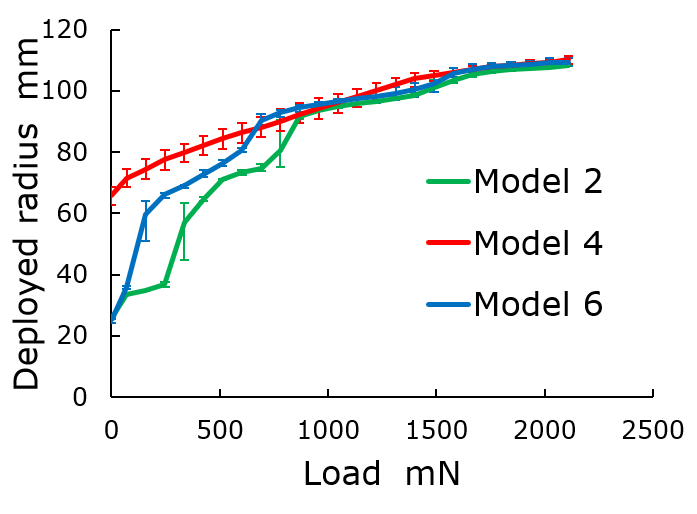}}
\end{center}
  \caption{Comparison of the deployment performance of planar membranes.}
  \label{rplanar}
\end{figure*}

\begin{figure}[h]
\centering
\footnotesize
\stackunder[5pt]{\includegraphics[width=\mosizeb\textwidth]{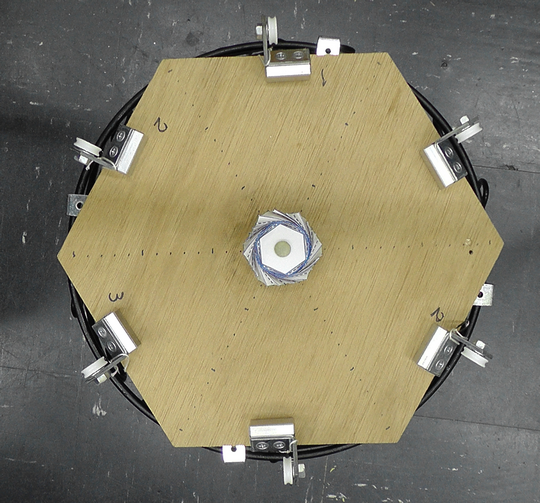}}{}
\hspace{0.1cm}
\stackunder[5pt]{\includegraphics[width=\mosizeb\textwidth]{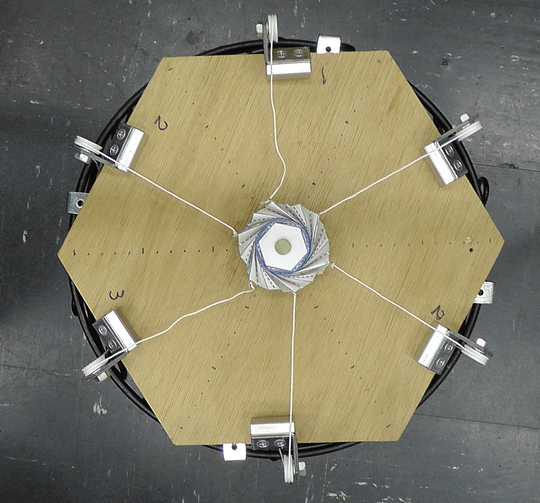}}{}
\hspace{0.1cm}
\stackunder[5pt]{\includegraphics[width=\mosizeb\textwidth]{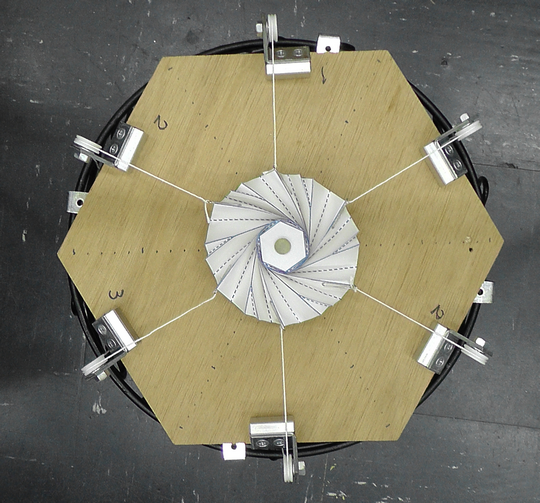}}{}
\hspace{0.1cm}
\stackunder[5pt]{\includegraphics[width=\mosizeb\textwidth]{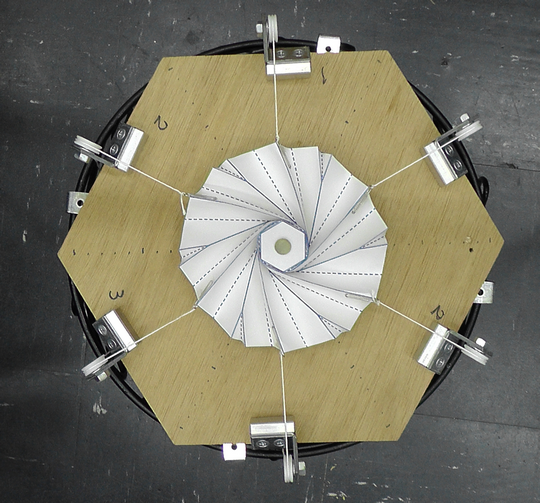}}{}
\hspace{0.1cm}
\stackunder[5pt]{\includegraphics[width=\mosizeb\textwidth]{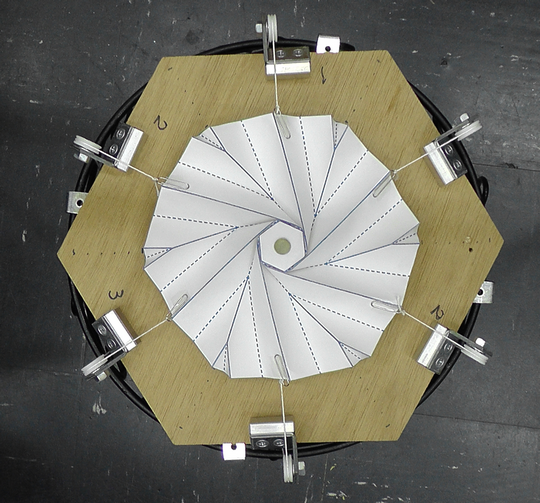}}{}
%\hspace{0.1cm}
\\
\stackunder[5pt]{\includegraphics[width=\mosizeb\textwidth]{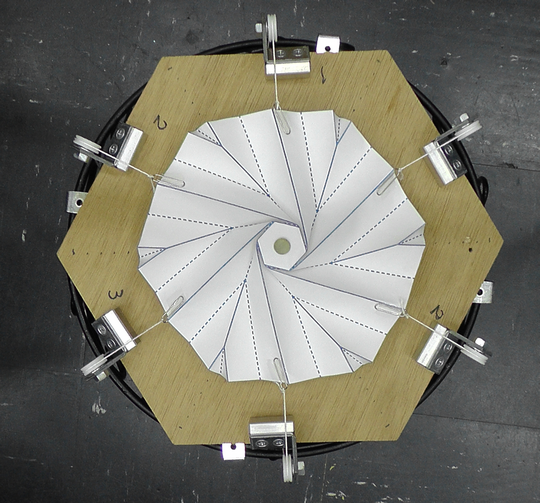}}{}
\hspace{0.1cm}
\stackunder[5pt]{\includegraphics[width=\mosizeb\textwidth]{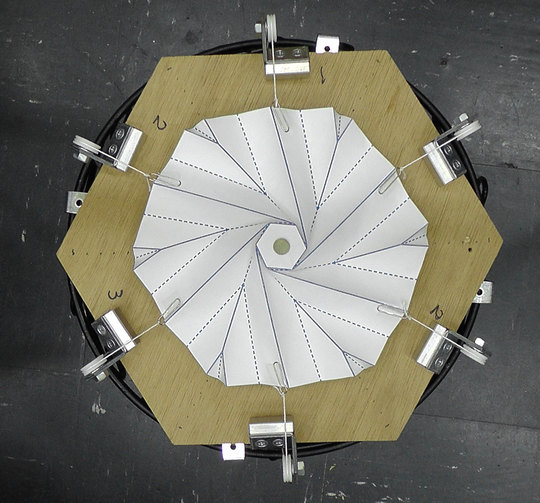}}{}
\hspace{0.1cm}
\stackunder[5pt]{\includegraphics[width=\mosizeb\textwidth]{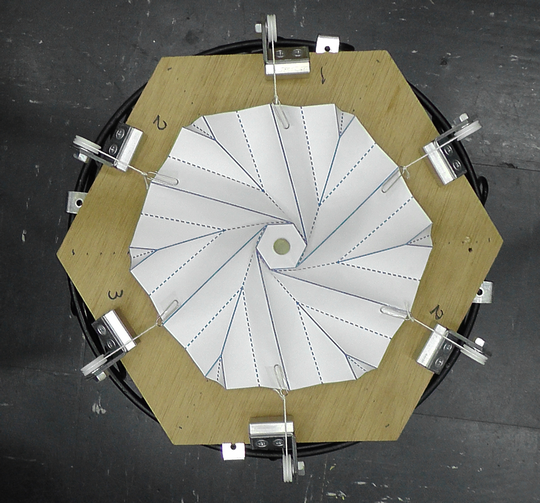}}{}
\hspace{0.1cm}%
\stackunder[5pt]{\includegraphics[width=\mosizeb\textwidth]{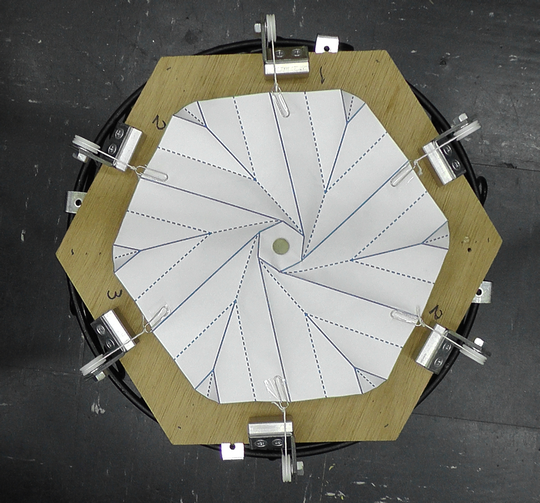}}{}
\hspace{0.1cm}%
\stackunder[5pt]{\includegraphics[width=\mosizeb\textwidth]{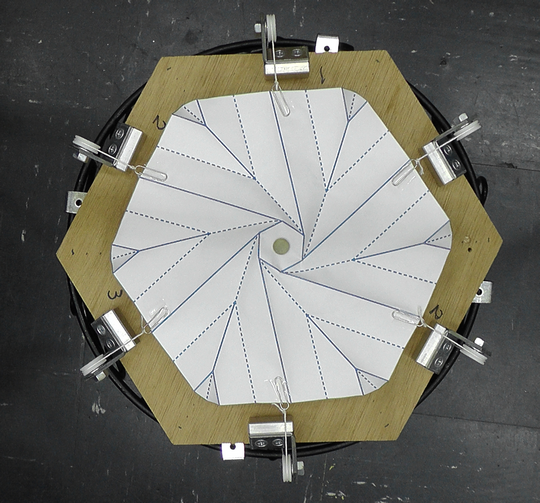}}{}
\caption{Deployment behaviour of Model 1}
\label{deploy1}
\end{figure}

\begin{figure}[h]
\centering
\footnotesize
\stackunder[5pt]{\includegraphics[width=\mosizeb\textwidth]{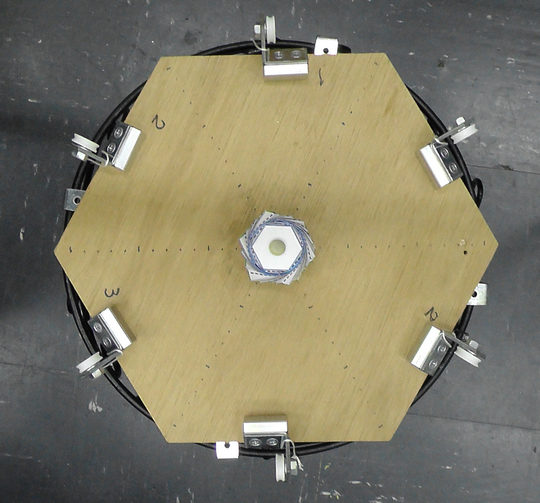}}{}
\hspace{0.1cm}
\stackunder[5pt]{\includegraphics[width=\mosizeb\textwidth]{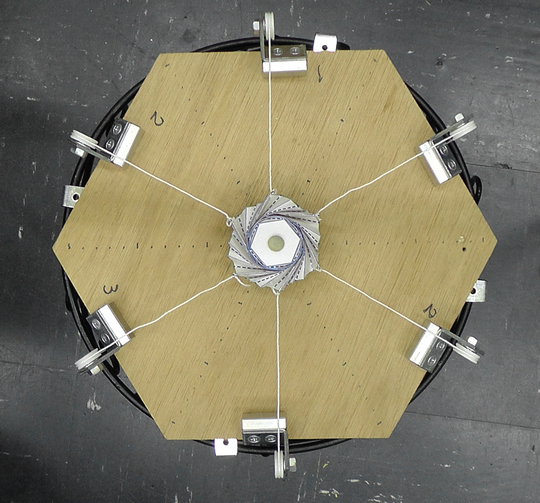}}{}
\hspace{0.1cm}
\stackunder[5pt]{\includegraphics[width=\mosizeb\textwidth]{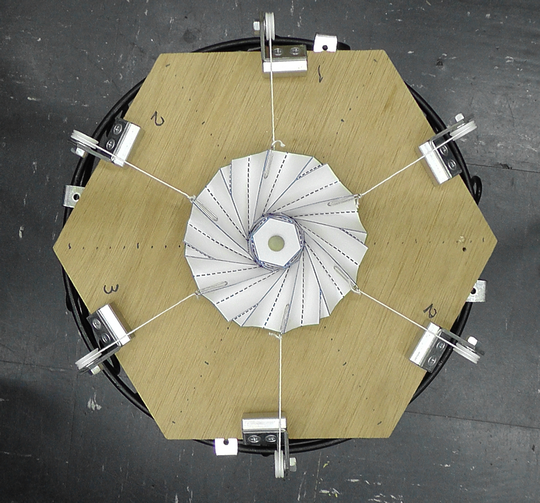}}{}
\hspace{0.1cm}
\stackunder[5pt]{\includegraphics[width=\mosizeb\textwidth]{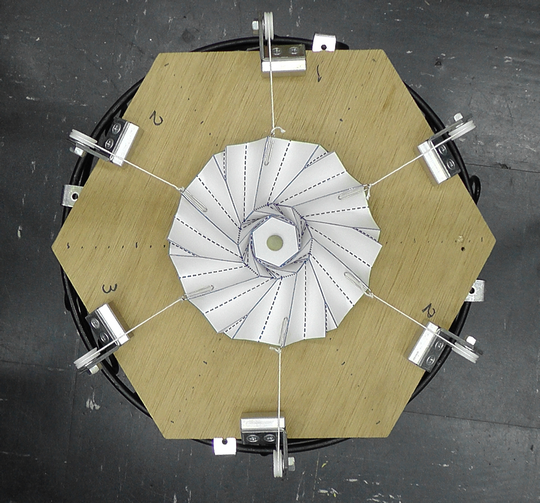}}{}
\hspace{0.1cm}
\stackunder[5pt]{\includegraphics[width=\mosizeb\textwidth]{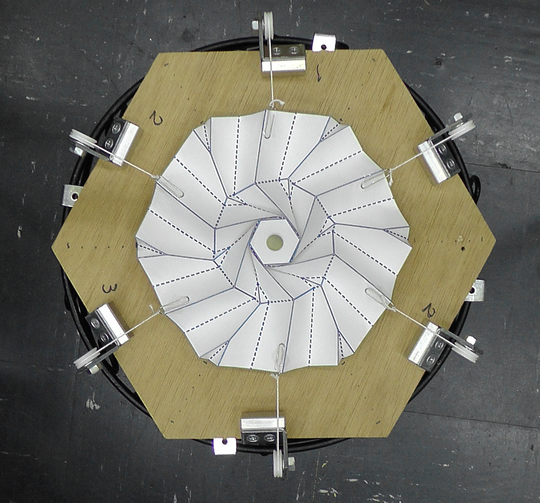}}{}
%\hspace{0.1cm}
\\
\stackunder[5pt]{\includegraphics[width=\mosizeb\textwidth]{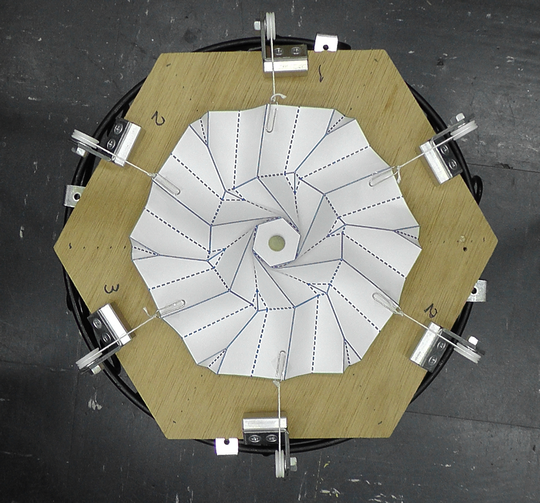}}{}
\hspace{0.1cm}
\stackunder[5pt]{\includegraphics[width=\mosizeb\textwidth]{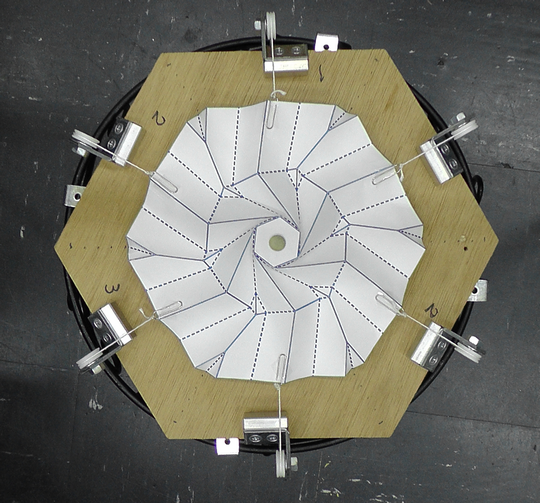}}{}
\hspace{0.1cm}
\stackunder[5pt]{\includegraphics[width=\mosizeb\textwidth]{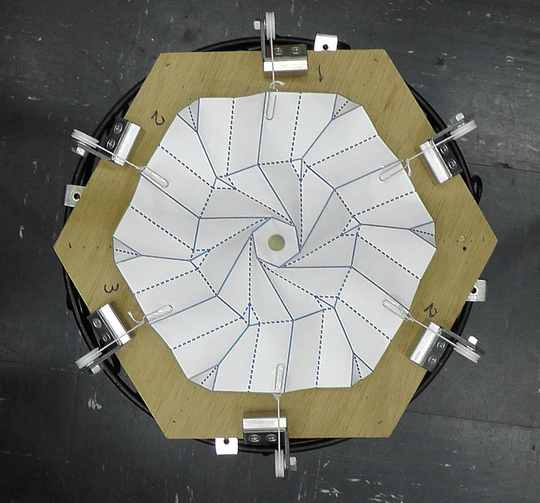}}{}
\hspace{0.1cm}%
\stackunder[5pt]{\includegraphics[width=\mosizeb\textwidth]{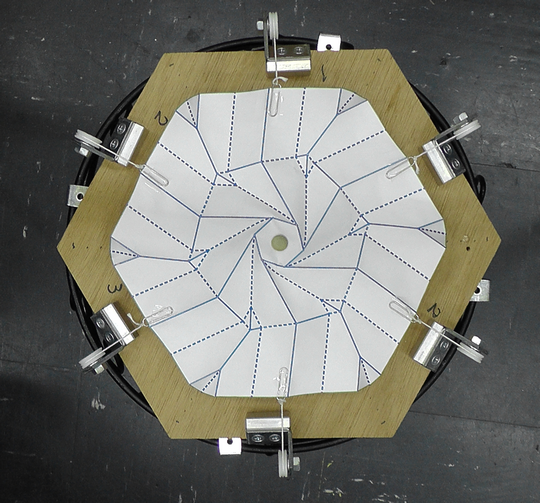}}{}
\hspace{0.1cm}%
\stackunder[5pt]{\includegraphics[width=\mosizeb\textwidth]{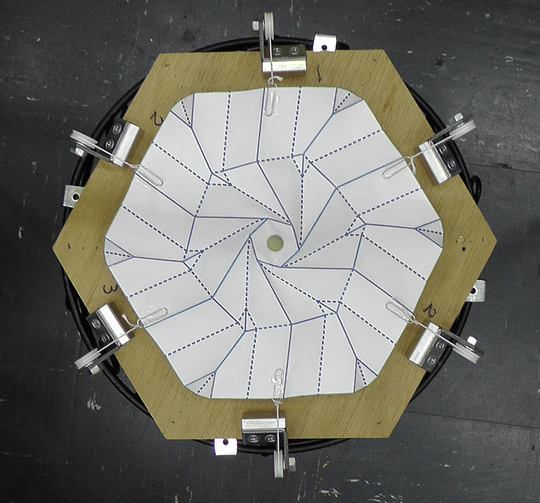}}{}
\caption{Deployment behaviour of Model 2}
\label{deploy2}
\end{figure}

\begin{figure}[h]
\centering
\footnotesize
\stackunder[5pt]{\includegraphics[width=\mosizeb\textwidth]{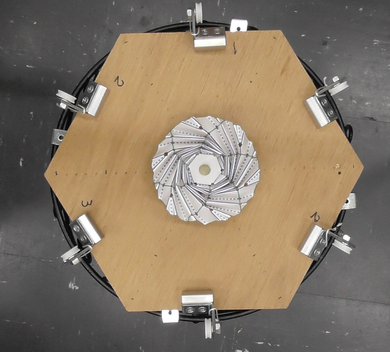}}{}
\hspace{0.1cm}
\stackunder[5pt]{\includegraphics[width=\mosizeb\textwidth]{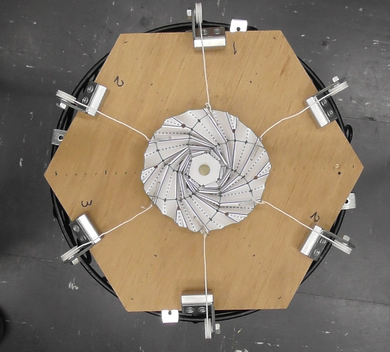}}{}
\hspace{0.1cm}
\stackunder[5pt]{\includegraphics[width=\mosizeb\textwidth]{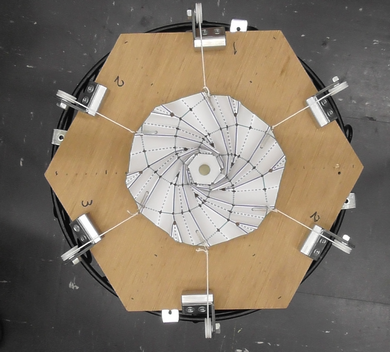}}{}
\hspace{0.1cm}
\stackunder[5pt]{\includegraphics[width=\mosizeb\textwidth]{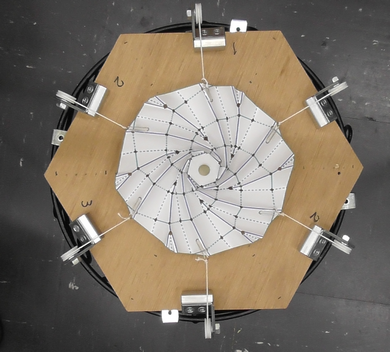}}{}
\hspace{0.1cm}
\stackunder[5pt]{\includegraphics[width=\mosizeb\textwidth]{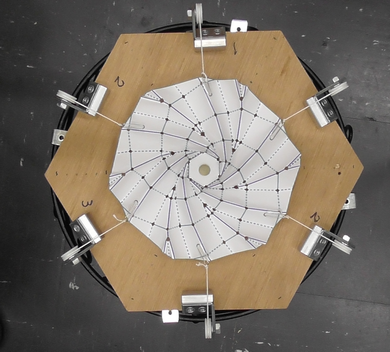}}{}
%\hspace{0.1cm}
\\
\stackunder[5pt]{\includegraphics[width=\mosizeb\textwidth]{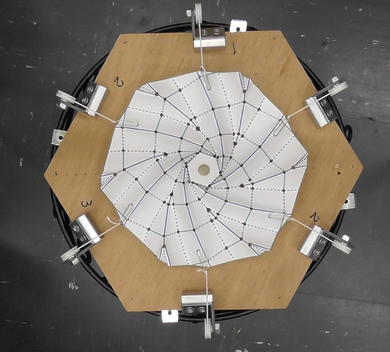}}{}
\hspace{0.1cm}
\stackunder[5pt]{\includegraphics[width=\mosizeb\textwidth]{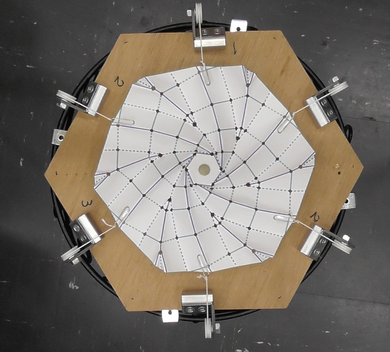}}{}
\hspace{0.1cm}
\stackunder[5pt]{\includegraphics[width=\mosizeb\textwidth]{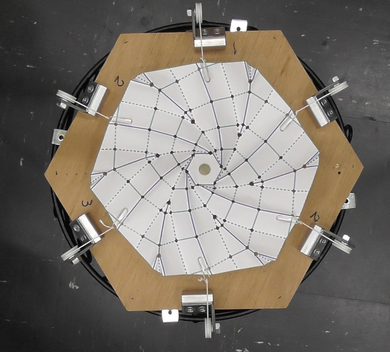}}{}
\hspace{0.1cm}%
\stackunder[5pt]{\includegraphics[width=\mosizeb\textwidth]{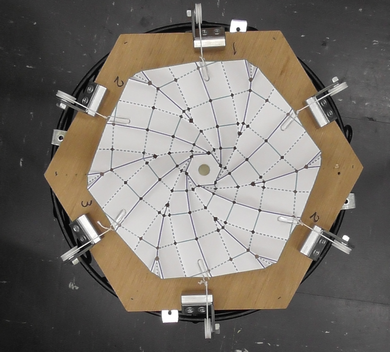}}{}
\hspace{0.1cm}%
\stackunder[5pt]{\includegraphics[width=\mosizeb\textwidth]{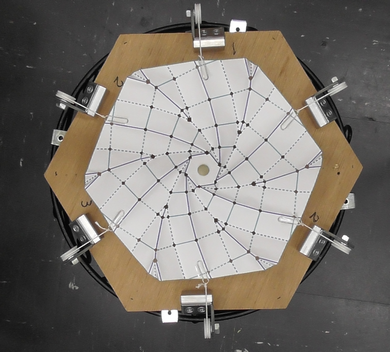}}{}
\caption{Deployment behaviour of Model 3}
\label{deploy3}
\end{figure}

\begin{figure}[h]
\centering
\footnotesize
\stackunder[5pt]{\includegraphics[width=\mosizeb\textwidth]{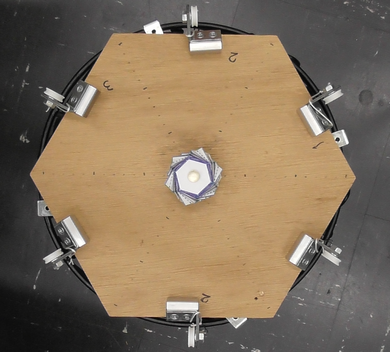}}{}
\hspace{0.1cm}
\stackunder[5pt]{\includegraphics[width=\mosizeb\textwidth]{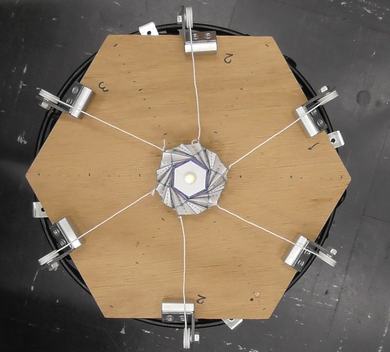}}{}
\hspace{0.1cm}
\stackunder[5pt]{\includegraphics[width=\mosizeb\textwidth]{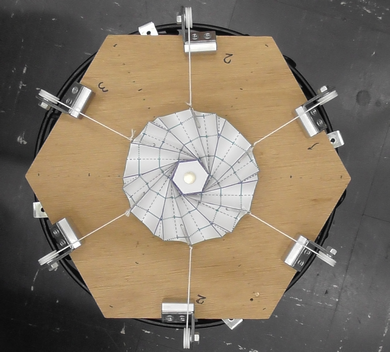}}{}
\hspace{0.1cm}
\stackunder[5pt]{\includegraphics[width=\mosizeb\textwidth]{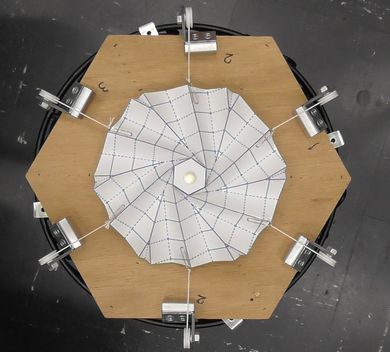}}{}
\hspace{0.1cm}
\stackunder[5pt]{\includegraphics[width=\mosizeb\textwidth]{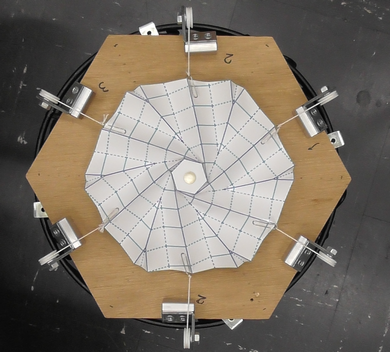}}{}
%\hspace{0.1cm}
\\
\stackunder[5pt]{\includegraphics[width=\mosizeb\textwidth]{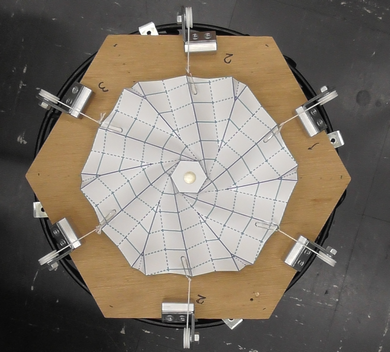}}{}
\hspace{0.1cm}
\stackunder[5pt]{\includegraphics[width=\mosizeb\textwidth]{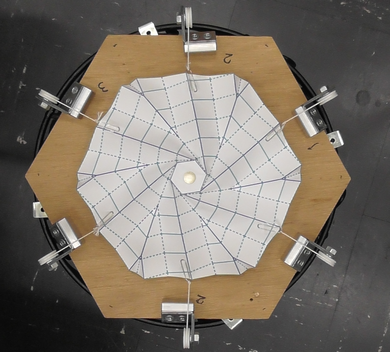}}{}
\hspace{0.1cm}
\stackunder[5pt]{\includegraphics[width=\mosizeb\textwidth]{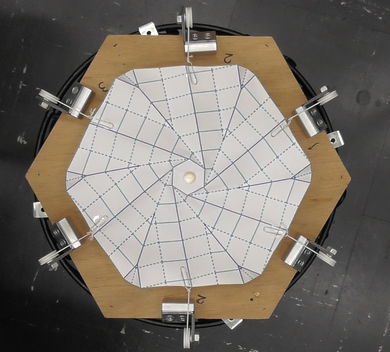}}{}
\hspace{0.1cm}%
\stackunder[5pt]{\includegraphics[width=\mosizeb\textwidth]{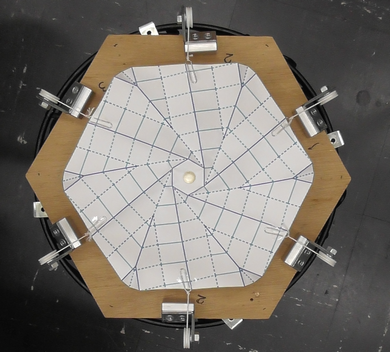}}{}
\hspace{0.1cm}%
\stackunder[5pt]{\includegraphics[width=\mosizeb\textwidth]{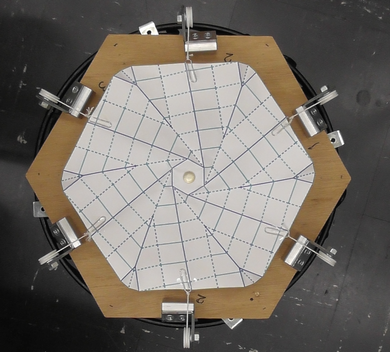}}{}
\caption{Deployment behaviour of Model 6}
\label{deploy6}
\end{figure}

\begin{itemize}
  \item the folding pattern of planar membranes follow a polynomial-like behaviour with positive gradient.
  \item folding layouts corresponding to model 3 and model 4 offer competitive performance, achieving improved radius of deployment for smaller applied loads.
\end{itemize}

We believe the above observations occur due to the fact of model 3 and model 4 are able to be folded without sharp folds, in which the absence of plastic deformation of paper is expected to induce in smoother deployment.

\begin{figure*}[t]
\begin{center}
     \subfigure[Model 1, Model 3, Model 5]{\includegraphics[width=\plasizec\textwidth]{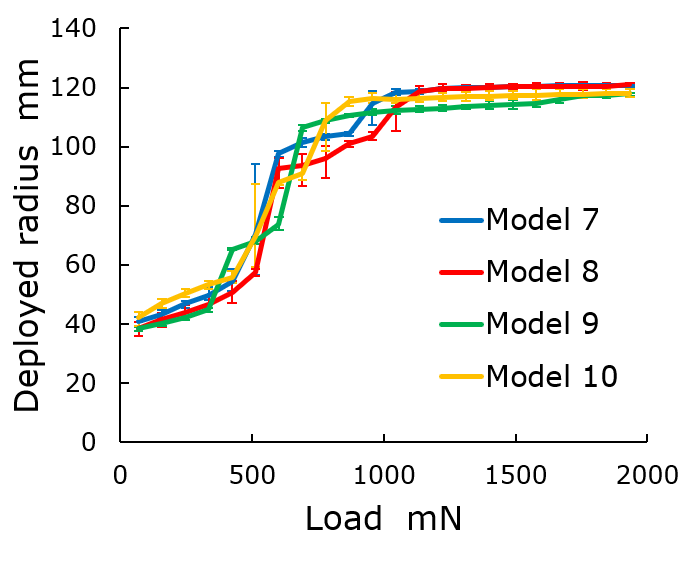}}
     \subfigure[Model 2, Model 4, Model 6]{\includegraphics[width=\plasizec\textwidth]{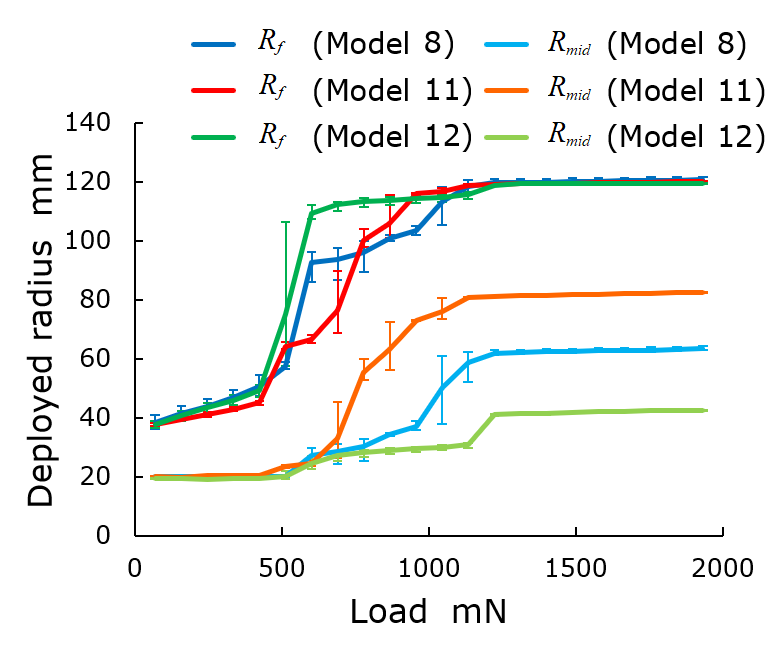}}
\end{center}
  \caption{Comparison of the deployment performance of curved membranes..}
  \label{rcurve}
\end{figure*}

Likewise, Fig. \ref{rcurve} shows the comparison of the deployment behaviour of curved membranes. In the same line of the above-mentioned results, the $x$-axis ($y$-axis) denotes the magnitude of the applied load (deployed radius). The vertical bars in Fig. \ref{rcurve} denote the lower and upper bounds on deployed radius over three independent unfolding experiments. For the sake of easing the visualization of our results, we show the deployment performance of multi-spiral folding patterns such as model 7 - model 10 in Fig. \ref{rcurve}-(a). We also show the deployment performance of model 8, model 11 and model 12 at the radius of the inscribed circle of the membrane surface ($R_f$), and at the radius of the inscribed circle of the polygon that forms the inner spiral fold pattern ($R_{mid}$). Note that model 7 - model 10 enable the formation of 1, 2, 3 and 4 spirals, respectively. And model 8, model 11 and model 12 are topologically similar (using 2 spirals), yet the boundary between spirals are in distinct locations.

Due to the above observations, Fig. \ref{rcurve}-(a) becomes relevant to evaluate the effect of number of spirals on the deployment behaviour, and Fig. \ref{rcurve}-(b) is relevant to evaluate the effect of the location of the boundary between spirals. By observing the overall trend of deployment behaviour of curved membranes in Fig. \ref{rcurve}, we can note that it belongs to the family of s-shaped logistic-like geometries as a function of the applied load. Here, reasonably small increments in the applied loads within the range $[500, 1000]$ \si{mN} are shown to translate into steep increments of deployed radius. Also, we observed that the final deployed radius of model 7 to model 12 lies around the range $[117, 120]$ mm. We believe the reason behind being unable to reach the equilibrium at a radius of 120 mm. is due to the effect of plastic deformation of creased paper. Also, by observing the results of Fig. \ref{rcurve}, we note the following facts:

\begin{itemize}
  \item There is no considerable difference among the deployment behaviours of multiple spirals.
  \item The deployment behavior of all curved membrane surfaces was such that the deployment progressed more rapidly within intermediate load domains (around [500 - 1000] mN) compared to smaller loads around (0 - 500] mN.
  \item The unfolding of the surface was confirmed in all cases with loads being close to 2000 mN., with observed differences in the order of centimeters.
\end{itemize}

Compared to the results from the planar layout (Fig. \ref{rplanar}), in which the unfolding progresses in a concave polynomial-like fashion, the curved surfaces shows a logistic-like unfolding behaviour with rapid progression above certain applied load. We believe the above occurs due to the torque generated during deployment is effectively distributed along the curved membrane, increasing the angle of rotation for any applied load and suppressing friction inhibiting the fast membrane expansion. Thus from the viewpoint of stability, the multi-spiral folding of the membrane shows the favorable deployment.

The above observations offer the foundational insights on the effect of multi-spirals in the deployment performance of flat and curved membranes. Studying the performance behaviour using tailored membrane materials and deployment approaches, as well as extending the applicability to structures with rigid origami configurations are potential to further advance towards the versatile membranes for solar sail applications.

\section{Conclusion}

In this paper, we have proposed the schemes to package flat and curved membranes of finite thickness by using folding with multiple spirals. The effect of the membrane thickness during storage is taken into account during folding. Whereas the spiral folding of planar membranes targets two-dimensional surfaces, the spiral folding of curved membranes targets three-dimensional surfaces modeled by a dome-shaped structure. The governing equations rendering the folding lines are determined by the juxtaposition of concentric spirals and by accommodating the folding lines to consider the membranes' thickness.

Our experiments using the tensile deployment based on force applied in the radial and circumferential directions of paper-based membranes have shown that (1) planar and multi-spiral folding approaches offered the improved deployment performance in comparison to other spiral approaches, and (2) deployment in curved surfaces progressed rapidly within a finite intermediate load domain.

Potential future work in our agenda involves the experimentation with other materials, the consideration of curvature variation in curved surfaces, and the modeling of modular folding membranes. We also aim to investigate the applications in developing efficient packaging schemes for tailored membranes, and designing protective membranes (transportation and health science). We believe our approach may find use in developing compact packaging mechanisms of membranes with finite thickness in space applications and manufacturing.

\bibliography{mybibfile}

\end{document}